\documentclass{pasj01}
\Received{$\langle$reception date$\rangle$}
\Accepted{$\langle$acception date$\rangle$}
\Published{$\langle$publication date$\rangle$}

\usepackage{xspace}
\usepackage{framed}
\usepackage{txfonts}
\usepackage{epstopdf}
\usepackage{xcolor}
\usepackage{rotating}
\usepackage{xspace}
\usepackage{ulem}
\usepackage{url}
\usepackage{comment}
\usepackage{multirow}

\newcommand{\bx}{\boldsymbol{x}}
\newcommand{\br}{\boldsymbol{r}}
\newcommand{\bk}{\boldsymbol{k}}
\newcommand{\bu}{\boldsymbol{u}}
\newcommand{\bq}{\boldsymbol{q}}

\newcommand{\deltab}{\delta_\mathrm{b}}
\newcommand{\deltabzero}{\delta_{\mathrm{b},0}}
\newcommand{\hMpci}{h\,\mathrm{Mpc}^{-1}}
\newcommand{\hiMpc}{h^{-1}\,\mathrm{Mpc}}

\newcommand{\hiGpc}{h^{-1}\,\mathrm{Gpc}}
\newcommand{\kny}{k_\mathrm{Ny}}
\newcommand{\Lbox}{L_\mathrm{box}}

\newcommand{\be}{\begin{eqnarray}}
\newcommand{\ee}{\end{eqnarray}}
\newcommand{\dfrac}[2]{\frac{\displaystyle #1}{\displaystyle #2}}

\begin{document}

\title{GINKAKU: Scalable Cosmological Structure Formation Simulation Code and Post-processing Pipeline}

\author{Takahiro \textsc{Nishimichi},\altaffilmark{1,2,3,4}\footnotemark[*]
Satoshi \textsc{Tanaka},\altaffilmark{5,2} and
Kohji \textsc{Yoshikawa}\altaffilmark{6}}

\altaffiltext{1}{Department of Astrophysics and Atmospheric Sciences, Faculty of Science, Kyoto Sangyo University, Motoyama, Kamigamo, Kita-ku, Kyoto 603-8555, Japan}
\altaffiltext{2}{Center for Gravitational Physics and Quantum Information, Yukawa Institute for Theoretical Physics, Kyoto University,
Kyoto 606-8502, Japan}
\altaffiltext{3}{Kavli Institute for the Physics and Mathematics of the Universe (WPI),
The University of Tokyo Institutes for Advanced Study (UTIAS),
The University of Tokyo, Kashiwa, Chiba 277-8583, Japan}
\altaffiltext{4}{RIKEN Center for Advanced Intelligence Project, 1-4-1 Nihonbashi, Chuo-ku, Tokyo 103-0027, Japan}
\altaffiltext{5}{Graduate School of Social Data Science, Hitotsubashi University, 2-1, Naka, Kunitachi 186-8601, Tokyo, Japan}
\altaffiltext{6}{Center for Computational Sciences, University of Tsukuba, 1-1-1 Tennodai, Tsukuba, Ibaraki 305-8577, Japan}
\email{nishimichi@cc.kyoto-su.ac.jp}


\maketitle

\begin{abstract}
We introduce GINKAKU, a new cosmological $N$-body code developed for the Dark Quest II (DQ2) simulation campaign and designed for controlled ensemble production across the cosmological model space demanded by next-generation galaxy surveys, including massive neutrinos and clustering dark energy. Built on top of the FDPS framework, GINKAKU couples a TreePM gravity solver with a linear-response treatment of external source terms representing components that are not evolved as $N$-body particles, formulated in the $N$-body gauge. This design consistently incorporates massive-neutrino perturbations, general-relativistic corrections in the $N$-body gauge, radiation perturbations at early times, and dark-energy clustering with non-unit effective sound speed at the linear level, while preserving Newtonian particle dynamics on subhorizon scales. The code is validated through internal convergence studies and through cross-comparisons with GADGET, PKDGRAV3 and RAMSES on shared initial conditions: differences between simulation codes in the nonlinear power spectrum can be brought below the $\sim 1\%$ level by tuning the internal accuracy parameters, and we identify a production-grade fiducial setting that achieves this control at modest computational cost. We apply GINKAKU to an initial set of DQ2 production runs --- eight cosmological models with $3{,}000^3$ particles in cubic boxes of up to $4\,\hiGpc$ --- processed by a renewed post-processing pipeline that reduces the inter-resolution spread of the halo mass function to $\lesssim 1\%$ and includes a halo-shape measurement extension supporting intrinsic-alignment statistics. The scale-dependent-growth cosmologies reproduce the expected nonlinear signatures of massive neutrinos and clustering dark energy, demonstrating its suitability for emulator-scale production. A total matter power spectrum emulator constructed from these runs is presented in an accompanying paper.
\end{abstract}

\section{Introduction} \label{sec:intro}
The large-scale structure of the Universe is an active research field in cosmology. It offers a means to study the fundamental physics governing the evolution of the Universe and its contents, including the still-unknown nature of dark matter and dark energy, which together make up about $95\%$ of its total energy budget. Large-scale observational programs have been conducted, are underway, or are planned in the near future to unveil the nature of these components within and beyond the standard cosmological model. These programs include the Sloan Digital Sky Survey\footnote{\url{https://www.sdss.org/}}, the Kilo-Degree Survey\footnote{\url{https://kids.strw.leidenuniv.nl/}}, the Dark Energy Survey\footnote{\url{https://darkenergysurvey.org}}, the Hyper Suprime-Cam\footnote{\url{https://hsc.mtk.nao.ac.jp/ssp/}} and the Prime Focus Spectrograph\footnote{\url{https://pfs.ipmu.jp/index.html}} surveys on the Subaru Telescope, the Dark Energy Spectroscopic Instrument Survey\footnote{\url{https://www.desi.lbl.gov/}}, the Vera C. Rubin Observatory\footnote{\url{https://www.lsst.org/}}, the Euclid mission\footnote{\url{https://www.euclid-ec.org/}} and the Nancy Grace Roman Space Telescope\footnote{\url{https://roman.gsfc.nasa.gov/}}.

In particular, a major goal of large-scale-structure studies is to subject the spatially flat $\Lambda$-cold dark matter framework (hereafter $\Lambda$CDM) to more detailed tests. Here $\Lambda$CDM denotes the standard minimal cosmological model with the simplest forms of dark components: cold dark matter and a cosmological constant. Although this model has successfully explained a wide range of observational results, including cosmic microwave background (CMB) anisotropies \citep{WMAP9cosmo,Planck2018cosmo,ACTDR4cosmo,SPT3G:2021}, galaxy clustering~\citep{cole:2005aa,parkinson12,SDSS_final_Alam}, and weak gravitational lensing~\citep{Hikage19,Hamana2020-bh,Dalal2023-HSCY3,Li2023-HSCY3,More2023-HSCY3,Amon2022-DESY3,Secco2022-DESY3,DESY3:3x2pt,KiDS1000:Heymans,Asgari2021-KiDS}, there are several observational tensions that may indicate its limitations. These include the Hubble tension (see e.g.,\ \cite{Riess22} for a representative direct distance-ladder measurement), and the $S_8$ tension~(e.g., \cite{KiDS1000:Heymans,Asgari2021-KiDS}), both indicating parameter values inconsistent with those derived from CMB assuming $\Lambda$CDM~\citep{Planck2018cosmo}. Extensive tests of $\Lambda$CDM from multiple observational perspectives are therefore crucial for assessing the validity of the model.

Realizing the statistical power of these surveys for testing $\Lambda$CDM requires theoretical predictions of comparable precision. Structure formation, however, is an inherently nonlinear process governed primarily by gravity. To trace the evolution of tiny primordial fluctuations, one must account for the effects of self-gravity within an expanding Universe. Analytical methods based on perturbative expansion are applicable only in the weakly nonlinear regime, ensuring that higher-order corrections remain subdominant~\citep{bernardeau02}. However, in reality, observable luminous astronomical objects, such as galaxies, form in highly overdense regions with density contrasts far exceeding unity. Consequently, numerical simulations become indispensable tools for studying the formation and evolution of such structures. On these scales, various non-gravitational ``baryonic'' processes become important, including phenomena such as gas cooling and feedback, and further complicate theoretical predictions.

A practical approach to mitigate the impact of baryonic physics when predicting the clustering properties of galaxies in a controlled theoretical framework is to treat galaxies as ``biased tracers'' of the underlying density field, which is largely governed by invisible dark matter (\cite{kaiser84}; see also \cite{Desjacques18} for a comprehensive review). We can decompose the prediction of galaxy statistics into a sequence of relations among different tracer fields. For instance, the halo model approach regards dark matter halos, which are gravitationally bound systems formed around density peaks, as the fundamental building blocks~\citep{seljak:2000uq,ma:2000lr,scoccimarro:2001fj,Cooray02}. These halos are themselves biased tracers of the underlying matter fluctuations, with the bias depending primarily on halo mass \citep{mo96,2001MNRAS.323....1S,Tinker10} and also on secondary halo properties such as age, concentration, ellipticity, and spin~\citep{gao:2005fk,2007MNRAS.377L...5G,2008ApJ...687...12D,2008MNRAS.389.1419L,faltenbacher:2010lr,2019MNRAS.482.1900H}. Since the formation of halos is driven primarily by gravity, gravity-only simulations can provide robust predictions for halo abundances and clustering properties. The remaining question concerns the connection between halos and galaxies. This connection can vary among different types of galaxies and is typically modeled statistically with adjustable parameters, which can be constrained by comparison with observations~\citep{peacock:2000qy,berlind:2002kx,2005ApJ...633..791Z,conroy06,2011ApJ...736...59Z,2018ARA&A..56..435W}.

These gravity-only simulations are typically realized by $N$-body methods, which are powerful tools for studying the dynamics of collisionless, self-gravitating systems.
They are among the most widely used means of resolving structures on small scales, where nonlinearity plays a prominent role in cosmological structure formation. However, the computational cost of calculating self-gravitational interactions becomes a bottleneck as the number of particles, $N_\mathrm{p}$, increases. In the naive direct method, this cost scales as $\mathcal{O}(N_\mathrm{p}^2)$. To address this challenge, more efficient algorithms have been developed over several decades. These include the Barnes-Hut tree method~(Tree; \cite{Barnes1986-rj}), the Particle-Mesh method~(PM; \cite{hockney81}), the Particle-Particle Particle-Mesh method~(P$^3$M; \cite{hockney81}), TreePM~\citep{Xu1995-wj,Bagla2002}, and Fast Multipole Method~(FMM; \cite{Greengard1987-ao,Dehnen2000-sb,Dehnen2002-ek}). These techniques enable simulations with enormous numbers of particles within practical wall-clock time.
Furthermore, acceleration techniques, implemented at either the hardware or software level, have significantly contributed to the realization of large-scale particle simulations. Examples include application-specific integrated circuits such as GRAPE~(GRAvity PipE; \cite{Ito1990-yi,Makino1998-yv}), the Phantom-GRAPE library\footnote{\url{https://bitbucket.org/kohji/phantom-grape/}}~\citep{Nitadori2006-ek,2012NewA...17...82T,2013NewA...19...74T}, which exploits Single Instruction / Multiple Data (SIMD) instructions, graphics processing units (GPUs; \cite{Nyland2004-fc,Elsen2006-zj}), and field-programmable gate arrays~(FPGAs; \cite{Lienhart2002-ne,Hamada2009-ru}).

Large-scale $N$-body simulations have become increasingly accessible, thanks to publicly available codes such as GADGET-2\footnote{\url{https://wwwmpa.mpa-garching.mpg.de/gadget/}}~\citep{gadget2} and GADGET-4\footnote{\url{https://wwwmpa.mpa-garching.mpg.de/gadget4/}}~\citep{gadget4} based on TreePM, PKDGRAV3\footnote{\url{https://bitbucket.org/dpotter/pkdgrav3/}}~\citep{potter2017} and SWIFT\footnote{\url{https://swift.strw.leidenuniv.nl/}}~\citep{Schaller2024-SWIFT} based on FMM, and the hybrid HACC code~\citep{Habib2016-qq}, among others. In parallel, the Framework for Developing Particle Simulators~(FDPS\footnote{\url{https://github.com/FDPS}}; \cite{2016PASJ...68...54I,2018PASJ...70...70N}) has made it easier to write efficient, scalable code for massively parallel supercomputers with flexible combinations of various algorithms. Thanks to these technological advancements, $N$-body simulations have reached the scale of more than one trillion \citep{potter2017, Ishiyama2021-wx}, and even more than ten trillion particles \citep{Ishiyama2022-uf} in the context of large-scale structure formation simulations.

Building on these technological advances, large ensembles of cosmological simulations have become a key infrastructure for a new class of statistical methods. In recent years, the concept of simulation-based inference (SBI) has gained prominence across various fields of science and engineering. In cosmology, a major development is the use of emulation, a technique that approximates computationally expensive simulators with faster statistical surrogate models known as emulators.
This approach enables repeated likelihood evaluations and parameter inference that would otherwise be computationally prohibitive~\citep{Heitmann06,Habib07}. Emulation has found widespread application in cosmology, with different research groups developing emulators for various summary statistics or random fields. Notable examples include the Coyote Universe~\citep{Coyote1,Coyote2,Coyote3,Coyote_ex,Kwan13,2015ApJ...810...35K}, PkANN~\citep{PkANN1,PkANN2}, Mira-Titan Universe~\citep{MiraTitan1,MiraTitan2,MiraTitan3,MiraTitan4}, MassiveNuS~\citep{MassiveNuS}, the Aemulus Project~\citep{Aemulus1,Aemulus2,Aemulus3,Aemulus4,Aemulus5}, Dark Quest~\citep{2019ApJ...884...29N,2020PhRvD.102f3504K}, EuclidEmulator~\citep{EuclidEmu1,EuclidEmu2}, field-level emulators~\citep{NECOLA,Kodi_Ramanah_2020} based on Quijote Simulations~\citep{Quijote}, BACCO simulations~\citep{BACCO,BACCO_bias,BACCO_baryon}, AbacusCosmos~\citep{AbacusCosmos,2018MNRAS.tmp.2206W} and its upgraded program, AbacusSummit~\citep{AbacusSummit}, the Goku emulators~\citep{Goku,GokuN}, Aletheia~\citep{Aletheia}, and the CSST cosmological emulator series~\citep{CSST-emu-I,CSST-emu-II,CSST-emu-III}. Recent developments have seen some of these emulators being effectively applied to cosmological inference problems, where they are confronted with observational data related to large-scale structures~\citep{2020MNRAS.492.2872W,Miyatake22-emu2,2022PhRvD.105h3517K,Aemulus5,Yuan22-clustering}.

In addition to emulation, other strategies for SBI have emerged in recent discussions. These include methods such as Bayesian initial-condition reconstruction based on Hamiltonian Monte Carlo~\citep{Jasche2013-BORG,2019A&A...621A..69R,2022MNRAS.509.3194P}, direct optimization of a parameterized posterior~\citep{2019arXiv190104454S}, initial condition reconstruction based on differentiable particle-mesh simulators~\citep{2021A&C....3700505M,2021arXiv210412864M,Li2022-pmwd,Lanzieri2022-JaxPM}, and the Cosmological Evidence Modelling approach proposed by~\citet{2019MNRAS.490.1870L,2022MNRAS.509.1779L}. While these innovative techniques have partly mitigated the challenges arising from the curse of dimensionality, it is essential to emphasize that a substantial number of simulations, conducted with precise numerical accuracy and encompassing sufficiently large volumes, remain crucial across all of these approaches---both as training data for emulators and as forward models within the inference frameworks above.

Motivated by this continuing need for large, high-fidelity simulation suites,
this work introduces GINKAKU and the associated numerical pipeline for the ongoing Dark Quest simulation campaign\footnote{\url{http://darkquestcosmology.github.io/}}. The primary objective of this project is to calibrate various fundamental summary statistics related to dark matter halos, employing a suite of $N$-body simulations. In its initial phase, Dark Quest~I (DQ1 hereafter)~\citep{2019ApJ...884...29N}, simulations were carried out for 101 distinct cosmological models within the six-parameter $w$CDM cosmology framework. L-GADGET2, a variant of GADGET-2 used for the original Millennium Simulation and tailored for very large $N$ but sharing the same TreePM gravity solver, was used in DQ1 with internal parameters calibrated in earlier work~\citep{nishimichi:2007lr,Valageas11a,Takahashi12}. These simulations employed $N_\mathrm{p}=2,048^3$ particles, contained within periodic comoving boxes with side lengths of $1$ or $2\,\hiGpc$. Based on these simulations, a software tool called DarkEmulator was developed specifically for halos with mass $\gtrsim 10^{12}\,h^{-1}M_\odot$. DarkEmulator not only serves as a valuable resource for modeling these halos but also, when combined with a model linking halos and galaxies, functions as a theoretical template for galaxy clustering and galaxy--galaxy lensing statistics~\citep{2020PhRvD.102f3504K,Miyatake22-emu1}.

The objective of the second phase of this campaign, Dark Quest II (DQ2), is to broaden the model space by incorporating crucial physical effects, such as the suppression of small-scale perturbations due to massive neutrinos and the clustering of dark energy. Interest in such non-standard dark-energy scenarios has been further sharpened by recent baryon acoustic oscillation (BAO) measurements from the Dark Energy Spectroscopic Instrument (DESI). When combined with cosmic microwave background and Type Ia supernova data from samples such as Pantheon+~\citep{Scolnic2022-PantheonPlus,Brout2022-PantheonPlus}, Union3~\citep{Rubin2023-Union3}, and the Dark Energy Survey five-year sample~\citep{DES2024-Y5SNe}, these measurements show a preference for an evolving dark-energy equation of state at the few-$\sigma$ level (DR1: \cite{DESI_DR1}; DR2: \cite{DESI_DR2}); these results motivate simulation infrastructure that can routinely accommodate non-standard dark-energy scenarios, including both modified background expansion histories and perturbations in the dark-energy component. Simultaneously, we aim to enhance mass resolution in preparation for next-generation galaxy surveys such as the Subaru Prime Focus Spectrograph Survey and the Euclid mission. Achieving these goals necessitates the use of more efficient and versatile numerical codes. To address this, we have developed a new $N$-body simulator called GINKAKU (Gravitational INtegrator for Kinetic Analysis of the darK Universe), built on the foundation of FDPS. In addition to the core $N$-body gravitational solver, we have implemented the linear response method (section~\ref{subsec:linear}) to handle various effects that are challenging to address within the particle method, such as massive neutrinos, general-relativistic corrections, and the clustering of dark energy. We also describe our post-processing procedures, including halo identification under different mass definitions and efficient, accurate measurement of the matter power spectrum.

Beyond these individual ingredients, the overall design philosophy of GINKAKU differs from that of campaigns optimized for a single realization at the largest available particle count (e.g., the Uchuu simulation; \cite{Ishiyama2021-wx}): it is built for controlled ensemble production across a broad range of cosmological models, balancing numerical accuracy, computational throughput, and physical extensibility. The large-scale runs presented in this paper demonstrate the scalability of the code and define its production configuration, while the full simulation and validation suite together provide the numerical foundation for the matter power spectrum emulator DarkEmulator2 described in the accompanying paper \citep[hereafter ``the accompanying paper'']{Tanaka2026-emu}.

Although the present paper is the first dedicated description of GINKAKU, the code has already been in active use for some time, and a number of recently published or submitted analyses rely on simulation products generated with it~\citep{Peron25,Ishikawa25,Enomoto25,Pu24}. Two of these deserve particular mention. The first is the Beyond-2pt mock-data challenge~\citep{Beyond2pt24}, a community-wide, parameter-masked (blind) analysis challenge designed to validate the modeling of emerging beyond-two-point galaxy-clustering statistics, for which GINKAKU provided the underlying mock data. The second is the small-scale analysis of the Hyper Suprime-Cam Year~3 cosmic-shear two-point correlations~\citep{Terasawa24}, which exploits DarkEmulator2 trained on GINKAKU simulations. The aim of the present paper is to document, in one place, the numerical methods, validation tests, and reference settings that underlie these existing applications and the forthcoming emulator-scale simulation campaigns.

The paper is structured as follows. In section~\ref{sec:preliminaries}, we summarize the basic equations used in our numerical implementation. We then describe both the initial-condition generator and the simulator in section~\ref{sec:code}. Section~\ref{sec:accuracy} presents a convergence study of the matter power spectrum with respect to the internal accuracy parameters, including a comparison with other publicly available simulation codes to further validate the new code. Section~\ref{sec:postprocess} then describes our post-processing procedures, presenting our halo catalogs and statistical-analysis codes along with a comparison between the DQ1 and DQ2 post-processing pipelines.
Section~\ref{sec:productruns} presents the DQ2 production runs with $N_{\mathrm{p}}=3{,}000^{3}$ particles, including the fiducial, massive-neutrino, and clustering-dark-energy models used to demonstrate the production configuration.
Section~\ref{sec:summary} summarizes the paper. The performance characteristics of the code are examined in Appendix~\ref{sec:scaling}: we study how it scales with problem size and processor count, including scaling comparisons against other codes and large-$N$ runs on the Fugaku supercomputer. Appendix~\ref{sec:nu_approx} compares the approximate treatments for neutrinos. Appendix~\ref{sec:ultimate} returns to the residual code-to-code differences identified in section~\ref{subsec:codes} and shows that they progressively vanish as the accuracy parameters are tightened, tracing the remaining offsets to the timestep scheme. Appendix~\ref{sec:mass_def} discusses the halo mass function for different mass definitions. Finally, Appendix~\ref{sec:SSM} demonstrates how our code can conduct simulations influenced by super-survey modes.

\section{Preliminaries} \label{sec:preliminaries}
We summarize the formulation underlying our simulations, including the gauge choice and the treatment of physical effects that cannot be directly tracked by the simulation particles.

\subsection{Newtonian equations} \label{subsec:newton}
We summarize the standard evolution equations in the Newtonian limit in an expanding universe to introduce our notation. We work with comoving coordinates, denoted as $\bx = \br /a$, where $\br$ represents the physical coordinates, and $a$ is the scale factor. Following the super-comoving formulation \citep{Doroshkevich1980,1997astro.ph.10043Q}, we make a convenient choice for the peculiar velocity and potential variable as follows:
\be
&&\bu = a^2\dot{\bx},\label{eq:u_velocity}\\
&&\Psi(\bx) = a\left[\phi(\bx)+\frac{1}{2}a\ddot{a}x^2\right].
\ee
Here, the dot represents the time derivative, $(\mathrm{d}/\mathrm{d}t)$, and $\phi$ represents the Newtonian gravitational potential. Finally, we opt for the time variable $\tau = \log a$. With these variables, the Newtonian equations can be summarized as:
\be
&&\frac{\mathrm{d}\bx}{\mathrm{d}\tau} = \frac{\bu}{a\mathcal{H}(\tau)},\\
&&\frac{\mathrm{d}\bu}{\mathrm{d}\tau} = - \frac{\nabla \Psi(\bx)}{\mathcal{H}(\tau)},\label{eq:Euler}\\
&&\nabla^2\Psi(\bx) =
\frac{3}{2}\Omega_{\mathrm{m},0}\mathcal{H}_0^2\,
\delta(\bx),\label{eq:Poisson_newt}
\ee
where $\mathcal{H} = \dot{a} = aH$ denotes the conformal expansion rate, with the present value denoted as $\mathcal{H}_0$, 
$\Omega_\mathrm{m,0}$ represents the present matter density in units of the critical density,
and $\delta = \rho/\bar{\rho}-1$ indicates the total matter overdensity.

\subsection{General relativistic corrections and the gauge choice} \label{subsec:GR}
Since we are aiming at providing theoretical templates across a broad dynamic range, it is crucial to adequately address gauge issues, particularly on large scales near the horizon. Furthermore, it should be noted that massive neutrinos exhibit relativistic behavior during the initial stages of structure formation, especially if their masses are light. Following \citet{LesgourguesPastor2006}, the redshift at which a neutrino species with mass $m_\nu$ transitions to non-relativistic is given by:
\be
1+z_\mathrm{nr} = \frac{m_{\nu}}{5.28\times 10^{-4}\,\mathrm{eV}}.
\ee
This transition occurs around the time at which we initialize the simulations (see section~\ref{subsec:IC}) for the neutrino-mass range of cosmological interest, $\sum m_\nu \lesssim 0.2\,\mathrm{eV}$ (e.g., \cite{Planck2018cosmo}). Therefore, a proper account of relativistic corrections is necessary to evolve the system with such ingredients.

\begin{figure}[t]
 \centering
 \includegraphics[width=\linewidth]{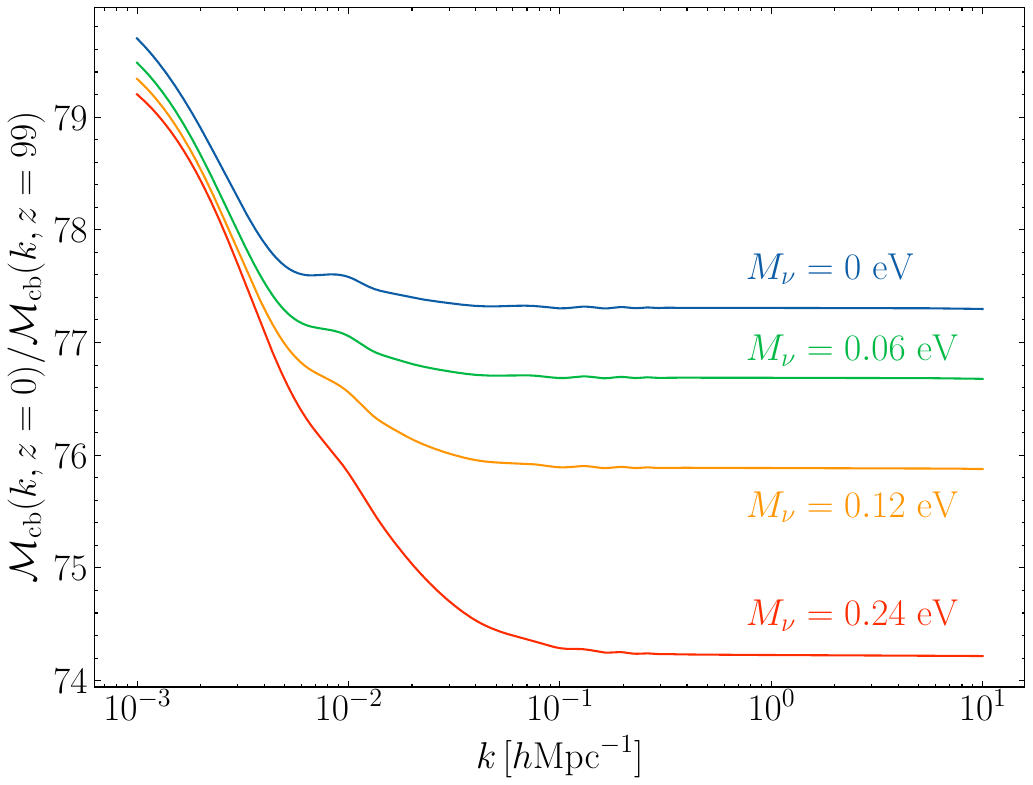}
 \caption{Scale dependence of the linear growth factor from $z=99$ to $z=0$. We plot the ratio of the CDM+baryon transfer function at these two redshifts for four models with different neutrino masses as indicated in the figure legend. The transfer function is computed in the $N$-body gauge as explained in the text. {Alt text: Four overlapping curves of the linear growth ratio plotted against wavenumber. The curves separate at intermediate scales as the neutrino mass increases, reflecting larger free-streaming suppression for heavier neutrinos.}\label{fig:linear_growth}}
\end{figure}

Standard Newtonian simulations are typically initialized with the linear matter transfer function computed by a Boltzmann solver in the synchronous gauge. The relation between such Newtonian setups and the full relativistic description has been clarified in a series of works \citep{2011PhRvD..83l3505C,2015PhRvD..92l3517F,2016JCAP...07..053A}, leading to several gauge-aware prescriptions for cosmological simulations. Among these, we adopt the $N$-body gauge \citep{2015PhRvD..92l3517F}, in which the relativistic deformation of space is absent, simplifying the interpretation of simulation snapshots\footnote{$N$-body motion gauge \citep{2016JCAP...09..031F} is another convenient choice, with no correction needed during Newtonian simulations, but at the cost of extra post-processing to shift the final locations of the particles to interpret them in the observable coordinate system.}. That is, the number-counting density estimate precisely corresponds to the comoving density. Furthermore, the Poisson equation remains unchanged from equation~(\ref{eq:Poisson_newt}). On the other hand, the geodesic equation for massive particles is influenced by a new correction term in addition to the right-hand side of the Newtonian Euler equation~(\ref{eq:Euler}). On sub-horizon scales this correction is at most of order $(\mathcal{H}/k)^2$ relative to the Newtonian gravitational force, and it vanishes identically in a universe containing only pressureless matter and a cosmological constant. Together with the other physical effects not captured by the simulation particles, it will be absorbed into the external source term $S_\mathrm{ext}$ introduced below in equation~(\ref{eq:Poisson}). Thus, the $N$-body gauge proves to be a suitable choice for $N$-body simulations, requiring only minor adjustments to standard Newtonian codes.

We recast the GR correction in the geodesic equation as an effective potential by modifying the Poisson equation as
\be
\nabla^2\Psi(\bx) =
\frac{3}{2}\Omega_{\mathrm{cb},0}\mathcal{H}_0^2\,
\delta_\mathrm{part}(\bx) + S_\mathrm{ext}(\bx).\label{eq:Poisson}
\ee
This adjustment ensures that the form of the geodesic equation remains identical to the Newtonian one. Now, the peculiar gravitational potential, $\Psi$, is interpreted as an effective potential sourced not only by the inhomogeneities of massive particles in the simulation box, $\delta_\mathrm{part}$, but also by any other physical effects not accounted for by simulation particles, denoted by $S_\mathrm{ext}$. This term encompasses both the geometric GR correction discussed above and the linear perturbations of physical components that are not represented as simulation particles, including massive or massless neutrinos \citep{2019JCAP...03..022T}, radiation \citep{2017MNRAS.466L..68B,2017MNRAS.470..303A}, and dark-energy perturbations \citep{2019JCAP...08..013D}. In the above, we have also replaced $\Omega_{\mathrm{m},0}\,\delta(\bx)$ in the original Newtonian Poisson equation~(\ref{eq:Poisson_newt}) with $\Omega_{\mathrm{cb},0}\,\delta_\mathrm{part}(\bx)$, because our simulation particles represent only the CDM-plus-baryon (``cb'') fluid.

Once this extra term is calculated and incorporated into the source term of the Poisson equation, we can use the standard Newtonian equations for updating velocities (i.e., the kick operation) and particle positions (the drift operation).

We utilize the CLASS Boltzmann solver \citep{class1,class2} to compute the transfer functions, denoted as $\mathcal{M}_i = \delta_i
^{\mathrm{L}}/\zeta$, where $\zeta$ represents the primordial curvature perturbations, and $\delta_i^{\mathrm{L}}$ is the linear overdensity of the species labeled by $i$. Here, the curvature perturbation is assumed to be Gaussian and is characterized by its power spectrum
\be
\langle \zeta_{\bk} \zeta_{\bk'} \rangle = (2\pi)^3\delta_\mathrm{D}^3(\bk+\bk')P_\zeta(k),
\ee
with $\delta_\mathrm{D}$ being the Dirac delta function and $k=|\bk|$.
These transfer functions are then converted to the $N$-body gauge using the relation given explicitly by \citet{Brando2021}:
\be
\mathcal{M}_i^{(N\mathrm{-body})}(k) = \mathcal{M}_i^{(S/N)}(k) + \frac{3 \mathcal{H} (1+w_i)}{k^2}\mathcal{M}_{\theta,\mathrm{tot}}^{(S/N)}(k),
\label{eq:gauge}
\ee
where $\mathcal{M}_{\theta,\mathrm{tot}}$ is the transfer function for the total velocity divergence of all species, and $w_i$ denotes the equation-of-state parameter for species $i$. 
The superscript $S/N$ indicates that the quantity is evaluated in the synchronous or conformal Newtonian gauge. In practice, we run CLASS in the synchronous gauge and perform gauge conversion using equation~(\ref{eq:gauge}). We have verified that starting with the conformal Newtonian gauge yields identical results up to numerical accuracy.

\begin{figure}[t]
 \centering
 \includegraphics[width=\linewidth]{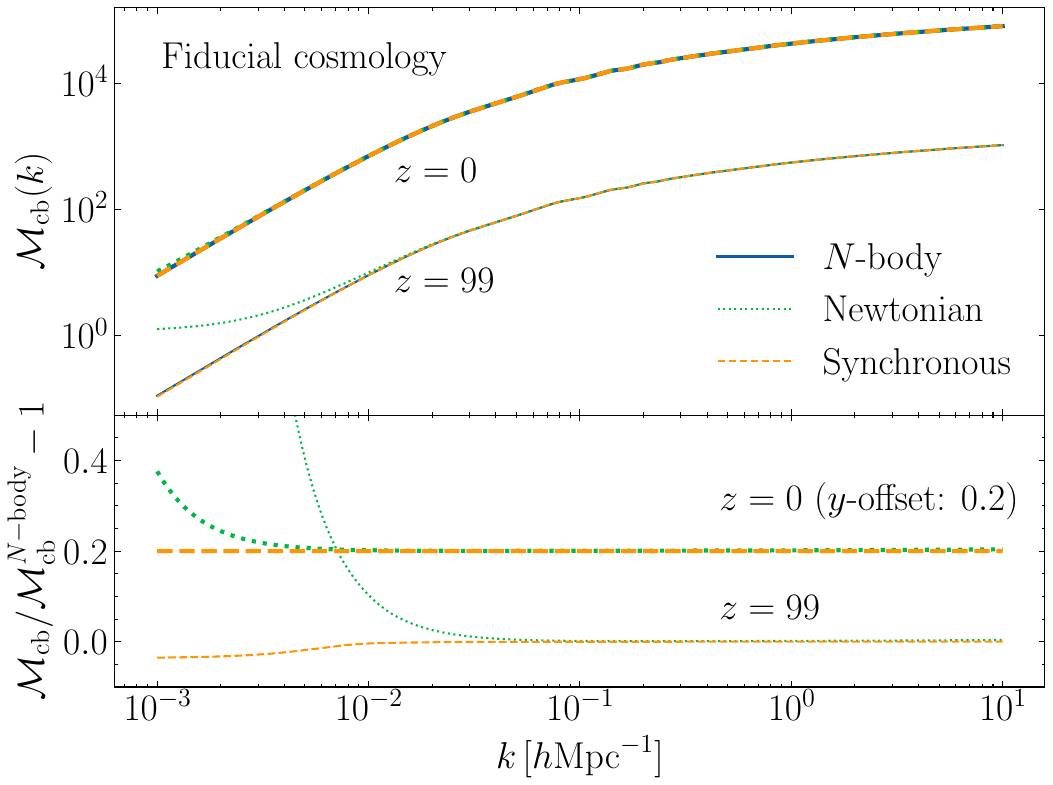}
 \caption{Gauge dependence of the cb transfer function. We show the cb transfer function in $N$-body (solid), conformal Newtonian (dotted), and synchronous (dashed) gauge at $z=0$ (thick) and $z=99$ (thin). We show the fractional difference from $N$-body gauge in the bottom panel. Note that the results at $z=0$ are vertically offset by $+0.2$ for clarity. We adopt here the fiducial cosmological model (see Table~\ref{tab:product_runs}). {Alt text: Two-panel transfer-function plot. The upper panel shows six curves for three gauges at two redshifts that overlap on small scales but diverge on large scales, with the synchronous and N-body gauges nearly indistinguishable at redshift zero. The lower panel shows four fractional differences from the N-body gauge, growing toward low wavenumber and larger at the higher redshift.}\label{fig:gauge}}
\end{figure}

We consistently employ the $N$-body gauge throughout this paper and subsequent ongoing analyses within the Dark Quest II project. As depicted in figure~\ref{fig:gauge}, differences between the gauges become pronounced primarily on large scales ($k\lesssim0.01\,\hMpci$), with greater significance observed at higher redshifts. Unless explicitly stated otherwise, the transfer functions and other statistical measures of the fluctuations are given in the $N$-body gauge.

\subsection{Linear implementation of the source term} 
\label{subsec:linear}
While incorporating these external-source effects at the nonlinear level is challenging, most of them primarily affect large scales, where a linear approximation suffices. Within this approximation, a straightforward approach to implementing these effects is by replacing $S_\mathrm{ext}(\bx)$ with its linear counterpart:
\be
\nabla^2\Psi(\bx) =
\frac{3}{2}\Omega_{\mathrm{cb},0}\mathcal{H}_0^2\,
\delta_\mathrm{part}(\bx) + S_\mathrm{ext}^{\mathrm{L}}(\bx).
\ee

To obtain $S_\mathrm{ext}^{\mathrm{L}}$, which represents the linearized version of the external source field, one can derive it from a Boltzmann solver. These solvers consistently incorporate all effects at the linear level, unlike in the time evolution followed in standard Newtonian $N$-body simulations. Therefore, in principle, one could extract it directly from a Boltzmann code by combining the relevant linear perturbation variables. However, we opt for a simpler method that requires no code modification. We observe that, once the cb transfer function is converted to the $N$-body gauge, its time evolution already reflects the cumulative effect of such sources, including radiation perturbations.

By comparing the cb transfer function at different epochs, we can reconstruct the total external source influencing the evolution with the help of the Newtonian evolution equation:
\begin{eqnarray}
&&\mathcal{M}_\mathrm{cb}''(k,\tau) + \left[1+\bigl(\ln \mathcal{H}(\tau)\bigr)'\right]\mathcal{M}_\mathrm{cb}'(k,\tau) \nonumber\\
&&\qquad= \frac{3}{2}\,\Omega_\mathrm{cb}(\tau)\mathcal{M}_\mathrm{cb}(k,\tau) + \frac{1}{a \,\mathcal{H}^2(\tau)} \mathcal{M}_S(k,\tau),
\label{eq:Mcb_evo}
\end{eqnarray}
where we denote by $'$ the derivative with respect to $\tau = \log a$. This equation is identical to the linearized evolution equation of density perturbations in the Newtonian limit if we ignore the second term on the right-hand side. The extra term $\mathcal{M}_S$ effectively represents the transfer function for the external source field, which affects the evolution of the cb density field.

With this, we can calculate the additional source term in the modified Poisson equation~(\ref{eq:Poisson}) as $S_\mathrm{ext}^{\mathrm{L}}=\mathcal{M}_S\zeta$. 
In practice, we first create a table of the cb transfer function, $\mathcal{M}_{\mathrm{cb}}$,
as a function of both $k$ and $\tau$, and then employ cubic spline interpolation of the tabulated numerical values to evaluate the derivative terms. Inserting these to equation~(\ref{eq:Mcb_evo}), we evaluate $\mathcal{M}_S$. Once $\mathcal{M}_S$ is determined, one can compute the correction to the Poisson equation using a specific random realization of the curvature perturbations, $\zeta$, which were used to set up the initial conditions:
\begin{eqnarray}
S_{\mathrm{ext},\bk}^{\mathrm{L}}(\tau) = \mathcal{M}_S(k,\tau)\,\zeta_{\bk}.
\end{eqnarray}
In practice, this term is incorporated into the Poisson solver for the Particle-Mesh (PM) force at every PM time step. 

In the context of structure formation involving massive neutrinos, a grid-based treatment of linear neutrino perturbations has been shown to accurately replicate the evolution of the total matter power spectrum, particularly when the sum of the neutrino masses is smaller than $\sim 0.5 \mathrm{eV}$ \citep{Brandbyge09}. However, a limitation arises in the evolution of neutrino perturbations, which is by construction at the linear order. This approach fails to capture the clustering of a small fraction of low-velocity neutrinos in the tail of the distribution function~\citep{2018MNRAS.481.1486B,2018JCAP...09..028B}, known to be confined within massive cluster regions and thus contributing to nonlinear growth. A potential solution to enhance accuracy in describing neutrino clustering on small scales could involve a hybrid approach, transitioning to particle realization mid-simulation \citep{2018MNRAS.481.1486B}.

The primary focus of this study is not on neutrino clustering itself, but rather on the impact of neutrinos on the fluctuations of the total matter field, dominated by the cb fluid, and the resulting impact on the clustering of biased tracers such as halos or galaxies. Therefore, linear methods suffice as long as we limit ourselves to models with low-mass neutrinos consistent with current observations. However, there exists a straightforward technique to enhance accuracy, partially recovering the nonlinear behavior of neutrino clustering, known as the linear response method \citep{2013MNRAS.428.3375A}. Operating within the framework of linear theory, this method computes the \textit{ratio} of the transfer functions, $\mathcal{M}_S / \mathcal{M}_\mathrm{cb}$, and multiplies it by the \textit{nonlinear} overdensity $\delta_\mathrm{cb}$ observed in the simulation to estimate the source field $S_\mathrm{ext}$.

The linear response method assumes that the Fourier phases of neutrino perturbations closely track those of the cb fluid, and that the amplitude of neutrino perturbations undergoes the same fractional amount of nonlinear growth as that of the cb fluid. Further rationale behind this method is detailed in the original paper. A practical advantage of the linear response method is that it does not require storing an additional field on the grid.

We employ the linear response approach in our simulation code. Practically, we define the ``boost'' factor as:
\be
B(k,\tau) = \frac{\frac{3}{2}\Omega_\mathrm{cb}(\tau) \mathcal{M}_\mathrm{cb}(k,\tau) + \frac{1}{a\,\mathcal{H}^2(\tau)}\mathcal{M}_S(k,\tau)}{\frac{3}{2}\Omega_\mathrm{cb}(\tau) \mathcal{M}_\mathrm{cb}(k,\tau)}. \label{eq:boost}
\ee
The denominator corresponds to the linear Newtonian gravitational source, while the numerator additionally includes the external-source contribution. This factor therefore estimates the relative impact of the external source field on the perturbations represented by the particle distribution in simulations. The boost factor is obtained from the transfer function table prepared prior to simulation and is multiplied by the Green's function of the Poisson solver to evaluate the effective potential. In the absence of the external source field, the boost factor equals unity, implying no correction to the evolution of the simulated structures.

In practical terms, the difference between the two methods described above -- the direct linear evaluation of the source term and the linear response method -- is not significant except in scenarios involving large neutrino masses. In most cases, the correction introduced by the boost factor becomes significant only on large scales where linear theory remains applicable. Under such circumstances, the relationship $(\mathcal{M}_S/\mathcal{M}_\mathrm{cb})\,\delta_\mathrm{cb} \simeq (\mathcal{M}_S/\mathcal{M}_\mathrm{cb})\,\delta_\mathrm{cb}^\mathrm{L} = S_\mathrm{ext}^{\mathrm{L}}$ holds, and the linear response method essentially recovers the direct linear method for the source field $S_\mathrm{ext}$. For the same reason, we can safely disregard the effect of $S_\mathrm{ext}$ in the computation of short-range tree force.

\begin{figure}[t]
 \centering
 \includegraphics[width=\linewidth]{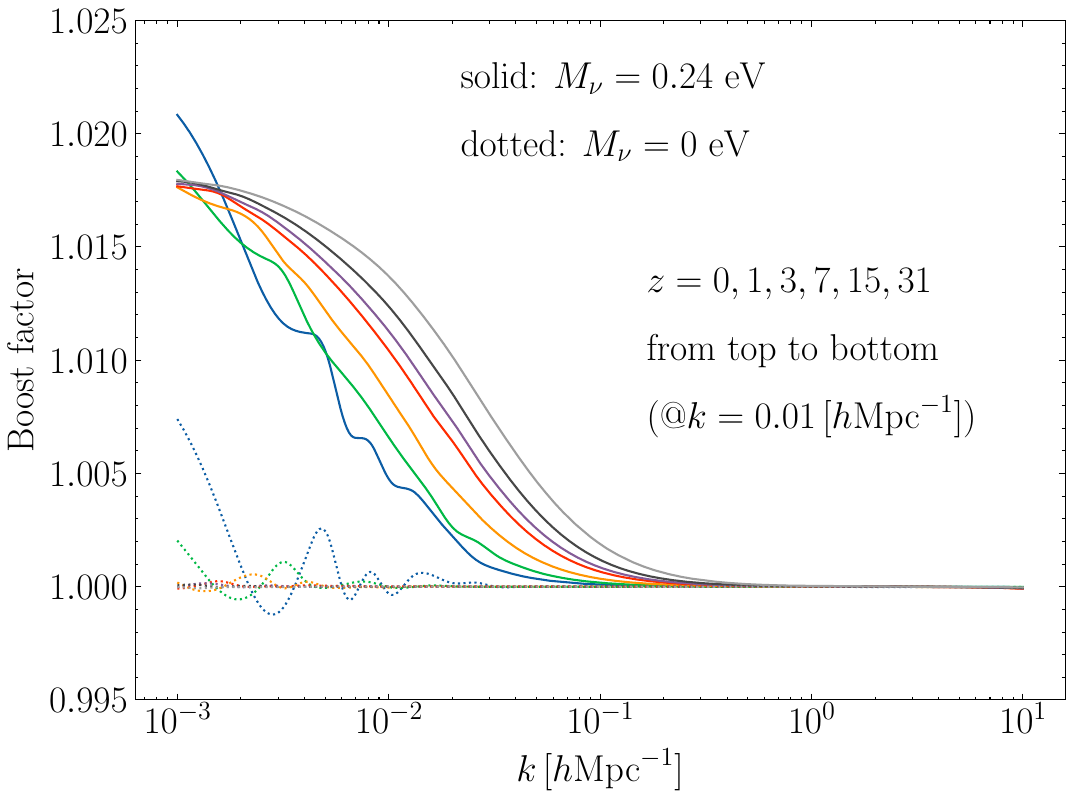}
 \caption{Boost factor at different redshifts for models with massive ($M_\nu=0.24\,\mathrm{eV}$, an exaggerated value well above current observational bounds, adopted here for demonstration purposes; solid) and massless (dotted) neutrinos. {Alt text: Plot of the boost factor against wavenumber for two cosmologies at several redshifts. The massive-neutrino curves rise above unity on large scales with redshift-dependent amplitude, while the massless-neutrino curves stay near unity with small large-scale oscillations.} \label{fig:boost_mnu}}
\end{figure}

We illustrate the boost factor for two neutrino-mass models at various redshifts in figure~\ref{fig:boost_mnu}. In the case of massive neutrinos, the boost factor surpasses unity on large scales (solid curves), indicating that neutrinos contribute to structure growth via gravitational instability on scales larger than their free-streaming scale. This introduces an additional source of gravitational forces on large scales, not accounted for when evaluating forces solely from the distribution of the simulation particles. The redshift dependence of the boost factor arises from the shift of the free-streaming scale to smaller scales over time.
The size of the bump observed in the lowest-$k$ regime roughly corresponds to the fraction of the mean neutrino density relative to the cb fluid. Conversely, the boost factor remains nearly unity across all depicted redshifts in models without massive neutrinos (dotted curves). The curves exhibit small oscillatory features on large scales, particularly at higher redshifts, attributed to residual perturbations in radiation. While these oscillations are minor, they persist on the largest scales. Over time, the integration of these small oscillations tends to cancel out, leaving a positive correction at $k\lesssim0.005\,\hMpci$. This explains the observed increase in linear growth even when $M_\nu=0\,\mathrm{eV}$ in figure~\ref{fig:linear_growth}. Similar oscillations are present in the solid lines for models with massive neutrinos at high redshifts ($z\gtrsim10$) by the same mechanism.

\begin{figure}[t]
 \centering
 \includegraphics[width=\linewidth]{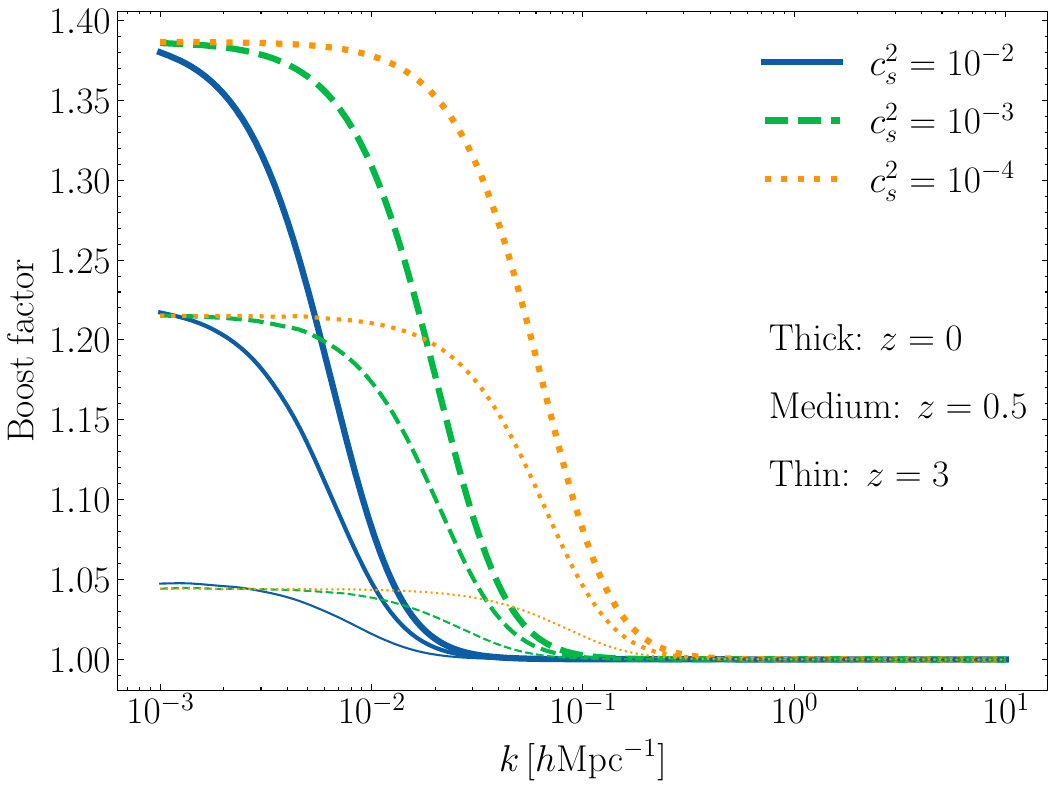}
 \caption{Boost factor at different redshifts for clustering dark energy models. Line style indicates the effective dark-energy sound speed (as labeled in the figure legend), with different sound speeds producing different magnitudes of dark-energy clustering; line width encodes the redshift. To exaggerate the impact, we consider flat cosmologies with a high dark energy density, $\Omega_\mathrm{de}=0.8$ and the equation-of-state parameter $w_\mathrm{de} = -0.6$, significantly deviating from cosmological constant ($w_\mathrm{de}=-1$). {Alt text: Plot of the boost factor against wavenumber for clustering dark-energy models at several redshifts. Departures from unity grow toward large scales and toward low redshift, and the scale of departure extends to larger wavenumber as the effective sound speed decreases.} \label{fig:boost_DE}}
\end{figure}

Figure~\ref{fig:boost_DE} shows the boost factor for models incorporating dark energy clustering, illustrating its contribution to the late-time growth of matter fluctuations. As expected, the effect becomes more pronounced at lower redshifts, corresponding to periods when the fraction of dark energy in the total energy budget becomes higher. The scale at which the boost factor deviates from unity depends on the square of the sound speed of dark energy, denoted as $c_\mathrm{s}^2$ in the figure legend. Models with a lower $c_\mathrm{s}^2$ can influence the growth of the cb fluid on smaller scales. Although these models represent somewhat extreme cases with $\Omega_\mathrm{de}=0.8$ and $w_\mathrm{de}=-0.6$, we will examine analogous extreme scenarios with the $N$-body code in section~\ref{subsubsec:clustering_de} (see also figure~\ref{fig:pk_ratio}). Through $N$-body simulations, we aim to investigate whether and how these large-scale corrections propagate to smaller scales, which are more relevant to galaxy observables, via nonlinear gravitational interactions (i.e., mode coupling between different scales).

\section{Code implementation} \label{sec:code}
We now describe in detail the actual implementation of the numerical codes developed to set up and perform simulations. Our notation for the internal parameters and their fiducial values is summarized in table~\ref{tab:sim_params}; each parameter is introduced where it is first defined in this section, and the convergence of the resulting simulations with respect to these parameters is assessed in section~\ref{sec:accuracy}.

\begin{table*}
\tbl{Internal parameters that control the accuracy of the simulations}{
\begin{tabular}{clcc}
\hline\noalign{\vskip3pt} 
\bfseries{Parameter} & Description & Equation & Fiducial value\\ 
\hline\noalign{\vskip3pt} 
$L_\mathrm{box}$ & Side length of the periodic comoving box & -- & --\\
$N_\mathrm{p}$   & Number of simulation particles & -- & --\\
$N_\mathrm{g}$ & Total number of PM grid points & -- & $2^3 N_\mathrm{p}$\\
$H_\mathrm{g}$ & Grid spacing & $L_\mathrm{box}/N_\mathrm{g}^{1/3}$ & --\\
$H_\mathrm{p}$ & Mean inter-particle distance per dimension & $L_\mathrm{box}/N_\mathrm{p}^{1/3}$ & --\\
$z_\mathrm{ini}$  & Starting redshift & -- & By solving Eq.~(\ref{eq:zini})\\
$\eta_\mathrm{tree}$ & Time-step parameter for the tree force & (\ref{eq:dtime_tree}) & $0.7$\\
$\eta_\mathrm{PM}$ & Time-step parameter for the Particle-Mesh force & (\ref{eq:dtime_PM}) & $0.125$\\
$\eta_\mathrm{global}$ & Optional parameter to multiply both the tree and PM time steps & -- & $1$\\ 
$\Delta\tau_\mathrm{max}$ & Maximum time-step size & (\ref{eq:dtime_PM}) & $0.03$\\
$\theta$ & Opening angle for the tree force & -- & $0.5$\\ 
$\epsilon_\mathrm{p}$ & Plummer softening length in units of $H_\mathrm{p}$ & -- & $0.0214$\\
$\epsilon_\mathrm{g}$ & Smoothing scale for the PM force in units of $H_\mathrm{g}$ & -- & $3$\\
$r_\mathrm{s}$ & Transition scale for the tree and PM force & $\epsilon_\mathrm{g}H_\mathrm{g}$ & --\\
\hline\noalign{\vskip3pt} 
\end{tabular}}
\label{tab:sim_params}
\end{table*}

\subsection{Initial condition generator} \label{subsec:IC}
Our initial condition generator is a custom implementation that builds on the Zel'dovich-approximation-based approach of N-GenIC~\citep{springel2015,gadget2} and the second-order Lagrangian perturbation theory (2LPT) scheme described in \citet{scoccimarro98}, as used in large-volume cosmological simulations by \citet{Valageas11a}. The use of 2LPT to displace particles from the regular pre-initial lattice, rather than the Zel'dovich approximation alone, suppresses the transient nonlinear decaying modes that arise from truncating the perturbation theory series at first order~\citep{crocce06a}. For the present work the generator has been substantially extended to support the scale-dependent linear growth factors required by the $N$-body gauge and by the presence of massive neutrinos. Rather than extending the perturbative solution to scale-dependent growth in a fully consistent manner, we adopt the simplified prescription described below.

We utilize the standard Einstein-de Sitter solution, albeit with the linear density field entering the first-order potential replaced by one with scale-dependent growth. Up to second order, the displacement field $\boldsymbol{\Phi}$ is given by (see e.g., \cite{bouchet1995})
\be
&&\boldsymbol{\Phi}(\bq,\tau) = \boldsymbol{\Phi}^{(1)}(\bq,\tau) + \boldsymbol{\Phi}^{(2)}(\bq,\tau),
\ee
where $\boldsymbol{\Phi}^{(1)}(\bq,\tau) = \nabla_{\bq}\phi^{(1)}(\bq,\tau)$ and $\boldsymbol{\Phi}^{(2)}(\bq,\tau) = \nabla_{\bq}\phi^{(2)}(\bq,\tau)$, with $\phi^{(1)}$ and $\phi^{(2)}$ being the first- and second-order Lagrangian potentials. These are given by
\be
&&\nabla^2_{\bq}\phi^{(1)}(\bq,\tau) = -\delta^{\mathrm{L}}_{\mathrm{cb}}(\bq,\tau),\label{eq:nablaphi1}\\
&&\nabla^2_{\bq}\phi^{(2)}(\bq,\tau) = -\frac{3}{7}\sum_{i>j}\left\{\phi_{,ii}^{(1)}(\bq,\tau)\phi_{,jj}^{(1)}(\bq,\tau)-\left[\phi_{,ij}^{(1)}(\bq,\tau)\right]^2\right\}.
\label{eq:nablaphi2}
\ee
Here, the subscripts $i$ and $j$ after a comma denote partial derivatives with respect to the $i$-th and $j$-th spatial coordinates, respectively. The linear overdensity of the cb fluid, serving as the source term of the first-order potential, is computed in Fourier space as $\delta_{\mathrm{cb},\bk}^\mathrm{L}(\tau)=\mathcal{M}_\mathrm{cb}(k,\tau)\zeta_{\bk}$, with the transfer function provided in the $N$-body gauge.

In the solutions above, it is important to note that the time dependence is not factorizable from the spatial dependence, unlike the standard case with scale-independent linear growth. When the linear growth factor is scale independent, that is, $\mathcal{M}(k,\tau)=D_+(\tau)\mathcal{M}_0(k)$, the solutions described correspond to setting the second-order growth factor to $D_2(\tau) = -3/7 D_+^2(\tau)$, a relation that is exact in the Einstein-de Sitter cosmology.

We then compute the peculiar velocity at the first and the second orders corresponding to the displacement fields, $\boldsymbol{\Phi}^{(1)}_{\bk}(\tau)$ and $\boldsymbol{\Phi}^{(2)}_{\bk}(\tau)$, respectively. This is done by
\be
&&\bu_{\bk}^{(1)}(\tau) = a\,\mathcal{H}(\tau)\,f(k,\tau)\,\boldsymbol{\Phi}^{(1)}_{\bk}(\tau),\\
&&\bu_{\bk}^{(2)}(\tau) = a\,\mathcal{H}(\tau)\,f_2(k,\tau)\,\boldsymbol{\Phi}^{(2)}_{\bk}(\tau),
\ee
with the scale-dependent linear and second-order growth rate being
\be
&&f(k,\tau) = \frac{\mathrm{d}\log \mathcal{M}_{\mathrm{cb}}(k,\tau)}{\mathrm{d}\log a},\\
&&f_2(k,\tau) = 2f(k,\tau),\label{eq:f2}
\ee
respectively. Our evaluation of the second-order growth rate, $f_2$, involves two approximations. First, because the time dependence of $\phi^{(1)}$ is not separable from its spatial dependence, we cannot factorize the growth factor out of the products $\phi^{(1)}\phi^{(1)}$ on the right-hand side of equation~(\ref{eq:nablaphi2}) when we take the time derivative of $\boldsymbol{\Phi}$ to evaluate $\bu$. Second, we use the EdS approximation again to arrive at equation~(\ref{eq:f2}).

Previous studies have examined the accuracy of the approximations for $D_2$ and $f_2$ for non-EdS cosmologies. It has been shown that in open universes or in flat universes with a nonzero cosmological constant $\Lambda$, these are weak functions of $\Omega_\mathrm{m}$ \citep{bouchet1992,bouchet1995}. We therefore expect a similar level of accuracy in our scale-dependent case with non-separable time and spatial dependence.
In practice, we set up initial conditions at redshifts where nonlinear corrections are small and the matter density parameter is close to unity (i.e., close to the EdS value $\Omega_\mathrm{m}=1$). For instance, $\Omega_\mathrm{cb}=0.992$ already at $z=10$ for our fiducial cosmology discussed below, so the approximate treatment above does not cause any practical problems. We have checked that the final results are stable against the choice of the starting redshift.

In what follows, we build on our previous DQ1 study~\citep{2019ApJ...884...29N} and select the initial redshift to minimize the sum of two competing effects: the error due to the truncation of the PT series at second order, and the inaccuracy of force computation when the particle distribution closely approaches the regular lattice. Clearly, the former is more severe at lower starting redshifts, while the latter is more pronounced at higher starting redshifts. The optimal starting redshift depends on the spatial resolution and cosmology. In DQ1, it was found that a linear displacement of $20$ to $30\%$ of the mean inter-particle distance ($H_\mathrm{p}$) per dimension in the pre-initial configuration was optimal. We follow this and determine the initial redshift, $z_\mathrm{ini}$, by solving
\be
\sigma_\mathrm{d}(z_\mathrm{ini}) =  0.25\,H_\mathrm{p},
\label{eq:zini}
\ee
where
\be
\sigma_\mathrm{d}(z) =  \sqrt{\frac{1}{6\pi^2}\int P_\mathrm{cb}^\mathrm{L}(k,z)\mathrm{d}k},
\ee
with the linear power spectrum for the cb fluid
\be
P_\mathrm{cb}^\mathrm{L}(k,z) = \mathcal{M}^2_\mathrm{cb}(k,z)P_\zeta(k),
\ee
and $H_\mathrm{p}=L_\mathrm{box}/N_\mathrm{p}^{1/3}$. We have explicitly verified that this criterion --- originally calibrated with L-GADGET2 in DQ1 --- carries over to GINKAKU as well: simulations started at redshifts higher than the value selected by equation~(\ref{eq:zini}) yield $z=0$ matter power spectra that agree to within $\lesssim 0.5\%$ on all scales below the particle Nyquist wavenumber, so adopting the DQ1 prescription for our new code is justified.

We note that third-order Lagrangian perturbation theory (3LPT) initial conditions have been discussed and implemented in recent literature \citep{Tatekawa_3LPT_1,Tatekawa_3LPT_2,Tatekawa_3LPT_3,Michaux21}. Interested readers can also refer to studies such as \citet{Aviles20,Elbers22} for discussion on LPT in the presence of massive neutrinos. Although 3LPT can further reduce transients, we retain 2LPT for simplicity in this study. For the cosmologies and starting redshifts considered in this work, the impact of this approximation on the final statistics is well within our target accuracy, as confirmed by the convergence tests in section~\ref{sec:accuracy}. As explained above, the optimal starting redshift increases with the mass resolution. Consequently, as we discuss the resolution dependence in the following sections, we simultaneously investigate the transient effect stemming from the use of 2LPT initial conditions. By observing the convergence of simulation results across different resolutions, we verify that the residual transient effect is negligible.

Another initial-condition effect worth noting is the artificial force originating from the regular lattice pre-initial particle distribution. This can be understood analytically by particle linear theory~\citep{Marcos06}: the growing modes of a regular particle lattice differ from those of the continuous fluid, so fluctuations seeded with the continuum growing mode do not evolve as expected at small scales. A method to correct for this has been proposed by \citet{Garrison16}.

Both of these effects suppress the matter power spectrum on small scales. By employing more simulation particles and increasing the initial redshift $z_\mathrm{ini}$ according to equation~(\ref{eq:zini}), we expect to mitigate these systematic effects as the simulations achieve higher resolution and start from earlier cosmic times. To verify convergence, we monitor how the matter power spectrum changes with varying numbers of simulation particles. In the accompanying paper~\citep{Tanaka2026-emu}, instead of directly correcting for systematics in each simulation, we propose a power spectrum emulator that learns and accounts for the resolution dependence in addition to the cosmology dependence. This emulator approach addresses small-scale systematics at the level of the final statistical predictions.

\subsection{$N$-body code} \label{subsec:nbody}
Our newly developed TreePM $N$-body solver builds upon two previous codes, GADGET-2~\citep{gadget1,gadget2} and GreeM~\citep{Yoshikawa_2005,2009PASJ...61.1319I,Ishiyama2012-ey,Ishiyama2022-uf}. To facilitate efficient parallelization on modern large-scale computational facilities, we utilize the Framework for Developing Particle Simulators ~(FDPS; \cite{2016PASJ...68...54I,2018PASJ...70...70N}). FDPS operates in a hybrid MPI-OpenMP mode, automatically optimizing inter-process communication and domain decomposition. It requires minimal user code to perform particle-based simulations for a variety of physical systems; one simply needs to define the particles, their interactions, and the time integrator. In our implementation, we draw heavily on the methodology of these codes, with internal accuracy parameters validated against GADGET-2.

We made slight modifications to the library to optimize memory usage by disabling the memory pool. The original PM implementation provided within FDPS always allocates the mesh arrays necessary for PM calculations to enhance calculation speed. However, this can result in memory usage up to twice that of GADGET-2 for the same number of grid points. In our modified implementation, we allocate PM mesh memory only when necessary, reducing memory requirements to typically $50\%$ of the original, comparable to GADGET-2. This adjustment has a minimal impact on computation speed, resulting in only about a $1\%$ degradation in the worst-case scenario.

The PM force is computed by a modified version of the Poisson solver implemented in FDPS, based on GreeM; the underlying PM algorithm is described in detail in \citet{Yoshikawa_2005}, and we focus here on the modifications introduced to incorporate the linear response method.

The PM solver first computes the density contrast using the Triangular Shaped Cloud (TSC) interpolation~\citep{hockney81} of the distribution of simulation particles. The density field is then converted to the gravitational potential field by solving the Poisson equation using the Fast Fourier Transform (FFT). In our modified implementation, we multiply the boost factor (i.e., equation~\ref{eq:boost}) into the Green's function used in the Poisson solver to incorporate the effects from external sources, as discussed in the previous section. The force field is then computed by taking the derivative of the potential via the four-point finite difference, and finally the forces at the locations of the simulation particles are obtained by the TSC interpolation.

The PM force is evaluated considering the simulation particles to have a finite extent following the ``S2'' shaped profile, which reflects the shape of the TSC kernel~\citep{hockney81}. It acts as a smoothing operation that defines the smooth \textit{reference} force to be solved by the PM grid. We choose the size of this smoothing kernel, $\epsilon_\mathrm{g}$, to be three in units of the PM grid spacing $H_\mathrm{g}$ by default; for separations larger than $\epsilon_\mathrm{g}H_\mathrm{g}$, the PM force follows the Newtonian $1/r^2$ behavior. This choice is calibrated for optimal accuracy in section~\ref{sec:accuracy}. The missing short-range contribution is compensated by the tree force as described below.

The calculation of the short-range force is facilitated by the Barnes-Hut tree algorithm, integrated within FDPS. We leverage the functionality of the \texttt{MonopoleWithCutoff} class, which computes the monopole moment of the force, representing simply the gravitational force from the center-of-mass location of each tree node. This computation incorporates a cutoff to prevent double counting when combined with the long-range PM force. To ensure efficiency, the algorithm traverses tree nodes only up to the smoothing scale of the PM solver, $\epsilon_\mathrm{g}H_\mathrm{g}$, from the target particle, beyond which no contribution is considered for the short-range force calculation. Our default value for the tree opening angle, $\theta$, is set to $0.5$, which we calibrate in detail in section~\ref{sec:accuracy}. Additionally, we implement Plummer softening with a constant softening length $\epsilon_\mathrm{p}$, specified in comoving coordinates. By default, $\epsilon_\mathrm{p}$ is set to $2.14\%$ of the mean inter-particle distance $H_\mathrm{p}$. Note that this differs from GADGET-2, which uses cubic-spline softening.

The short-range force calculation is accelerated by using SIMD instructions.
We adopt the implementation from the Phantom-GRAPE library\footnote{\url{https://github.com/ccsoni/Phantom-GRAPE}} \citep{Nitadori2006-ek,2012NewA...17...82T,2013NewA...19...74T}. While Phantom-GRAPE computes gravitational force from a precomputed look-up table using the lower 4 bits of the exponent and the upper 5 bits of the mantissa of the IEEE 754 floating-point representation of the squared distance between particles,
we evaluate forces directly using double-precision SIMD intrinsics. This approach not only enhances calculation accuracy but also leads to a more stable estimation of the tree time steps (see equation~\ref{eq:dtime_tree}). 
The performance of this SIMD implementation depends on processor-level factors such as vectorization efficiency, cache hierarchy, and memory bandwidth.
On the systems we have tested (Intel-based CPUs at the Center for Computational Astrophysics, National Astronomical Observatory of Japan), we do not experience degradation compared to the original Phantom-GRAPE implementation.

We employ a time integrator and time-step criteria similar to those used in GADGET-2.
However, since the current version of FDPS does not support individual time steps, we employ a global time step for all simulation particles unlike in GADGET-2. The global time step applies the criterion of the most-accelerated particle to all particles, which is more conservative than an individual time step approach~\citep{gadget2}; implementing the latter to reduce the computational overhead is deferred to future work. Even though we use the same time step for all simulation particles, we can still improve efficiency by adopting different time steps for the tree and PM calculations, which will be detailed in what follows.

We employ the ``Kick-Drift-Kick'' (KDK) leapfrog scheme for time integration, where the relevant operators for updating the $i$-th particle are given as \citep{1997astro.ph.10043Q}
\be
&&K(\Delta \tau):\left\{
\begin{array}{l}
\boldsymbol{x}_i \to \boldsymbol{x}_i, \\
\displaystyle\boldsymbol{u}_i \to \boldsymbol{u}_i + \boldsymbol{a}_i\int_\tau^{\tau+\Delta \tau}\frac{\mathrm{d}\tau}{\mathcal{H}(\tau)},
\end{array}
\right. \label{eq:kick}\\
&&D(\Delta \tau):\left\{
\begin{array}{l}
\boldsymbol{u}_i \to \boldsymbol{u}_i, \\
\displaystyle\boldsymbol{x}_i \to \boldsymbol{x}_i + \boldsymbol{u}_i\int_\tau^{\tau+\Delta \tau}\frac{\mathrm{d}\tau}{a(\tau)\mathcal{H}(\tau)},
\end{array}
\right. \label{eq:drift}
\ee
for the kick and drift operator, respectively, expressed in the super-comoving time and coordinate variables introduced in section~\ref{subsec:newton}.
Here, the acceleration of the $i$-th particle is given by
\be
\boldsymbol{a}_i = -\boldsymbol{\nabla}\Psi(\boldsymbol{x}_i).
\ee
In practice, as discussed above, this is evaluated using the TreePM method, which splits the potential into short- and long-range contributions. The size of the time step, $\Delta\tau$, is determined by the following criteria. For the long-range PM force, the time step is given by
\be
\Delta \tau_\mathrm{PM} = \min\left( \frac{\eta_\mathrm{PM}\,H_\mathrm{g}}{\sigma_u/(a\mathcal{H})},\Delta \tau_\mathrm{max}\right),
\label{eq:dtime_PM}
\ee
where $\sigma_u$ is the 3D RMS velocity dispersion (in the velocity convention defined by equation~\ref{eq:u_velocity}) and $\eta_\mathrm{PM}$ and $\Delta\tau_\mathrm{max}$ are accuracy parameters. These two parameters approximately correspond to \texttt{MaxRMSDisplacementFac} and \texttt{MaxSizeTimestep} in GADGET-2, with values of $0.125$ and $0.03$, respectively\footnote{In GADGET-2, the numerator of the analogous time-step criterion uses $\min(H_\mathrm{p}, r_\mathrm{s})$, whereas we consistently use the FFT grid spacing $H_\mathrm{g}$.}.

Our time step for the tree force is given by
\be
\Delta \tau_\mathrm{tree} = \min\left(\sqrt{\frac{\eta_\mathrm{tree} \epsilon_\mathrm{p} H_\mathrm{p}}{|\boldsymbol{a}_i|/(a\mathcal{H}^2)}},\Delta\tau_{\mathrm{PM}}\right),
\label{eq:dtime_tree}
\ee
where $\eta_\mathrm{tree}$ is another accuracy parameter, corresponding to twice the parameter \texttt{ErrTolIntAccuracy} in GADGET-2\footnote{The factor of two comes from absorbing the $1/2$ prefactor that appears in the corresponding GADGET-2 expression.}. A value around $0.7$ has been found to produce convergent results, which we adopt as the fiducial value. Unlike GADGET-2, which employs individual adaptive time steps, we apply the same tree time step to all particles, determined by the one with the highest acceleration. This limiter controls the additional displacement of particles due to acceleration over a time step, relative to the Plummer softening length.
We synchronize the tree and the PM time steps whenever $\tau + \Delta\tau_\mathrm{tree}$ exceeds the next PM synchronization time.

In summary, our full operation for one PM time step is expressed as
\be
U(\Delta \tau_\mathrm{PM})
&=& K_\mathrm{PM}\left(\frac{\Delta \tau_\mathrm{PM}}{2}\right)\nonumber\\
&&\hspace{-1.2cm}\times\,
\prod_n \left[K_\mathrm{tree}\left(\frac{\Delta \tau_{\mathrm{tree},n}}{2}\right) D\Bigl(\Delta \tau_{\mathrm{tree},n}\Bigr)K_\mathrm{tree}\left(\frac{\Delta \tau_{\mathrm{tree},n}}{2}\right)\right]\nonumber\\
&&\hspace{-1.2cm}\times\,
K_\mathrm{PM}\left(\frac{\Delta \tau_\mathrm{PM}}{2}\right),
\ee
where the index $n$ in the product runs over tree steps until it reaches the next synchronization point with the PM step. Note that the time step for the tree force, $\Delta\tau_{\mathrm{tree},n}$, is updated at each tree sub-step within a PM step from the current acceleration field, following equation~(\ref{eq:dtime_tree}).

Given the phase-space distribution of particles, the cosmology enters the dynamics through two channels: the expansion history, $\mathcal{H}$, incorporated into the kick and drift operators, and the boost factor in equation~(\ref{eq:boost}). The boost factor is applied to the peculiar potential, $\Psi$, which is initially computed from the simulation particle distribution (i.e., $\nabla^{-2}(3/2)\Omega_\mathrm{cb,0}\mathcal{H}_0^2\delta_\mathrm{part}$), to obtain the complete peculiar potential, including effects from external sources. Because both quantities are supplied as precomputed tables, the code can handle a wide range of cosmological models without modification, requiring only the appropriate tables as input.

\section{Accuracy assessment}
\label{sec:accuracy}
Our benchmark metric is the nonlinear power spectrum, with a particular focus on the evolution of the cb fluid to the present. We test the accuracy of the fiducial parameter setting described in the previous section, and define a higher-precision setting (``accurate'') based on these tests. The full list of accuracy settings used in this section, together with a short summary of the reference run used in the convergence tests, is collected in table~\ref{tab:ginkaku_models}.

A fundamental limitation arises from the finite number of particles: even with exact gravitational force calculations, initial fluctuations on scales smaller than the interparticle distance cannot be represented. This scale is characterized by the \textit{particle} Nyquist wavenumber:
\be
\kny \equiv \pi \frac{N_\mathrm{p}^{1/3}}{\Lbox}.\label{eq:kNy}
\ee
Note that this wavenumber is different from that of the FFT grid used in the PM force calculation or that used in power spectrum measurement. Our simulations start from grid pre-initial conditions, where small particle displacements from the regular grid introduce a distinct feature in the power spectrum at $\kny$, which gradually diminishes over time. However, this feature may introduce artificial forces, potentially affecting the early-time evolution of the matter density field on these scales. At late times, in our typical setup, simulation particles are not perfectly relaxed within virialized objects at $z=0$. Even after subtracting the standard $1/\bar{n}$ Poisson shot-noise contribution, residual errors from the finite number of particles can remain at the few-percent level on small scales.

\begin{figure}[t]
 \centering
 \includegraphics[width=\linewidth]{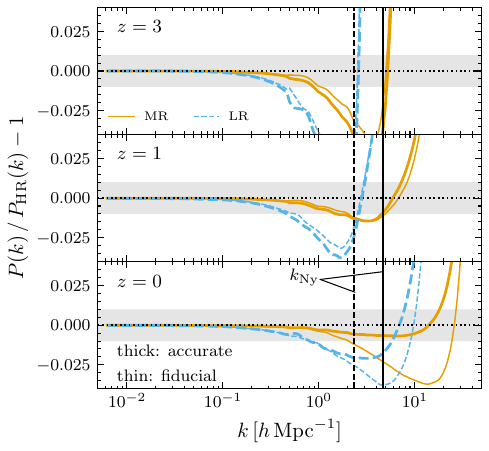}
 \caption{Convergence of the cb power spectrum from simulations with different numbers of particles. We show the fractional difference relative to the highest resolution setting (HR: $\Lbox=1\hiGpc$, $N_\mathrm{p}=3{,}000^3$). The MR ($N_\mathrm{p}=1{,}500^3$; solid) and LR ($N_\mathrm{p}=750^3$; dashed) simulations are started from the same initial random field as HR, within the same cubic comoving volume, but with fewer particles. Two different sets of accuracy parameters are employed: ``fiducial'' (thin) and ``accurate'' (thick), see table~\ref{tab:ginkaku_models}. The vertical lines represent the Nyquist wavenumber $k_{\mathrm{Ny}}$ for the MR and LR simulations. {Alt text: Fractional difference of the cold-dark-matter-plus-baryon power spectrum against wavenumber, comparing two lower-resolution simulations to the highest-resolution reference. Differences grow toward the Nyquist wavenumber, reaching a few percent for the two accuracy settings shown.} \label{fig:conv}}
\end{figure}

These finite-resolution effects are illustrated in figure~\ref{fig:conv}, where the dynamics starting from identical initial random fluctuations are followed with varying numbers of particles. Vertical lines in the figure indicate the location of $\kny$ for the two lower-resolution simulations, referred to as ``MR'' and ``LR'', while the fractional differences from the highest-resolution simulation, labeled ``HR,'' are shown. These resolutions roughly correspond to our target for studies of the largest-scale structures traced by halos down to minimum masses $\sim 10^{11}$--$10^{12}\,h^{-1}M_\odot$. The plot presents results with two accuracy settings for the force and time integration, labeled ``fiducial'' and ``accurate,'' represented by thin and thick curves, respectively.

At $z=3$, the small-scale power spectrum is enhanced near the Nyquist wavenumber, reflecting the grid pre-initial particle configuration. This can be interpreted as a residual excess of particle pairs at distances close to their pre-initial spacing, which persists despite the passage of time. Slightly below this wavenumber, there is a reduction in power by more than three percent compared to the highest resolution simulation. The suppression deepens for the stricter accuracy setting (i.e., thick lines).

As time progresses, the loss in power gradually diminishes, with the ratio showing a sharp upturn as it moves towards smaller scales. By $z=0$, the ``accurate'' setting agrees better with the highest-resolution reference than the ``fiducial'' setting. However, achieving one percent accuracy relative to the highest resolution setting remains challenging. Notably, the Nyquist wavenumber marks the scale where the resolution-induced power loss is largest. Therefore, in subsequent subsections, we focus on controlling the numerical error from internal accuracy parameters to within one percent up to the Nyquist wavenumber. As a result, we present the power spectrum ratio only up to $k_\mathrm{Ny}$ in the plots in the following subsections.

\begin{figure*}[htbp]
 \centering
 \includegraphics[width=\linewidth]{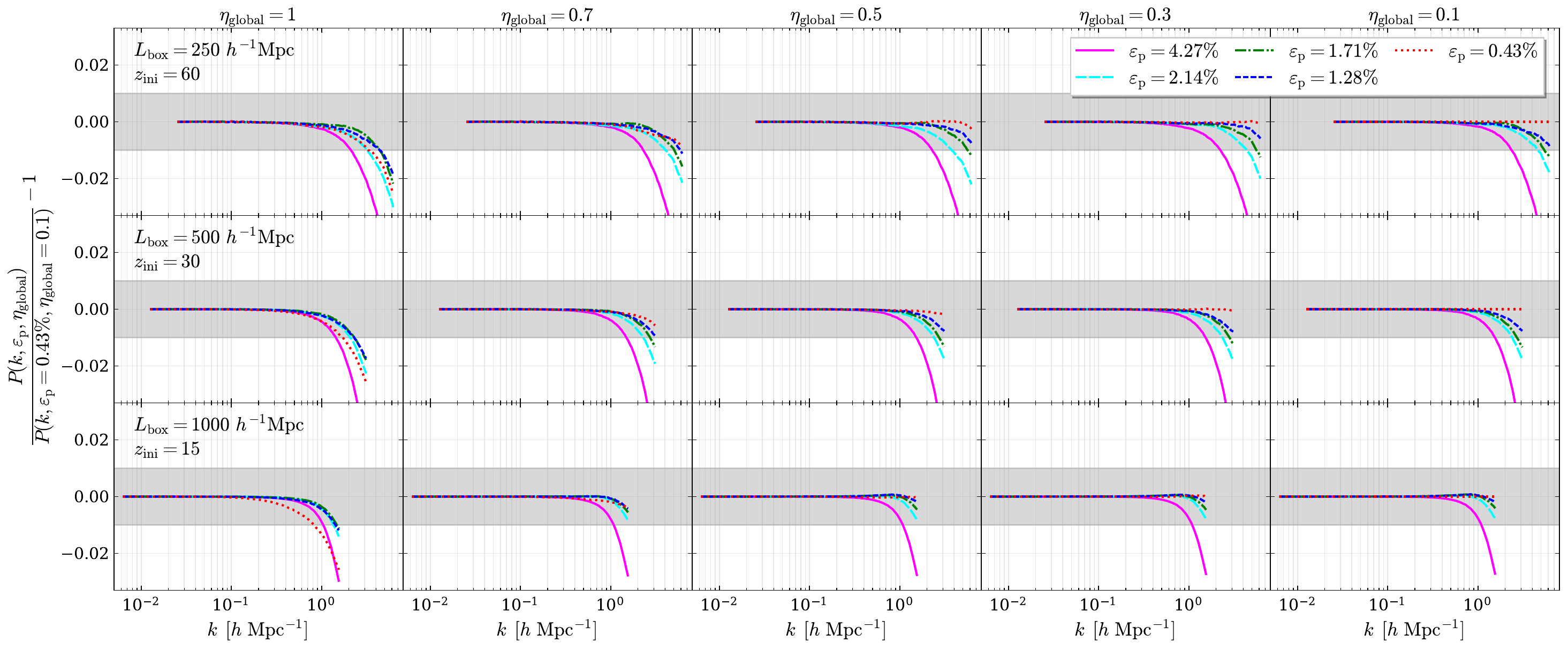}
 \caption{Convergence of the power spectrum
 from simulations with different softening length parameters $\epsilon_\mathrm{p}$ and time step sizes controlled by the parameter $\eta_\mathrm{global}$ for different resolutions ($N_\mathrm{p}=500^3$ particles in $L_{\mathrm{box}} = 250$, $500$ and $1000 \, h^{-1} \mathrm{Mpc}$ at $z=0$). We adopt the result with the smallest softening length and the most stringent time steps as the reference. The parameter $\eta_\mathrm{global}$ rescales both PM and tree time steps simultaneously (see section~\ref{subsubsec:soft_time}); the effective $\eta_\mathrm{tree}$ values from left to right are $0.70$, $0.34$, $0.18$, $0.063$ and $0.007$. {Alt text: Grid of power-spectrum ratio panels arranged by box size and time-step setting. Curves of different softening length show non-monotonic behavior at the fiducial time step but converge monotonically as the time steps are tightened across columns.} \label{fig:conv_ext}}
\end{figure*}

\begin{table*}
\tbl{Summary of the three accuracy settings for GINKAKU used in this section. The power spectrum accuracy is quoted relative to the reference run and assessed at $k = k_\mathrm{Ny}$. Relative costs are for $N_\mathrm{p}=500^3$ simulations; the fiducial setting is taken as the baseline.}{
\begin{tabular}{lccccc}
\hline\noalign{\vskip3pt}
\bfseries{Setting} & $\epsilon_\mathrm{p}$ [\%] & $\eta_\mathrm{tree}$ & Accuracy at $k_\mathrm{Ny}$ & Relative cost & Intended use \\ [2pt]
\hline\noalign{\vskip3pt}
fiducial & $2.14$ & $0.70$ & $\lesssim 1\%$ at $z=0$; may exceed $1\%$ at $z=3$ & $1\times$ & DQ2 production \\
accurate & $1.71$ & $0.23$ & $<1\%$ at $z=0$, with mild excess at $z=3$ & $\sim 2\times$ & Validation / cross-code comparison \\
reference\footnotemark[{$\dag$}] & $0.43$ & $0.007$ & Convergence reference for this work & $\gtrsim 20\times$\footnotemark[{$\ddag$}] & Convergence tests only \\ [2pt]
\hline\noalign{\vskip3pt}
\end{tabular}}
\label{tab:ginkaku_models}
\begin{tabnote}
\footnotemark[{$\dag$}] The reference setting is achieved via $\eta_\mathrm{global}=0.1$, which multiplies both the PM and tree time steps simultaneously (see section~\ref{subsubsec:soft_time}); the corresponding effective $\eta_\mathrm{tree}$ is $0.007$.\\
\footnotemark[{$\ddag$}] For the $L_\mathrm{box}=250\,\hiMpc$ box the reference run requires approximately $30\times$ the wall-clock time of the fiducial.
\end{tabnote}
\end{table*}

\subsection{Internal convergence study}
We now investigate the dependence of simulation accuracy on internal parameters. Several fundamental factors directly impact the final accuracy: the softening length, time stepping, tree opening criteria, and the transition scale of the two algorithms used to evaluate gravitational force. Previous studies on the impact of these parameters for GADGET can be found in, e.g., \citet{crocce06a,Coyote1,Reed13,Smith14,2015MNRAS.449.1454P,Baldauf_2015,2023A&A...671A.100E}, particularly in the context of the matter power spectrum or the halo mass function. While the impact of each parameter may be small when close to the fiducial value recommended in the GADGET user manual, the overall conclusion could differ when multiple parameters are varied simultaneously.

\subsubsection{Softening and time stepping}
\label{subsubsec:soft_time}

We first consider the gravitational softening length. The dependence of the power spectrum on the softening length is not straightforward to interpret. Softening is introduced to overcome the unphysical behavior induced by close encounters between particles that are significantly more massive than the actual fundamental mass components by many orders of magnitude. A large acceleration is exerted between the particle pair over a certain time step, resulting in unphysical scattering. Accurately tracking the orbits of such close particle pairs is computationally expensive. Softening, in which particles are represented as objects of finite extent rather than mathematical points, is a standard remedy. The effective size of these extended particles, the softening length, determines the force resolution; below this threshold, the force does not follow Newton's law, and one cannot resolve the growth of structures.

However, finding an optimal softening length is somewhat ambiguous. While reducing the softening formally improves the force resolution, it is established that if the softening is taken too small relative to the time step, spurious collisionality from numerical two-body scattering can contaminate the small-scale dynamics (e.g., \cite{power03,ludlow19,dehnen11}), though the direct consequence of this effect on the matter power spectrum has not been firmly established. Moreover, the impact of the softening on the power spectrum is tightly coupled to the time step setting --- a dependence that has not always been fully explored in prior studies, where the softening is often varied at a fixed time step criterion. To properly characterize these intertwined effects, we carry out a joint investigation in the two-dimensional parameter space of softening length and time step size below.

This joint exploration is presented in figure~\ref{fig:conv_ext}. To control the size of the time steps independently of the per-particle limiter, we introduce a parameter $\eta_\mathrm{global}$ that simultaneously rescales the PM and tree time steps obtained from the limiters introduced in section~\ref{subsec:nbody}. Because the tree time step scales as $\Delta\tau_\mathrm{tree}\propto\sqrt{\eta_\mathrm{tree}}$ (equation~\ref{eq:dtime_tree}), rescaling it by a factor $\eta_\mathrm{global}$ is equivalent to using an effective $\eta_\mathrm{tree}^\mathrm{eff}=\eta_\mathrm{global}^2\,\eta_\mathrm{tree}$; for example, $\eta_\mathrm{global}=0.1$ corresponds to $\eta_\mathrm{tree}^\mathrm{eff}=0.007$. The leftmost panels illustrate the dependence of the power spectrum on the softening length when the time step is set to the fiducial value. Here, we observe a nonmonotonic relation between the power spectrum and the softening length, $\epsilon_\mathrm{p}$. The suppression of the power spectrum compared to the reference is minimal when $\epsilon_\mathrm{p}$ is set to $1.71\%$ of the mean inter-particle distance for the three box sizes shown; this observation motivates our choice of $\epsilon_\mathrm{p}=1.71\%$ for the ``accurate'' setting defined below. However, when shorter time steps are employed, the trend changes. In the rightmost panels, with ten times smaller time steps, the power spectrum monotonically increases as the softening length decreases within the range explored here.

Our joint exploration spans softening lengths down to $\epsilon_\mathrm{p}=0.43\%$ in units of $H_\mathrm{p}$, roughly an order of magnitude below the $\sim5\%$ ansatz commonly adopted in the literature. Over this entire range, when the time steps are tightened commensurately, the power spectrum consistently increases as the softening decreases, with no indication of spurious two-body collisionality suppressing or artificially boosting the power. We therefore adopt the configuration with the smallest softening scale ($\epsilon_\mathrm{p}=0.43\%$) and the most stringent time steps ($\eta_\mathrm{global} = 0.1$) as the reference in the plots, since the power spectrum approaches this limit monotonically as the time steps are tightened across all three box sizes\footnote{Although our time step criterion in equation~(\ref{eq:dtime_tree}) already carries an explicit $\epsilon_\mathrm{p}$ dependence of the form $\Delta\tau \propto \sqrt{\epsilon_\mathrm{p}}$ following \citet{power03}, the necessity for finer time steps when $\epsilon_\mathrm{p}$ is reduced indicates that this $\sqrt{\epsilon_\mathrm{p}}$ scaling does not fully capture the true $\epsilon_\mathrm{p}$ dependence of the optimal time step. We retain the existing criterion to facilitate straightforward comparison with previous studies, and leave the development of a more accurate joint prescription for $\epsilon_\mathrm{p}$ and $\eta_\mathrm{tree}$ to future work.}.

This optimal configuration is, however, computationally prohibitive, with a wall-clock cost $\gtrsim 20$ times that of the fiducial setting\footnote{$\eta_\mathrm{global}=0.1$ already means ten times more time steps, and a smaller $\epsilon_\mathrm{p}$ requires even more.}. To strike a balance between accuracy and cost, we define two default settings whose parameters, accuracy targets, and intended uses are summarized in table~\ref{tab:ginkaku_models}. The ``fiducial'' setting (long-dashed lines in figure~\ref{fig:conv_ext}) follows the parameter values introduced in the previous section (table~\ref{tab:sim_params}) and reaches accuracy comparable to the DQ1 GADGET-2 setup, as we demonstrate later in figure~\ref{fig:comp_dq1dq2pkd}; near the Nyquist wavenumber, however, its deviation from the reference run may exceed $1\%$ (i.e., the right edge of each line in figure~\ref{fig:conv_ext}), which must be kept in mind when interpreting the fiducial results. The ``accurate'' setting instead tightens only the tree time steps via $\eta_\mathrm{tree}$, leaving $\eta_\mathrm{PM}$ at its fiducial value; this choice is supported by the analysis in section~\ref{subsubsec:PM} (figure~\ref{fig:PM_timestep}), which shows that $\eta_\mathrm{PM}$ has a negligible impact.

\begin{figure}[t]
 \centering
 \includegraphics[width=\linewidth]{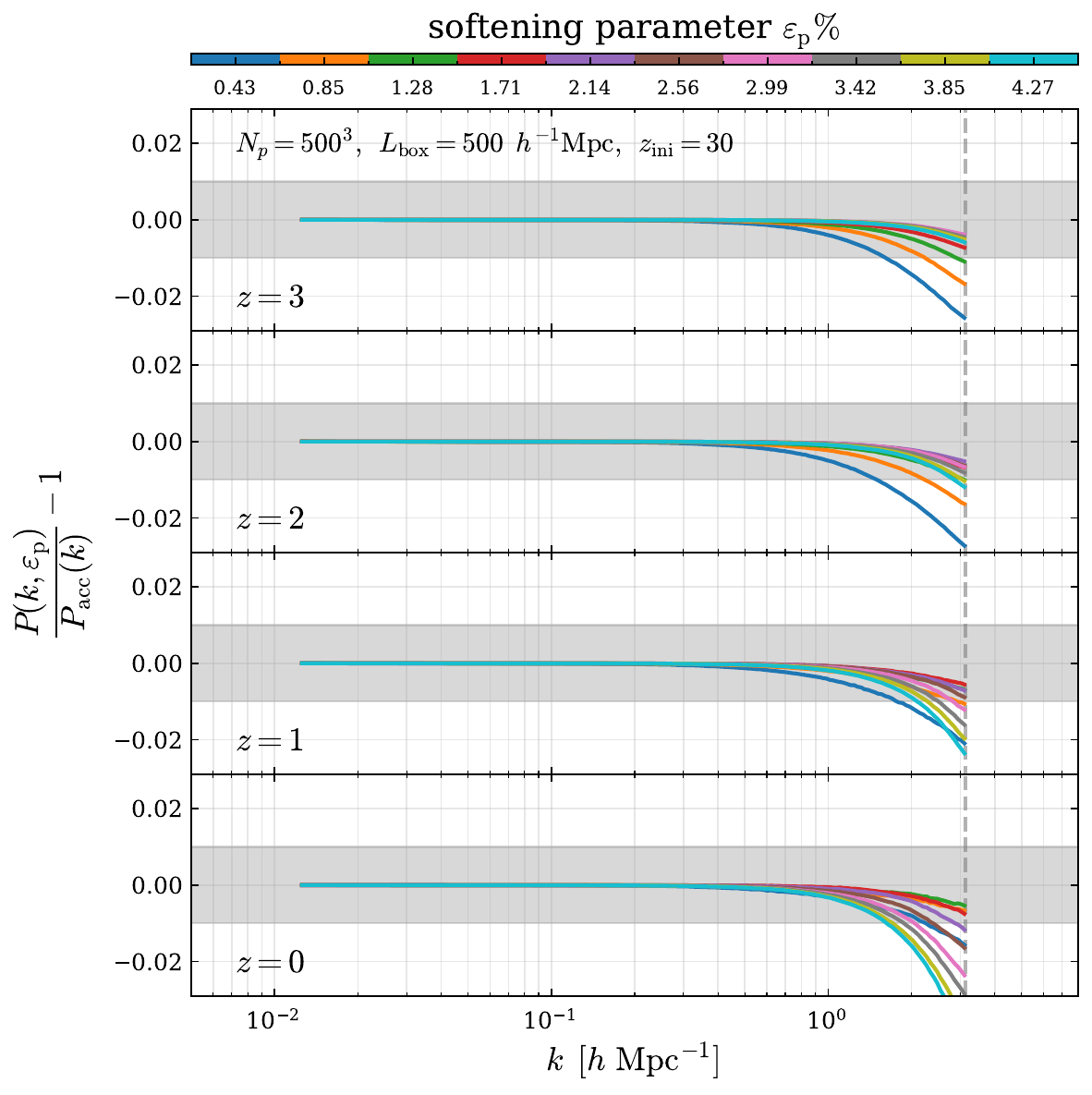}
 \caption{Dependence of the power spectrum on the softening parameter $\epsilon_\mathrm{p}$, as depicted in the color bar (in percent). The other accuracy parameters are fixed to the fiducial setting. We present the power spectrum normalized by the result of the ``accurate'' setting. The simulations employ $N_\mathrm{p}=500^3$ and $L_{\mathrm{box}} = 500\, h^{-1} \mathrm{Mpc}$, with results shown at $z=3$, $2$, $1$ and $0$. The vertical line marks the Nyquist wavenumber $k_{\mathrm{Ny}}$ for this resolution. {Alt text: Power-spectrum ratio against wavenumber at four redshifts, with curves color-coded by softening length. Shorter softening produces suppressed power at higher redshift and enhanced power at low redshift, indicating a redshift-dependent trade-off.} \label{fig:sft_fid_vs_acc}}
\end{figure}

Figure~\ref{fig:sft_fid_vs_acc} compares the two configurations across redshifts: we adopt the accurate setting as the denominator and vary $\epsilon_\mathrm{p}$ in the numerator while keeping the other parameters at their fiducial values. Although $\epsilon_\mathrm{p}=2.14\%$ may not be optimal at $z=0$, it yields stable results across redshift; shorter softening lengths suppress the small-scale power at higher $z$, with $\epsilon_\mathrm{p}=1.28\%$ giving a $1\%$ suppression at the Nyquist wavenumber at $z=3$. We therefore retain $\epsilon_\mathrm{p}=2.14\%$ as a reasonable redshift-averaged compromise at the chosen time-step size.

\subsubsection{Tree force}
\label{subsubsec:tree_opening}

\begin{figure*}[t]
 \centering
 \includegraphics[width=\linewidth]{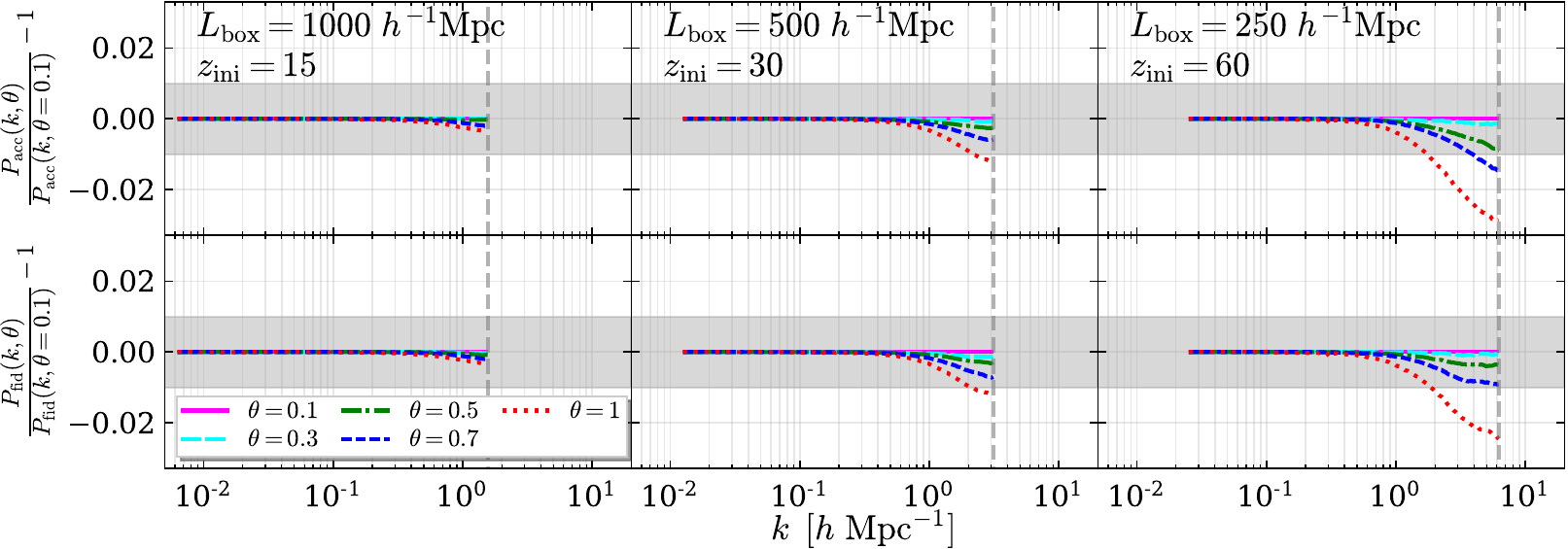}
 \caption{Dependence on the tree opening angle $\theta$. We consider both the accurate and fiducial settings for the other accuracy parameters and show the fractional difference to the most stringent tree opening angle, $\theta=0.1$, respectively in the upper and lower panels. The vertical lines show the Nyquist wavenumber $k_{\mathrm{Ny}}$ for each resolution. {Alt text: Two-row grid of power-spectrum ratio panels comparing tree opening angles at the accurate setting in the upper row and the fiducial setting in the lower row. The deviation from the smallest opening angle grows with mass resolution and depends little on the other accuracy parameters.} \label{fig:tree_theta}}
\end{figure*}

Having established the softening length and the time steps, we now focus on the accuracy of the gravitational force. Here, we discuss the impact of the tree opening angle, $\theta$, used in calculating the short-range force. Unlike the softening length, the interpretation of $\theta$ is straightforward: a smaller $\theta$ should always lead to higher accuracy, albeit at a greater computational cost. Since the impact of $\theta$ may depend on other accuracy parameters, we consider both fiducial and accurate configurations for the remaining parameters.

In figure~\ref{fig:tree_theta}, we present the results for both configurations in the upper (accurate) and lower (fiducial) panels. We adopt $\theta=0.1$ as the reference and display results for different values of $\theta$ using different line styles. By comparing the upper and lower panels, we observe that the plotted fractional difference is essentially independent of the time steps and softening lengths. In contrast, the deviation from the reference does depend on the spatial and mass resolution of the simulation, which vary with $H_\mathrm{p}$ at fixed $N_\mathrm{p}$. The right panels, which utilize the highest resolution, exhibit the strongest dependence on the tree opening angle. From this analysis, we conclude that $\theta=0.5$ is sufficient to ensure $1\%$ accuracy up to the Nyquist wavenumber, for simulations with mass and spatial resolutions comparable to those tested here. Therefore, we adopt $\theta=0.5$ for both fiducial and accurate configurations.

It is important to note that the impact of this parameter can be more pronounced when we adopt a different transition scale for the tree and PM force. We will revisit this topic in section~\ref{subsubsec:transition}.

\subsubsection{PM force}
\label{subsubsec:PM}

Next, we turn our attention to the accuracy of the PM force calculation, which is primarily controlled by the total number of PM grid points, denoted as $N_\mathrm{g}$. The PM force accuracy depends primarily on the relevant physical scale measured in units of the grid spacing $H_\mathrm{g}$. On scales comparable to this separation, inaccuracies in the force are anticipated due to anisotropies inherent in the regular lattice configuration of the grid points. The accuracy of the PM force is governed by the ratio $\epsilon_\mathrm{g}=r_\mathrm{s}/H_\mathrm{g}$, where $r_\mathrm{s}$ is the transition scale for the tree and PM force. A higher value of $\epsilon_\mathrm{g}$ indicates greater accuracy for a fixed $r_\mathrm{s}$. However, the dependence of the \textit{total} force on this parameter is nontrivial, as a higher $r_\mathrm{s}$ simultaneously extends the domain of the tree force. In such cases, the overall force accuracy may deteriorate if the tree force accuracy is insufficient. Consequently, a thorough investigation in the two-dimensional parameter space ($\epsilon_\mathrm{g}$, $r_\mathrm{s}$), potentially with additional dependence on $\theta$, is necessary to fully understand this aspect, as we examine in section~\ref{subsubsec:transition}.

In previous studies, the dependence of the matter power spectrum on the parameter \texttt{PMGRID} in GADGET-2 -- representing the number of grid points per dimension for the PM force calculation, equivalent to $N_\mathrm{g}^{1/3}$ in our notation -- has been extensively examined. Notably, \citet{Smith14} showed that this parameter can induce changes in the power spectrum exceeding $1\%$, a larger effect than many of the other accuracy parameters. However, identifying the optimal value for this parameter remains challenging, as the reported systematic effect displays a non-monotonic dependence on \texttt{PMGRID}. \citet{Baldauf_2015} suggested that setting \texttt{PMGRID} to $2N_\mathrm{p}^{1/3}$ might be superior to $N_\mathrm{p}^{1/3}$, based on comparisons with Effective-Field-Theory calculations. Furthermore, \citet{Baldauf_2016} noted a substantial dependence of the displacement field on this parameter, even on very large scales. It is important to note that these findings are based on experiments where the ratio $\epsilon_\mathrm{g}=r_\mathrm{s}/H_\mathrm{g}$ was kept constant when varying \texttt{PMGRID} (or $N_\mathrm{g}$).

\begin{figure*}[t]
 \centering
 \includegraphics[width=\linewidth]{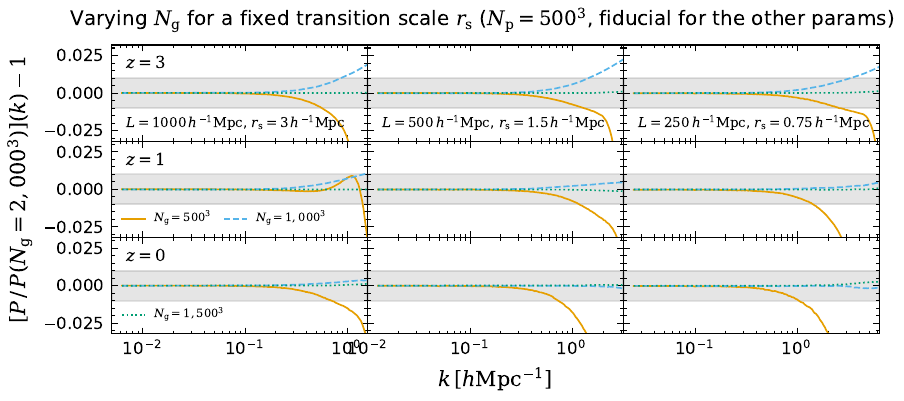}
 \caption{Convergence of the matter power spectrum with respect to the number of PM grid points, $N_\mathrm{g}$. The different line styles represent simulations with $N_\mathrm{g} = 500^3$ (solid), $1,000^3$ (dashed) and $1,500^3$ (dotted), each compared to the spectrum with $N_\mathrm{g}=2,000^3$, which serves as the reference with the highest PM force accuracy. The transition scale between the tree and PM forces is fixed to $r_\mathrm{s} = 1.5 H_\mathrm{p}$, corresponding to varying the ratio $\epsilon_\mathrm{g} = r_\mathrm{s}/H_\mathrm{g}$, with $\epsilon_\mathrm{g} = 1.5$, $3.0$ and $4.5$, compared with the reference run with $\epsilon_\mathrm{g}=6.0$. Our fiducial choice of $N_\mathrm{g} = 1,000^3$ ($\epsilon_\mathrm{g} = 3.0$) converges to the reference except at $z=3$, where the difference exceeds $1\%$ near the Nyquist wavenumber. {Alt text: Power-spectrum ratio against wavenumber for three coarser particle-mesh grid resolutions compared with the densest grid as the reference. Coarser grids deviate from the reference on small scales, with deviations shrinking monotonically as the grid resolution increases.}\label{fig:PM_GRID}}
\end{figure*}

We investigate the convergence of the power spectrum with respect to parameters that influence the accuracy of the PM force. Specifically, we vary the number of PM grid points, $N_\mathrm{g}$, while keeping the number of simulation particles fixed at $N_\mathrm{p}=500^3$ for three different box sizes to assess resolution-dependent convergence properties. In this analysis, we fix the transition scale $r_\mathrm{s}$ to be $1.5 H_\mathrm{p}$, varying $\epsilon_\mathrm{g}$ simultaneously to maintain the scale coverage of the tree force, thereby isolating the effect of PM force accuracy. Under these conditions, we expect the accuracy to improve monotonically with increasing $N_\mathrm{g}$, as smaller grid separations $H_\mathrm{g}$ better resolve scales around $r_\mathrm{s}$, the smallest scale covered by the PM calculation. Our fiducial choice for the number of grid points is $N_\mathrm{g} = 2^3 N_\mathrm{p}$, as recommended by \citet{Baldauf_2015}. The results are shown in figure~\ref{fig:PM_GRID}, where we adopt the most stringent case with $N_\mathrm{g}=2,000^3$ (or equivalently, $\epsilon_\mathrm{g}=6$ instead of the fiducial value of $3$) as the reference.

Figure~\ref{fig:PM_GRID} demonstrates that the power spectrum is not fully converged when $N_\mathrm{g}=500^3$, where the transition scale corresponds to $\epsilon_\mathrm{g} = 1.5$. This is qualitatively expected: the accuracy of the PM force is poor near the grid separation $H_\mathrm{g}$, and $1.5 H_\mathrm{g}$ is not sufficiently large. For $N_\mathrm{g} = 1,000^3$, the fractional difference from the reference spectrum remains within the $\pm1\%$ band, except on small scales near the Nyquist wavenumber at $z=3$. This corresponds to our fiducial choice of $\epsilon_\mathrm{g}=3$. Finally, with $N_\mathrm{g} = 1,500^3$ (i.e., $\epsilon_\mathrm{g}=4.5$), the ratio becomes nearly indistinguishable from the reference, indicating convergence in terms of PM force accuracy.

From this comparison, $\epsilon_\mathrm{g}=4.5$ might be considered a conservative choice. However, this setting is approximately $1.5$ times more time-consuming and requires around $3.4$ times more memory for the PM calculation compared to the fiducial choice of $\epsilon_\mathrm{g}=3$. Furthermore, at $z=3$, the slightly higher power obtained with the fiducial grid setting follows the trend toward higher resolution shown in figure~\ref{fig:conv}, suggesting that this excess partially compensates for the finite-$N_\mathrm{p}$ effects discussed at the beginning of this section. For these reasons, we adopt $\epsilon_\mathrm{g}=3$ in our fiducial setting.

\begin{figure*}[t]
 \centering
 \includegraphics[width=\linewidth]{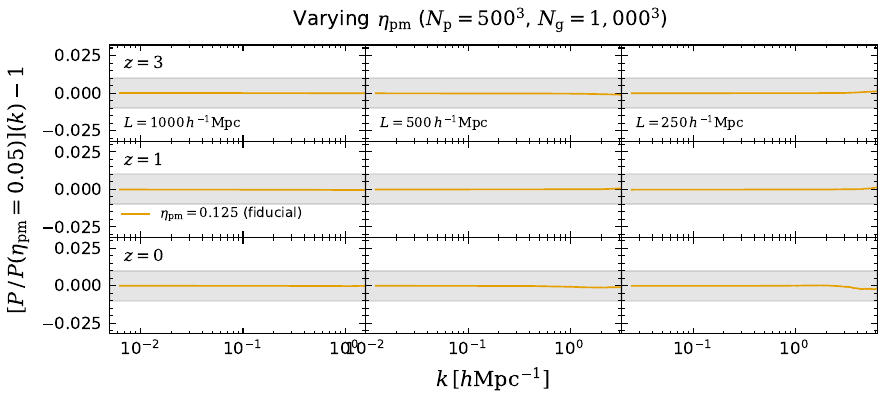}
 \caption{
 Impact of refining the PM force time step on the matter power spectrum. The solid line represents the fiducial simulation setting with the PM force time step criterion parameter $\eta_\mathrm{PM} = 0.125$, compared to a run with a more stringent value of $\eta_\mathrm{PM} = 0.05$. Refining the PM time step demonstrates a negligible effect on the matter power spectrum, with fractional differences remaining well within the $\pm 1\%$ band across all scales probed for the three redshifts and three resolutions considered here.
 {Alt text: Power-spectrum ratio against wavenumber comparing two particle-mesh time-step settings, at three redshifts and three resolutions. The fractional differences remain within one percent across all panels, confirming convergence with respect to the particle-mesh time step.}\label{fig:PM_timestep}}
\end{figure*}

Finally, we summarize in figure~\ref{fig:PM_timestep} a test in which the PM time step is reduced from the fiducial $\eta_\mathrm{PM}=0.125$ to $\eta_\mathrm{PM}=0.05$. The fractional impact on the matter power spectrum stays well within the $\pm 1\%$ band across all redshifts and resolutions considered, confirming that the fiducial PM time step is sufficient. This result also justifies our choice in section~\ref{subsubsec:soft_time} to tighten only the tree time steps (via $\eta_\mathrm{tree}$) in the ``accurate'' setting while leaving $\eta_\mathrm{PM}$ unchanged.

\subsubsection{Force transition scale}
\label{subsubsec:transition}

\begin{figure*}[t]
 \centering
 \includegraphics[width=\linewidth]{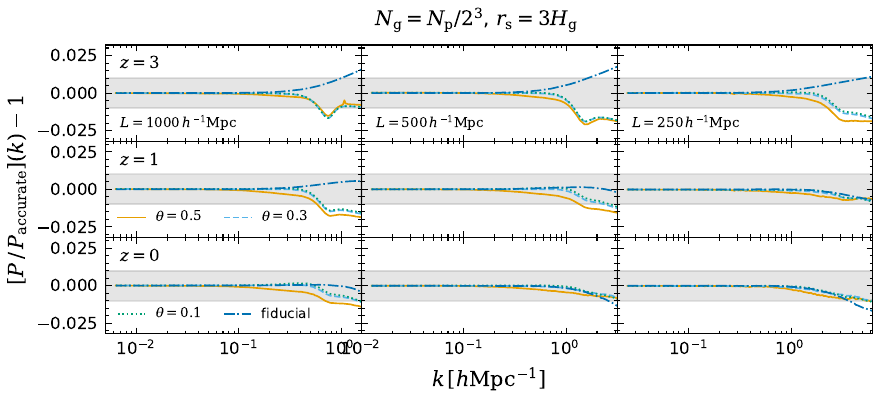}
 \includegraphics[width=\linewidth]{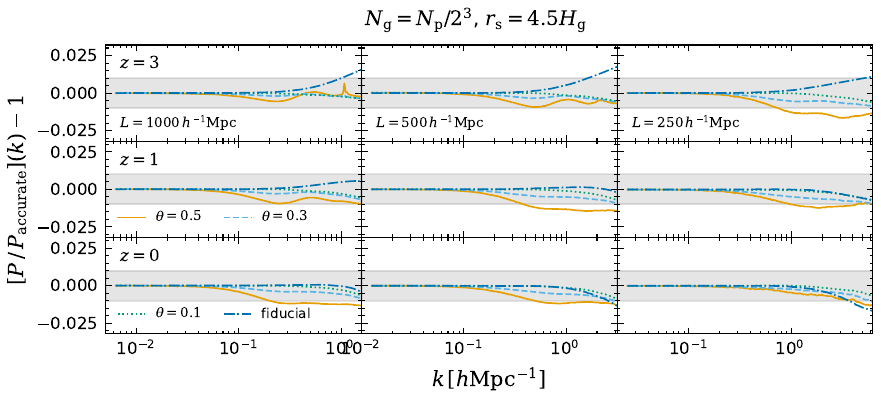}
 \caption{
 Impact of reducing the number of PM grid points, $N_\mathrm{g}$, to $N_\mathrm{p}/2^3$ from the fiducial setting of $2^3 N_\mathrm{p}$ on the matter power spectrum. The upper (lower) panels display results for $\epsilon_\mathrm{g} = 3$ ($4.5$). For comparison, the fiducial case with $N_\mathrm{g} = 2^3 N_\mathrm{p}$ is also shown as dot-dashed lines. We plot the fractional differences relative to the ``accurate'' case, which also has $N_\mathrm{g}=2^3N_\mathrm{p}$. With fewer grid points, power suppression is observed on small scales. This is mitigated when the opening angle for tree force, $\theta$, decreases from 0.5 (solid) to 0.1 (dotted). The results with $\theta=0.3$ (dashed) are almost on top of those with $\theta=0.1$. However, even $\theta = 0.1$ cannot fully recover the fiducial/accurate cases when $\epsilon_\mathrm{g}=3$. This improves by reducing the PM force error through the use of $\epsilon_\mathrm{g} = 4.5$.
 {Alt text: Two stacked rows of power-spectrum ratio panels comparing a coarse particle-mesh grid with the fiducial setting, for two force-splitting scales. A tighter tree opening angle reduces the small-scale suppression in both rows, but only the wider splitting fully recovers fiducial accuracy.}\label{fig:PM_GRID_half}}
\end{figure*}

Thus far, we have assessed the force accuracy of both the tree and PM calculations independently to examine how the power spectrum converges. Ideally, the overall force accuracy should be independent of the scale at which the two forces transition. From a computational standpoint, using fewer FFT grid points relative to the number of particles, i.e., $N_\mathrm{g}\lesssim N_\mathrm{p}$, is preferable for achieving high scalability on modern supercomputing systems with numerous CPU nodes. Therefore, it is desirable to maintain high accuracy regardless of the number of grid points used for the PM force calculation. In figure~\ref{fig:PM_GRID_half}, we present a similar accuracy test as in the previous subsection but with $N_\mathrm{g} = N_\mathrm{p}/2^3$, at a quarter of the per-dimension grid resolution used in the fiducial setting, as a representative memory-saving configuration for large supercomputer runs.

The upper panel of figure~\ref{fig:PM_GRID_half} shows the results for the simulations with $N_\mathrm{g} = N_\mathrm{p}/2^3$ and $\epsilon_\mathrm{g}=3$, using the ``accurate'' setting (with $N_\mathrm{g}=2^3N_\mathrm{p}$) as the reference. For comparison, we also display the ``fiducial'' setting (again, with $N_\mathrm{g}=2^3N_\mathrm{p}$) with a dot-dashed line. The results with $N_\mathrm{g}=N_\mathrm{p}/2^3$ show suppressed power spectra on small scales, which improves as $\theta$ decreases. This indicates that the tree force is responsible for a wider range of scales than previous cases, requiring greater accuracy than the fiducial setting of $\theta=0.5$. However, even with $\theta=0.1$, for which the results are nearly converged with respect to $\theta$, the accurate or fiducial settings with $N_\mathrm{g}=2^3N_\mathrm{p}$ are not fully recovered.

We therefore explore a larger transition scale, $\epsilon_\mathrm{g}=4.5$, in the lower panel. As expected, the suppression in the power spectrum for $\theta=0.5$ extends to smaller wavenumbers compared to the case with $\epsilon_\mathrm{g}=3$, as the domain of the tree force is now even broader. With $\theta=0.1$, we can nearly recover the accurate setting, indicating that $\epsilon_\mathrm{g}=3$ is insufficient to suppress the PM force error in the upper panel. These dependencies must be carefully considered when performing simulations with small $N_\mathrm{g}/N_\mathrm{p}$ ratios. For production runs that adopt $N_\mathrm{g}<N_\mathrm{p}$, we therefore recommend combining $\epsilon_\mathrm{g}\gtrsim 4.5$ with a tighter tree opening angle ($\theta\lesssim 0.3$). This conclusion is complementary to that of section~\ref{subsubsec:soft_time}: at the standard $N_\mathrm{g}=2^3N_\mathrm{p}$, the trade-off was driven by the softening length and the time step, whereas at $N_\mathrm{g}<N_\mathrm{p}$ it shifts onto $\theta$ and $\epsilon_\mathrm{g}$.

\subsection{Comparison with other codes}
\label{subsec:codes}

\begin{table*}
\tbl{Runs for the convergence check simulations. $N_\mathrm{p}$ is the number of $N$-body particles, $L_{\mathrm{box}}$ is the side length of the simulation box, $\epsilon_\mathrm{p}$ is the softening length in units of the mean particle separation, $\theta$ is the opening angle for tree force, $\eta_\mathrm{global}$ is an additional time-step parameter that multiplies the code's default step size, $\Delta t$.
}{
\begin{tabular}{ccccccccc}
\hline\noalign{\vskip3pt}
\bfseries{Model} & \bfseries{Code} & $N_\mathrm{p}$ &
$L_{\mathrm{box}} \, [h^{-1} \mathrm{Mpc}]$ &
$\epsilon_\mathrm{p}$ \, [\%] & $\theta$ & $\eta_\mathrm{global}$ & \bfseries{Run} & \bfseries{Remark} \\ [2pt]
\hline\noalign{\vskip3pt}
GAD & & & & $5$ &  $0.5$ & $0.1, 0.2, 0.5, 1$ & $12$ &\\
\vspace{-0.2cm}
& L-GADGET2\footnotemark[{$\ast$}] & $500^3$ & $250, 500, 1000$ & & & \\
DQ1 setting & & & & $5$ &  $0.5$ & $1$ & $3$ \\
\hline
PKDGRAV3 & PKDGRAV3 & $500^3$ & $250, 500, 1000$ & - & \footnotemark[{$\dag$}] & $1$ & $3$ & \footnotemark[{$\ddag$}] \\ [2pt]
\hline\noalign{\vskip3pt}
\end{tabular}}
\label{tab:convergence_runs}
\begin{tabnote}
\footnotemark[{$\ast$}] L-GADGET2 (see Introduction) is expected to give numerically equivalent results to GADGET-2 for identical accuracy parameters.\\
\footnotemark[{$\dag$}] The opening angle varies with redshift in PKDGRAV3. We set $\theta=0.4$ at $z>20$, $\theta=0.55$ at $2<z<20$ and  $\theta=0.7$ at $z<2$.\\
\footnotemark[{$\ddag$}] We set the FMM accuracy to the fifth order and the adaptive time-stepping parameter $\eta$ to 0.15, which appears in $\Delta t_i = \eta\sqrt{\epsilon_\mathrm{p}H_\mathrm{p}/|\boldsymbol{a}_i|}$.
\end{tabnote}
\end{table*}

\begin{figure*}[t]
 \centering
 \includegraphics[width=\linewidth]{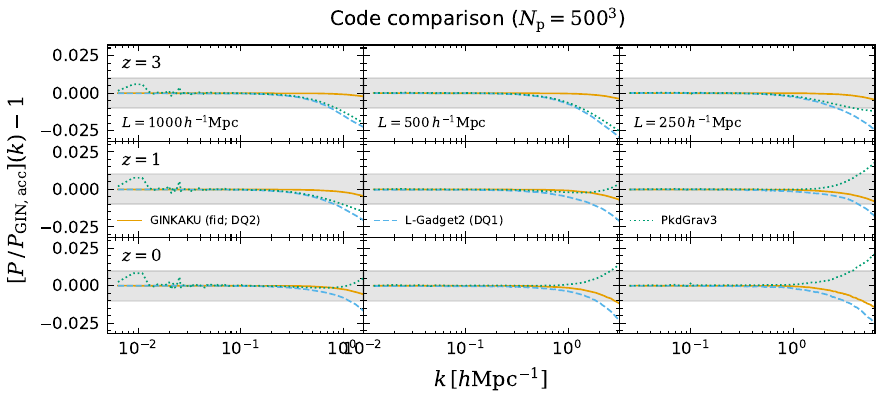}
 \caption{Comparison of the matter power spectra from simulations using different codes. We use the result from GINKAKU with the ``accurate'' setting as the reference and show the fractional differences relative to it. The results from L-GADGET2 (dashed; DQ1 default accuracy parameters) and PKDGRAV3 (dotted) are plotted. Additionally, we show the result from GINKAKU with the fiducial accuracy setting (solid), which is used for the DQ2 project. {Alt text: Code-to-code power-spectrum comparison against wavenumber for three box sizes. At $z=0$, PkdGrav3 and L-Gadget2 bracket the Ginkaku fiducial result. Deviations from the Ginkaku accurate reference stay within plus-or-minus one percent across most scales and redshifts.}\label{fig:comp_dq1dq2pkd}}
\end{figure*}

\begin{figure}[t]
 \centering
 \includegraphics[width=\linewidth]{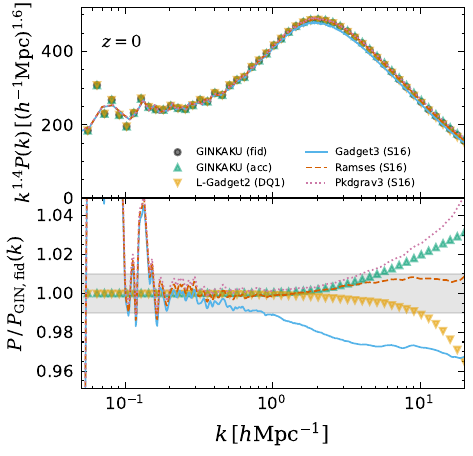}
 \caption{Power spectrum comparison. We perform two simulations with GINKAKU (fiducial: circles; accurate: upward triangles) and one with L-GADGET2 (DQ1 fiducial parameter setting; downward triangles). These are compared with the power spectrum data files by \citet{schneider2016} (S16) available at \url{https://www.ics.uzh.ch/~aurel/euclid.htm}: GADGET-3 (solid), RAMSES (dashed) and PKDGRAV3 (dotted). The upper panel shows the power spectrum scaled by $k^{1.4}$ and the lower panel shows the ratio to the GINKAKU simulation performed with the fiducial parameters. Note that the wavenumbers are binned differently in S16. Therefore, in the ratio plot, we resample the wavenumbers by cubic spline interpolation (100 points per decade) and then apply a second-order Savitzky--Golay filter with a window length of 31 sample points to reduce the high-frequency noise. {Alt text: Two-panel comparison of the matter power spectrum at redshift zero from several codes, presented as the scaled spectrum in the upper panel and as the ratio to a Ginkaku reference in the lower panel. The codes cluster within about one percent across most scales, with systematic offsets attributable to differing accuracy settings.} \label{fig:schneider}}
\end{figure}

Having established the convergence of the matter power spectrum by varying the accuracy parameters in our new simulation code, GINKAKU, we now compare its performance with other widely used $N$-body codes. Specifically, we consider L-GADGET2 (TreePM) and PKDGRAV3 (a Fast Multipole Method code). For these comparisons, we examine three different resolution levels by adopting box sizes of $L_\mathrm{box} = 1,000$, $500$ and $250\,\hiMpc$, each with $N_\mathrm{p} = 500^3$ particles. All simulations in this comparison share the same random number seed for the initial conditions, so that the cosmic variance nearly cancels in the code-to-code ratios presented below. In what follows, we first benchmark GINKAKU against L-GADGET2 and PKDGRAV3 on shared initial conditions of our own (figure~\ref{fig:comp_dq1dq2pkd}), and then position our code within the broader comparison project of S16 by reproducing their initial-condition setup (figure~\ref{fig:schneider}).

We present the fractional difference of the power spectrum in figure~\ref{fig:comp_dq1dq2pkd}, using the ``accurate'' setting of a GINKAKU run, denoted as $P_\mathrm{acc}$. In the figure legend, ``DQ1 setting'' refers to an L-GADGET2 run (dashed lines) with the internal parameters used in DQ1, while ``DQ2 setting'' represents a GINKAKU run (solid lines) with the fiducial parameters adopted in this paper. The dotted lines illustrate the results from PKDGRAV3 using the parameters described in \citet{schneider2016}. Different colors in the figure legend indicate the results for various box sizes.

Overall, the difference remains within $\pm 1\%$, with a tendency of PKDGRAV3 (L-GADGET2) to produce a larger (smaller) power spectrum on small scales\footnote{We see a small mismatch at the largest scales ($k\sim0.01\hMpci$) for the $L_\mathrm{box}=1\hiGpc$ box with PKDGRAV3. This is presumably related to differences in the treatment of long-range periodic forces. We have confirmed that large-scale modes near the fundamental wavenumber, $2\pi/L_\mathrm{box}$, closely follow the linear growth factor for both GINKAKU and L-GADGET2 runs, with the discrepancy $\lesssim 0.1\%$. We do not investigate this further here, as the mismatch is below one percent and the main focus is on the nonlinear evolution down to the Nyquist wavenumber.}. This is consistent with the comparison study presented in \citet{schneider2016}. The two GINKAKU results tend to lie in between these two codes when nonlinearity becomes important (i.e., at lower redshifts and for smaller box sizes). As already discussed in the previous subsections, more stringent accuracy settings give a lower power spectrum at $z=3$, moving further away from the higher-resolution trend illustrated in figure~\ref{fig:conv}.

To further test the accuracy of our simulation code, we use the initial condition files released by \citet{schneider2016}, hereafter S16. These files contain a much larger number of particles, $N_\mathrm{p}=2,048^3$, in a periodic comoving box of $L_\mathrm{box}=500\,\hiMpc$. This allows us to perform an accuracy test over a wider dynamic range extending into a more strongly nonlinear regime. The matter power spectra at $z=0$ measured in S16 with their GADGET-3, RAMSES, and PKDGRAV3 simulations are publicly available. Using the same initial conditions, we perform three simulations: one with L-GADGET2 using the parameters from DQ1, and the other two with GINKAKU using the ``fiducial'' and ``accurate'' settings.

We compare the six power spectra in figure~\ref{fig:schneider}, where the spectra from S16 are shown by lines, while our simulations are presented by symbols. The lower panel provides a zoomed-in view, showing the ratio to our fiducial GINKAKU run. First, the suppressed power spectrum observed for GADGET-3 in S16 (solid line) can be largely mitigated by refining the accuracy parameters, as indicated by our L-GADGET2 simulation with the parameters from DQ1 (downward triangles)\footnote{L-GADGET2 and GADGET-3 are not identical codes but share the same underlying TreePM gravity solver inherited from GADGET-2; the difference in the small-scale power is therefore likely dominated by the accuracy parameters rather than by fundamental differences in the gravity solver.}. This improvement has already been noted in \citet{BACCO}. Compared to the GADGET runs, the GINKAKU fiducial run (circles in the upper panel and the baseline in the lower panel) yields a slightly larger power spectrum on small scales, and is more consistent with RAMSES (dashed line). Finally, the other GINKAKU run with the accurate setting (upward triangles) is consistent with PKDGRAV3 (dotted line).

From these results, we confirm that GINKAKU can achieve accuracy comparable to that of widely used public codes. Depending on the requirements, the accuracy can be fine-tuned by adjusting the relevant parameters. Both our fiducial and accurate settings fall within the range of variation among the codes tested in the comparison project by S16. Based on these findings, the DQ2 production runs presented in section~\ref{sec:productruns} adopt the fiducial setting, while the accurate setting is reserved for validation and cross-code comparisons where tighter convergence at small scales is required.

\section{Post-processing}
\label{sec:postprocess}

We describe the updated post-processing pipeline, with several improvements over DQ1, covering halo identification and mass definition, the halo mass function, halo-shape measurement for intrinsic-alignment studies, and a noise-reduced estimator for two-point spatial statistics.

Throughout this section, we work within the fiducial $\Lambda$CDM cosmology with a neutrino mass sum of $M_\nu = 0.06\,\mathrm{eV}$ and the remaining parameters consistent with \citet{planck-collaboration:2015fj}. The simulations used for demonstration are performed with GINKAKU using the fiducial internal accuracy parameters. We present the results for the LR, MR and HR runs with different numbers of particles, which are already described in the previous section. These simulations are used for the resolution study and to calibrate the connection between simulations with different resolutions. Additionally, we perform $100$ random realizations for the fiducial cosmology with $N_\mathrm{p}=1,024^3$ and $L_\mathrm{box} = 1,024\,\hiMpc$. These simulations are used to validate the measurement code for the power spectra and the correlation functions.

\subsection{Structure finding}
\label{subsec:finder}
Following DQ1, we use the public phase-space temporal structure finder, ROCKSTAR \citep{Behroozi:2013}, to identify halos and subhalos. To support various cosmological models beyond flat $\Lambda$CDM, we made minor modifications to the original ROCKSTAR code. First, consistent with our simulation code, the modified version reads the expansion history from a numerical table instead of performing analytical evaluations. The value of the Hubble expansion rate $H(z)$ is interpolated from the table when necessary, such as when determining whether a particle is gravitationally bound to a candidate halo. Second, the virial density is computed numerically instead of using the approximate formula by \citet{1998ApJ...495...80B}, which is implemented in the original code and is valid only in flat $\Lambda$CDM. We solve the spherical-collapse equation for this, involving the expansion history $H(z)$ and hence introducing the cosmology dependence of the virial density. As in the original ROCKSTAR code, our code outputs all self-bound structures down to the mass resolution, without distinguishing between distinct halos and their substructures.

\begin{figure}[t]
 \centering
 \includegraphics[width=\linewidth]{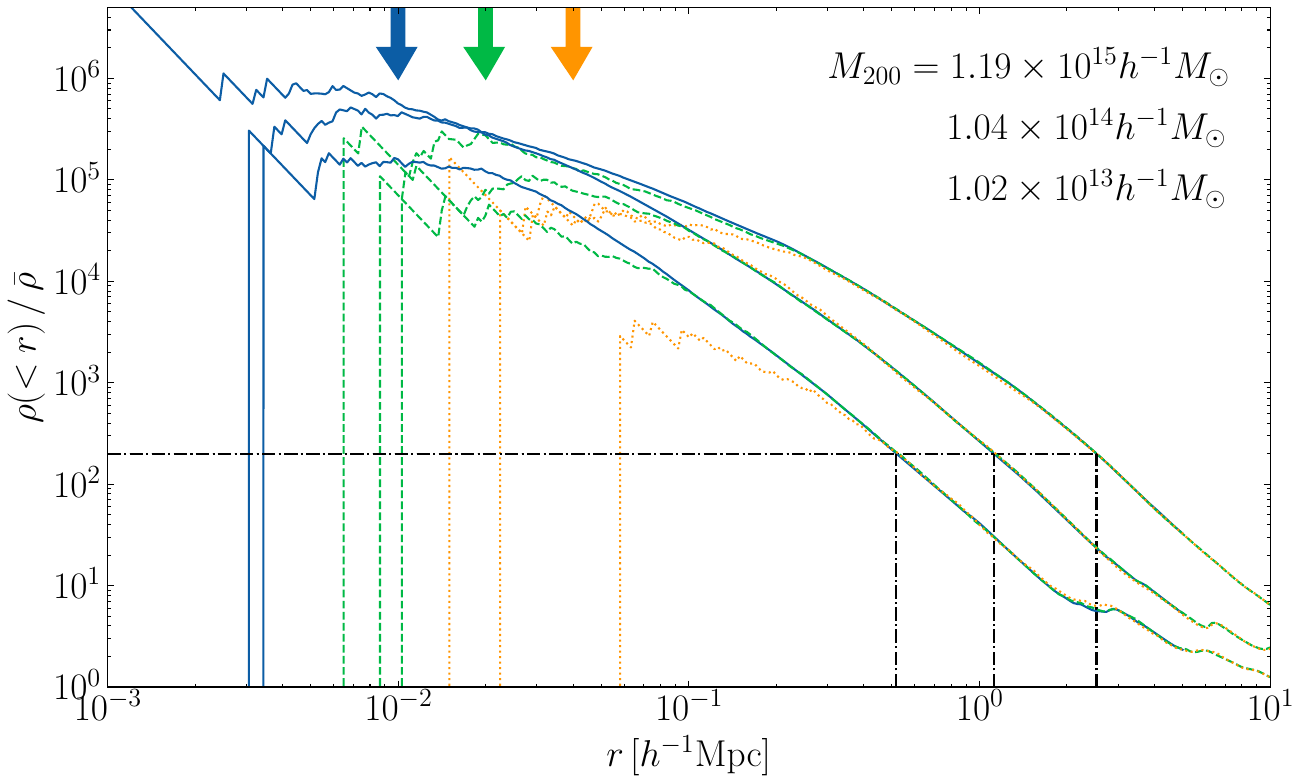}
 \caption{Our mass determination scheme based on the interior overdensity. We show the interior overdensity as a function of radius from the halo center, considering $\Delta=200$ as the density threshold. We select three halos at different mass scales, as indicated in the figure legend (values from the HR simulation), matched among three simulations with different resolutions (HR: solid, MR: dashed, and LR: dotted; the corresponding softening lengths are indicated by the arrows). Because the catalog stores the radial profile at a fine sampling rate, the mass at any threshold $\Delta$ can be evaluated quickly and accurately. {Alt text: Plot of interior overdensity against radius for three halos at three simulation resolutions. Curves at all resolutions overlap closely outside the softening scales, indicating that the mass enclosed within a given density threshold is robustly determined.} \label{fig:halomass}}
\end{figure}

While we previously focused only on ``central'' halos with the spherical overdensity (SO) mass definition, using an overdensity threshold of $\Delta = 200$ times the mean density in DQ1, we now retain all structures found by ROCKSTAR in the new simulation series. We postpone the central-satellite split for later analyses, as the definition of central halos (i.e., the employed overdensity threshold) can vary depending on the observables under consideration. This approach supports various conventional halo definitions and allows us to derive the halo mass function for different values of $\Delta$ from the master catalog (see also Appendix~\ref{sec:mass_def}).

To achieve this, we measure the spherical density profile around each structure, incorporating all mass represented by the simulation particles (both CDM and baryons), regardless of their gravitational binding state, in addition to the standard ROCKSTAR outputs. The masses determined from the profiles are the most relevant for gravitational lensing observations. Some example profiles are illustrated in figure~\ref{fig:halomass}. We employ $320$ logarithmic radial bins spanning four decades (80 sampling points per decade) to cover the range from the softening length to the halo edge. Given an overdensity threshold $\Delta$, we determine the mass $M_\Delta$ by identifying the radius $R_\Delta$ at which the interior density is $\Delta$ times the mean density. We then smoothly interpolate the sampled densities using a cubic spline function, ensuring that the radial binning is sufficiently fine to avoid significant errors in mass determination. In fact, the variation in the resulting mass is below one percent when only every other radial bin is used.

In determining the halo mass from the density profile, we can safely neglect the contribution from massive neutrinos. \citet{2004JCAP...12..005R} demonstrated that the neutrino fraction is significantly smaller in halo regions compared to the cosmic mean, using ``$N$-$1$-body'' simulations. Specifically, the neutrino fraction is about an order of magnitude smaller in the outskirts of halos for $M_\nu=0.15\,\mathrm{eV}$ and decreases rapidly toward the halo center. However, while we neglect the clustered neutrino mass inside halos, we do include the homogeneous neutrino contribution in the reference mean density when checking the overdensity threshold to find the halo edge radius, $R_\Delta$, so that the mass is defined in units of the \textit{total} matter density.

After this procedure, the raw mass estimate from counting simulation particles within $R_\Delta$ is corrected to reduce errors in the halo mass function due to limited mass resolution, which affects low-mass halos resolved by only a small number of particles. Motivated by a similar formula introduced in \citet{crocce10} for Friends-of-Friends halos in MICE simulations, a simple empirical correction was applied in DQ1:
\be
\tilde{M}_{200} = \left(1+n_\mathrm{p}^{-0.55}\right)M_{200},
\label{eq:mass_correction1}
\ee
where $M_{200}$ and $\tilde{M}_{200}$ are the original and corrected mass, respectively, with $n_\mathrm{p}$ being the number of simulation particles included in $R_\Delta$. As presented later in subsection~\ref{subsec:hmf}, we further refine this to make a smoother connection between simulations with different resolutions. The updated formula is
\be
\tilde{M}_\Delta = \left(1+0.55 n_\mathrm{p}^{-0.6}\right)M_\Delta,
\label{eq:mass_correction2}
\ee
where we have generalized the correction to an arbitrary overdensity threshold $\Delta$, added a coefficient ($0.55$) in front of the $n_\mathrm{p}$-dependent term, and slightly steepened the exponent from $-0.55$ to $-0.6$.

Finally, the split between central and satellite halos is performed as follows. For a given density threshold $\Delta$, we first calculate the radius, $\tilde{R}_\Delta = \left(3 \tilde{M}_\Delta / 4\pi \Delta \bar{\rho}_\mathrm{m}\right)^{1/3}$, corresponding to the corrected halo mass. If a halo (index $i$) is found within a sphere of radius $\tilde{R}_\Delta^{(j)}$ of another halo (index $j$) and has a smaller maximum circular velocity ($V_\mathrm{max}^{(i)} < V_\mathrm{max}^{(j)}$), we mark halo $i$ as a satellite. Therefore, our central/satellite split is adaptive: for the same physical halo, the central/satellite label can switch depending on $\Delta$.

\begin{figure}[t]
 \centering
 \includegraphics[width=\linewidth]{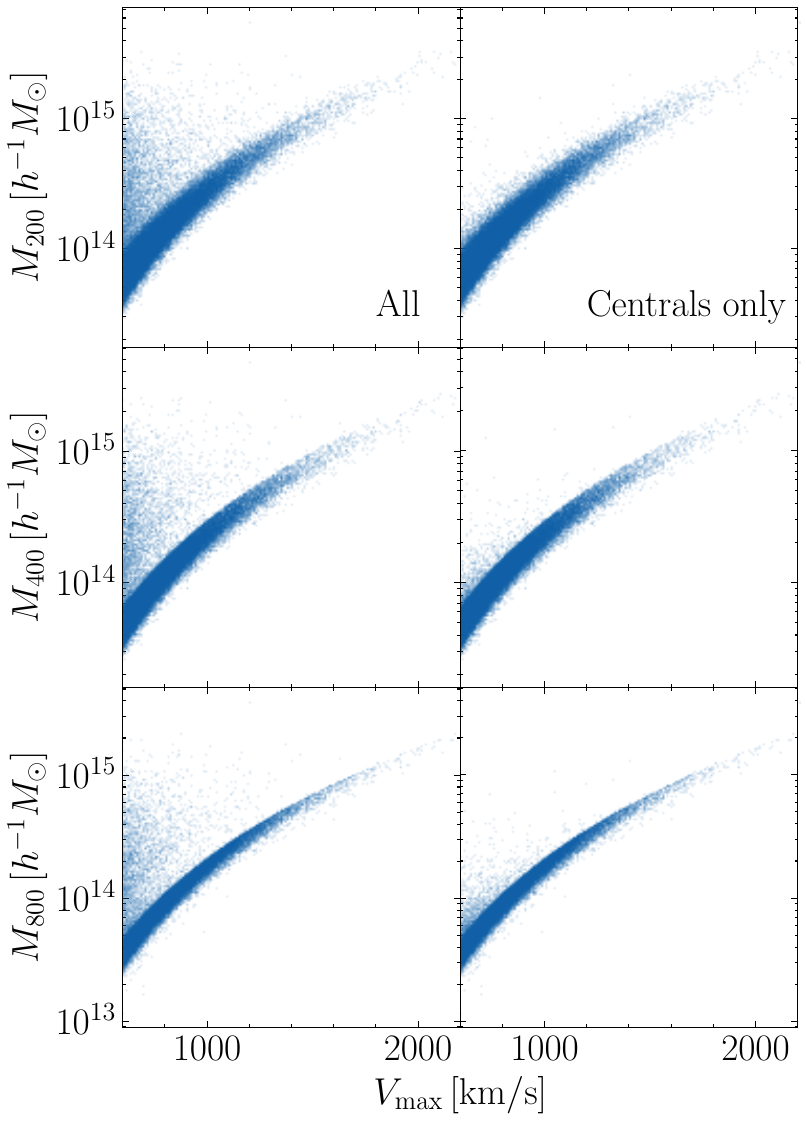}
 \caption{Halo masses for three different density threshold values ($\Delta = 200, 400$ and $800$ from top to bottom, as indicated by the $y$-axis labels) as a function of the maximum circular velocity $V_\mathrm{max}$ in the HR simulation at $z=0$. The left panels display all the structures found by ROCKSTAR, while the right panels show only central halos, determined based on the procedure explained in the main text. {Alt text: Six-panel scatter plot of halo mass against the maximum circular velocity, with rows for three overdensity thresholds. The left column shows all structures and the right column only central halos, with substructure outliers absent from the right panels.} \label{fig:halomass_vmax}}
\end{figure}

This procedure differs from DQ1, where we used $M_{200}$ instead of $V_\mathrm{max}$ to determine the parent-child relationship of halos. Additionally, halos with $M_{200}$ directly measured from the counting of nearby particles (both bound and unbound) that are larger than $1.3$ times the value of $M_{200}$ in the original ROCKSTAR catalog are discarded from the central list before checking the neighbors. This additional cut was necessary to avoid spurious massive halos, as the ROCKSTAR output is the bound mass, typically smaller than the mass measured directly. A large difference between the two masses indicates that the halo is likely a substructure of a more massive system. Since simulated halos are aspherical with non-concentric isodensity contours, a more significant structure (with a clearer density peak) can have a smaller SO mass (including unbound particles) than a nearby, less significant neighbor, leading to unintentional satellite marking. The impact of changing the empirical factor from $1.3$ is summarized in the Appendix~E of \citet{2019ApJ...884...29N}. Using $V_\mathrm{max}$ is advantageous as it can be more robustly determined than the mass, even with fewer particles, and is a better proxy for the significance of density peaks and, consequently, galaxies. Additionally, our new procedure no longer requires an adjustable parameter for the central-satellite separation.

Figure~\ref{fig:halomass_vmax} shows the distribution of halos identified in the HR simulation in the mass-$V_\mathrm{max}$ plane for three different density thresholds. The left panels show all structures found by ROCKSTAR, while the right panels show only the central halos according to our central-satellite split. Massive halos ($\gtrsim 10^{14}\,h^{-1}M_\odot$) with low $V_\mathrm{max}$ ($\lesssim 1000$~km/s), which appear far above the main locus in the left panels, are substructures located within more significant host halos and are mostly removed in the right panels. Another notable feature is the smaller scatter of the SO masses at a given $V_\mathrm{max}$ for larger values of $\Delta$, consistent with $V_\mathrm{max}$ being a better tracer of central density peaks.

\begin{figure}[t]
 \centering
 \includegraphics[width=\linewidth]{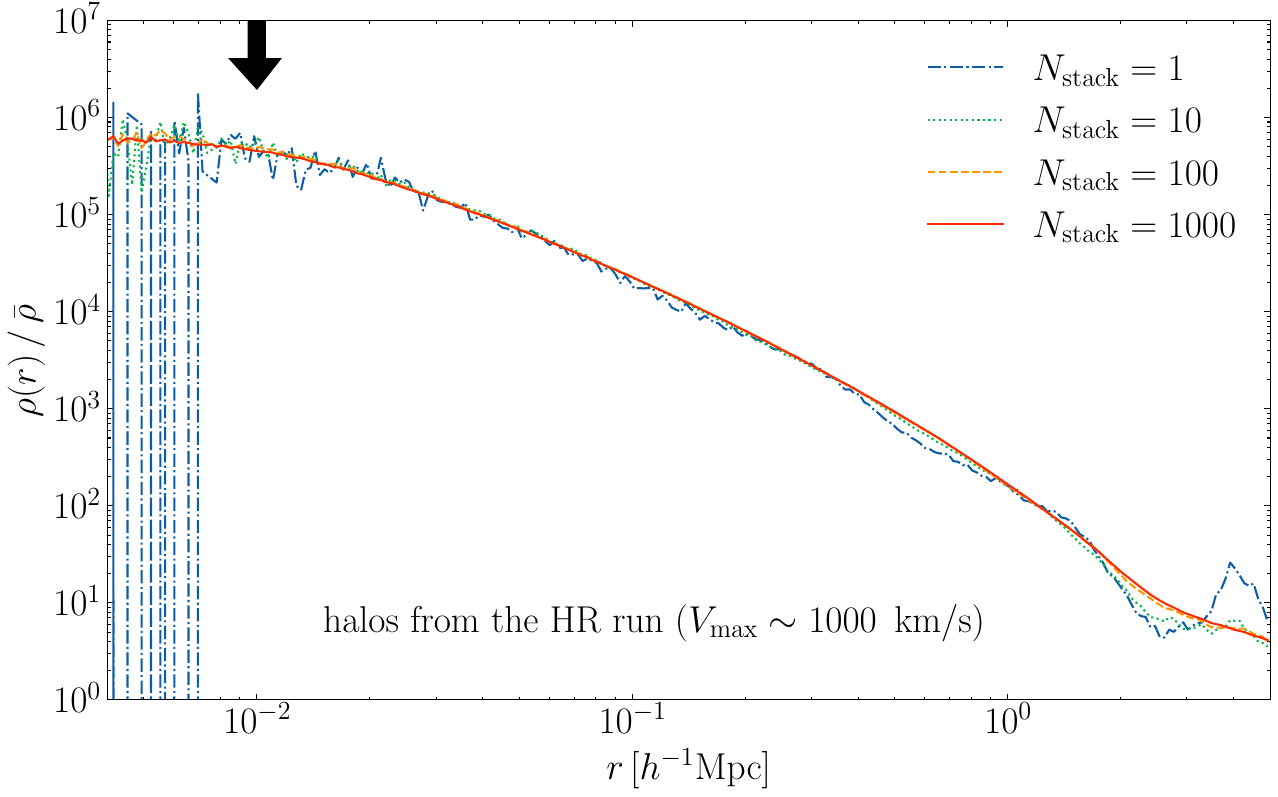}
 \caption{Averaged radial density profile of halos. We randomly select $1$, $10$, $100$ and $1000$ halos with $988\,\mathrm{km/s} < V_\mathrm{max} < 1014\,\mathrm{km/s}$ from the HR simulation at $z=0$. The vertical arrow indicates the softening length. {Alt text: Plot of the averaged radial density profile of halos in a narrow circular-velocity range, for four stacking sample sizes. The profile becomes progressively smoother as more halos are stacked, with the softening scale marked by an arrow.} \label{fig:stacked_profile}}
\end{figure}

\begin{figure}[t]
 \centering
 \includegraphics[width=\linewidth]{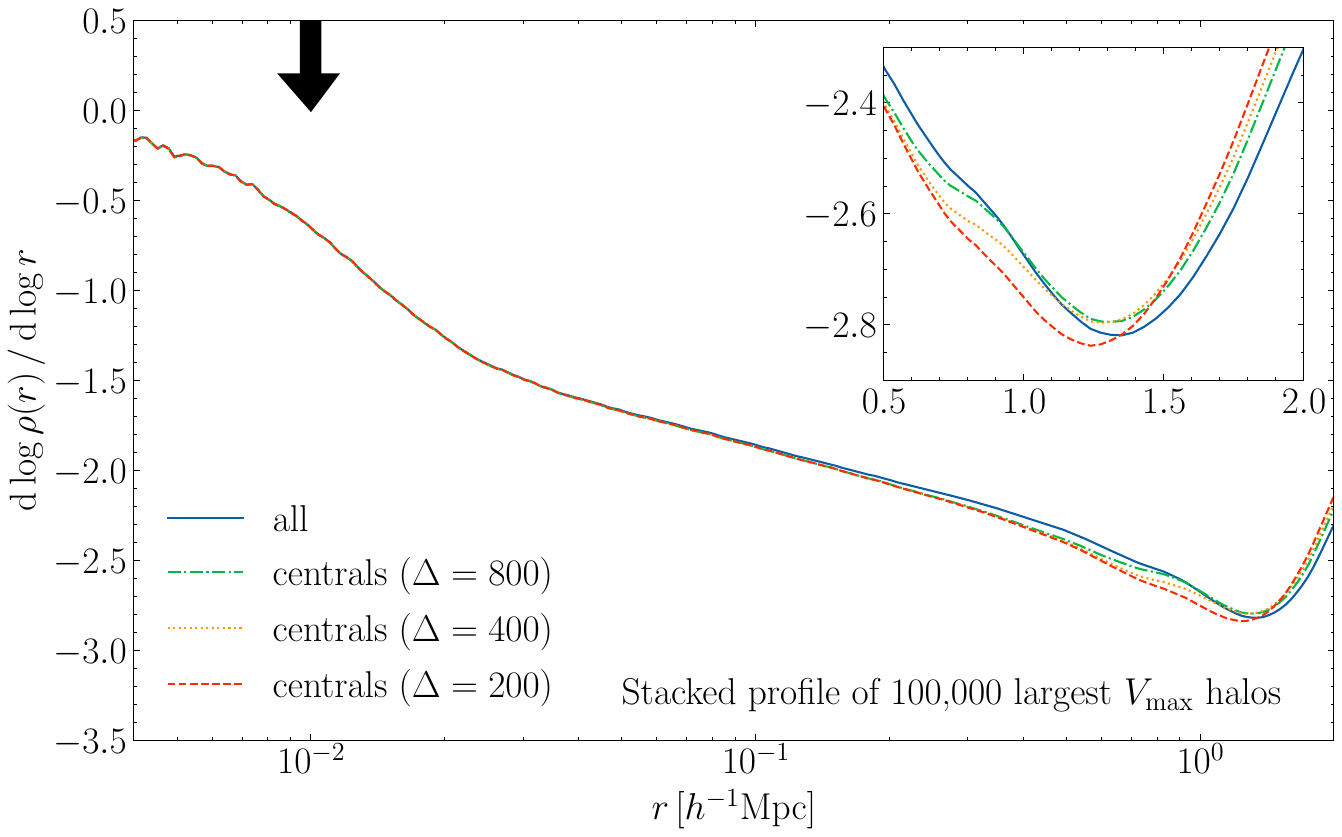}
 \caption{Logarithmic slope of the averaged radial density profile of central halos in the HR simulation at $z=0$ for different density thresholds. We consider the $100{,}000$ halos with the largest $V_\mathrm{max}$ before the central-satellite split (solid), and after discarding satellites with three values of $\Delta$ (dashed: 200, dotted: 400 and dot dashed: 800). The inset zooms in to show how the location of the steepest slope changes with the definition of central halos. We use a fifth order Savitzky--Golay filter with a window length of $31$ sample points to evaluate a smooth derivative. The vertical arrow indicates the softening length. {Alt text: Plot of the logarithmic slope of the stacked radial density profile for different central-halo definitions. An inset zoom shows that the location of the steepest slope shifts depending on how satellites are removed at different overdensity thresholds.}\label{fig:logslope}}
\end{figure}

Once the post-processed halo catalog is produced, studying the mass profile and related properties becomes straightforward. Figure~\ref{fig:stacked_profile} demonstrates how a smooth prediction of the radial density profile from the central region to the outer edge can be obtained by stacking an increasing number of halos. The logarithmic slope of the stacked profile is depicted in figure~\ref{fig:logslope} for different definitions of central halos. The so-called splashback feature, which is the dip in the slope marking the first apocenter of mass elements after accretion onto halos \citep{Diemer14}, is sensitive to how the central halos are selected (or, equivalently, how satellite halos are removed). Our halo database is useful for quantifying these interesting features for different halo definitions and cosmologies. Additionally, the stacked profile available on small scales can be combined with FFT-based estimates to make predictions for the halo-mass cross-correlation function over a wide range of scales (see subsection~\ref{subsec:spectra}).

\subsection{Halo mass function}
\label{subsec:hmf}
\begin{figure*}[ht]
 \centering
 \includegraphics[width=\linewidth]{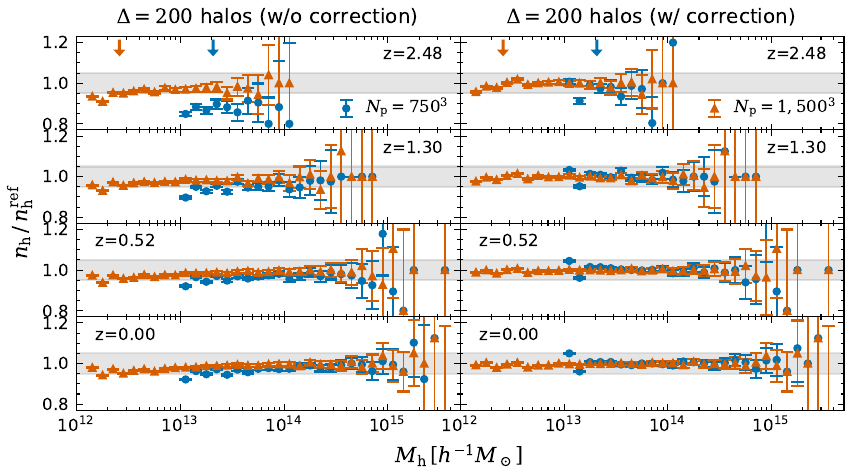}
 \caption{Comparison of the halo mass function measured from simulations with different resolutions. We plot the ratio of the halo mass function measured from simulations started with exactly the same initial conditions in a $(1\hiGpc)^3$ box, but traced by different numbers of particles. We adopt the highest resolution simulation with $3,000^3$ particles as the reference, and plot the results of lower resolution simulations (circles: $750^3$ particles and triangles: $1,500^3$ particles) relative to this reference. We use the spherical overdensity masses with an overdensity threshold of $\Delta = 200$ relative to the cosmic mean. We show the results without (left) and with correction (right) according to equation~ (\ref{eq:mass_correction2}). The shaded regions indicate the $\pm 5\%$ band relative to the reference. Also, the two vertical arrows at the top of each panel indicate the masses corresponding to $100$ particles for the $1,500^3$ (left) and $750^3$ (right) runs. {Alt text: Eight-panel comparison (four rows by two columns) of the halo mass function ratio against halo mass for two lower-resolution simulations relative to the highest-resolution reference. Rows from top to bottom correspond to redshifts 2.48, 1.30, 0.52, and 0.00. The left column uses raw masses and the right column applies the empirical correction, which brings all resolutions within the five-percent band.} \label{fig:HMF_th200}}
\end{figure*}

Building on the halo catalogs and mass corrections introduced above, we now examine how finite resolution shapes the measured halo mass function. As mentioned at the beginning of section~\ref{sec:accuracy}, the finite number of particles in a simulation ultimately imposes limitations on the accuracy of the results, once the convergence against the internal parameters is established. While the particle Nyquist wavenumber, defined in equation~(\ref{eq:kNy}), provides a rough estimate of the upper wavenumber limit for reliable predictions in Fourier space, this limit does not directly translate to halo masses. 

One consequence of finite mass and force resolution is the potential misestimation of the halo mass function. This can result from delayed structure formation on small scales, as seen in the power spectrum at high wavenumbers. Furthermore, halo identification and mass determination can become less accurate at lower resolution, even if the nonlinear dynamics are accurately followed. In this subsection, we examine these issues using the same set of three simulations as above (i.e., LR, MR and HR runs). Prior to this, we verified that tightening the internal parameters of the simulation code (relative to the fiducial parameters) does not significantly improve the halo mass function unless the mass resolution is increased by using more particles (not shown). As discussed in the previous subsection, we instead mitigate this limitation by applying an empirical correction (equation~\ref{eq:mass_correction2}) to the halo masses.

To assess the impact of mass resolution on the halo mass function, the left panel of figure~\ref{fig:HMF_th200} shows the halo mass function measured from the MR and LR runs, normalized to the HR run. We consider halos with a density threshold of $\Delta=200$, and focus on mass bins resolved by more than $50$ simulation particles, using raw mass estimates from the radial density profile. A noticeable decline in the ratio near the low-mass end is evident in the two lower-resolution simulations. The right panel demonstrates that applying our empirical mass correction (equation~\ref{eq:mass_correction2}) improves the agreement in the mass function across different resolutions. After the correction, the halo mass function agrees to within the $\pm5\%$ band down to the low-mass end considered here (50 particles). The vertical arrows at the top of each panel mark the mass scale corresponding to $100$ simulation particles, above which the halo mass function can be considered reliable; we deliberately extend our analysis down to $50$ particles to expose how the function degrades below this threshold.

This behavior holds across different overdensity thresholds, as shown in Appendix~\ref{sec:mass_def}: the empirical correction consistently improves convergence for various density thresholds, supporting the robustness of this approach across a range of halo mass definitions.

While the demonstration here focuses on the fiducial cosmological model, it is important to assess whether this approach holds for other cosmological models. Ideally, we would verify the empirical correction across a variety of cosmological parameters. Although we have not conducted such tests explicitly, figure~\ref{fig:HMF_th200} and the plots in the Appendix present results at different redshifts, which provide indirect evidence for the robustness of the method. Although this is not a substitute for a full cosmology-dependent calibration, the success of the correction across redshifts suggests that it is not strongly tied to a single nonlinear amplitude or halo population.

\subsection{Halo shapes}
\label{subsec:halo_shape}

Recent studies have focused on the additional information carried by the shapes of galaxies, which are affected by the cosmological tidal field~\citep{Chisari_2013,Schmidt_2015,Okumura_2019,Akitsu_2021,Harvey_2021,Okumura_2022,2023PhRvD.108h3533K,van_Dompseler_2023,Xu_2023}; see also \citet{Lamman_2024} for a recent review. This can be observed as the correlation between orientations of different galaxies or between the orientation and the overdensity field, i.e., the intrinsic alignment of galaxies. To understand this, alignments between halos are a good starting point.

The ROCKSTAR~\citep{Behroozi:2013} halo finder, which is the baseline of our halo catalogs, stores information on the lengths of the major and minor axes. However the default output does not support the directions of these axes. For studies of halo intrinsic alignment, the primary quantity of interest is the direction of the major axis of each halo. We therefore developed post-processing routines to measure these directions and related quantities. We use the spatial coordinates of the halo member particles recorded by ROCKSTAR to quantify the halo shapes and orientations. Specifically, we measure the shapes at various distances from the halo center to enable predictions for different galaxy populations and estimators (see, e.g., \cite{Shi_2021}). Also, we consider three different weights to define the mass tensor for determining halo shapes~\citep{Jing2002-xk,Suto2016-hi}. These are useful as different galaxy shape definitions or estimators may trace the underlying mass distribution at different scales. Furthermore, we also track halo--subhalo relations in a hierarchical manner to see how individual subhalos or subsubhalos align within the hierarchy. This can be used to make predictions for the intrinsic alignment signal using the halo model approach \citep{Cooray02,Fortuna_2020}, by separating the correlation signals within a halo and between distinct halos (i.e., one- and two-halo terms for the two-point correlation function).

For all radii and weighting schemes, we model the shape of halos as a triaxial ellipsoid:
\be
\left(\dfrac{x_1}{A_1}\right)^2 + \left(\dfrac{x_2}{A_2}\right)^2 + \left(\dfrac{x_3}{A_3}\right)^2 = 1, 
\label{eq:ellipsoid}
\ee
where the three axis lengths are given by $A_1 \leq A_2 \leq A_3$, and $x_1, x_2, x_3$ denote the coordinates defined along the three axes, with the origin set at the center of the ellipsoid. The ellipticity is defined by
\be
e = \dfrac{A_3 - A_1}{2\left(A_1 + A_2 + A_3\right)}.
\label{eq:ellipticity}
\ee
We determine the three axes and their orientations by measuring the inertia moment tensor, which we refer to as the mass tensor. Three different weighting schemes are employed to define the mass tensor. The simplest choice is unweighted:  
\be
I_{\alpha \beta} = \sum^{N_e}_{i=1} x^i_\alpha x^i_\beta, 
\label{eq:I_uw}
\ee
where $x^i_\alpha$ and $x^i_\beta$ $(\alpha,\,\beta = 1,\,2,\,3)$ represent the position of the $i$-th particle along the $\alpha$-th and $\beta$-th axes, and the summation is taken over the $N_e$ particles within the ellipsoid. Another definition incorporates a weight based on the particle's distance from the halo center:
\be
\hat{I}_{\alpha \beta} = \sum^{N_e}_{i=1} \dfrac{x^i_\alpha x^i_\beta}{ |\boldsymbol{x}^{i}|^2}.
\label{eq:I_sw}
\ee
Finally, we consider a weighting scheme that uses the ellipsoidal distance:
\be
\tilde{I}_{\alpha \beta} = \sum^{N_e}_{i=1} \dfrac{x^i_\alpha x^i_\beta}{|R^{i}_{e}|^2},
\label{eq:I_ew}
\ee
where
\be
|R^{i}_{e}|^2 = \left(\dfrac{x^i_1}{A_1}\right)^2 + \left(\dfrac{x^i_2}{A_2}\right)^2 + \left(\dfrac{x^i_3}{A_3}\right)^2. 
\ee

To measure these three tensors, the center of the ellipsoid and the three axes must be specified beforehand to identify the particles included in the summation. Consequently, an iterative procedure is required until convergence. We initialize the process using the center positions provided in the ROCKSTAR output and assume that the halos are spherical. For a given number of particle, $N_e$, we define the outer boundary, compute the mass tensor, and update the center-of-mass position based on the particles within the boundary. This process is repeated with the updated center and the axes for each of the three definitions. By default, we use $20$ values of $N_e$, corresponding to interior masses $0.05\, i\, M_\mathrm{vir}$ $(i=1,\,2,\,\cdots,\,20)$, to construct the radial profile of shapes and orientations.

To account for the hierarchical structure of halos and subhalos, we include not only the member particles explicitly listed in the ROCKSTAR catalog but also those associated with all \textit{nested} substructures. For example, when we measure the shape of a subhalo, we recursively include the substructures it contains within $R_\mathrm{vir}$, together with substructures at progressively deeper levels of the hierarchy. In DQ1, we primarily used the direct member particles of the (sub)halo of interest (e.g., in \cite{Okumura_2019,Kurita_2020}), but this approach often resulted in noticeable ``holes'' in the particle list. These gaps are naturally filled by including the member particles of the nested substructures, leading to more reliable estimates of halo shapes and orientations.

\begin{figure}[t]
 \centering
 \includegraphics[width=\linewidth]{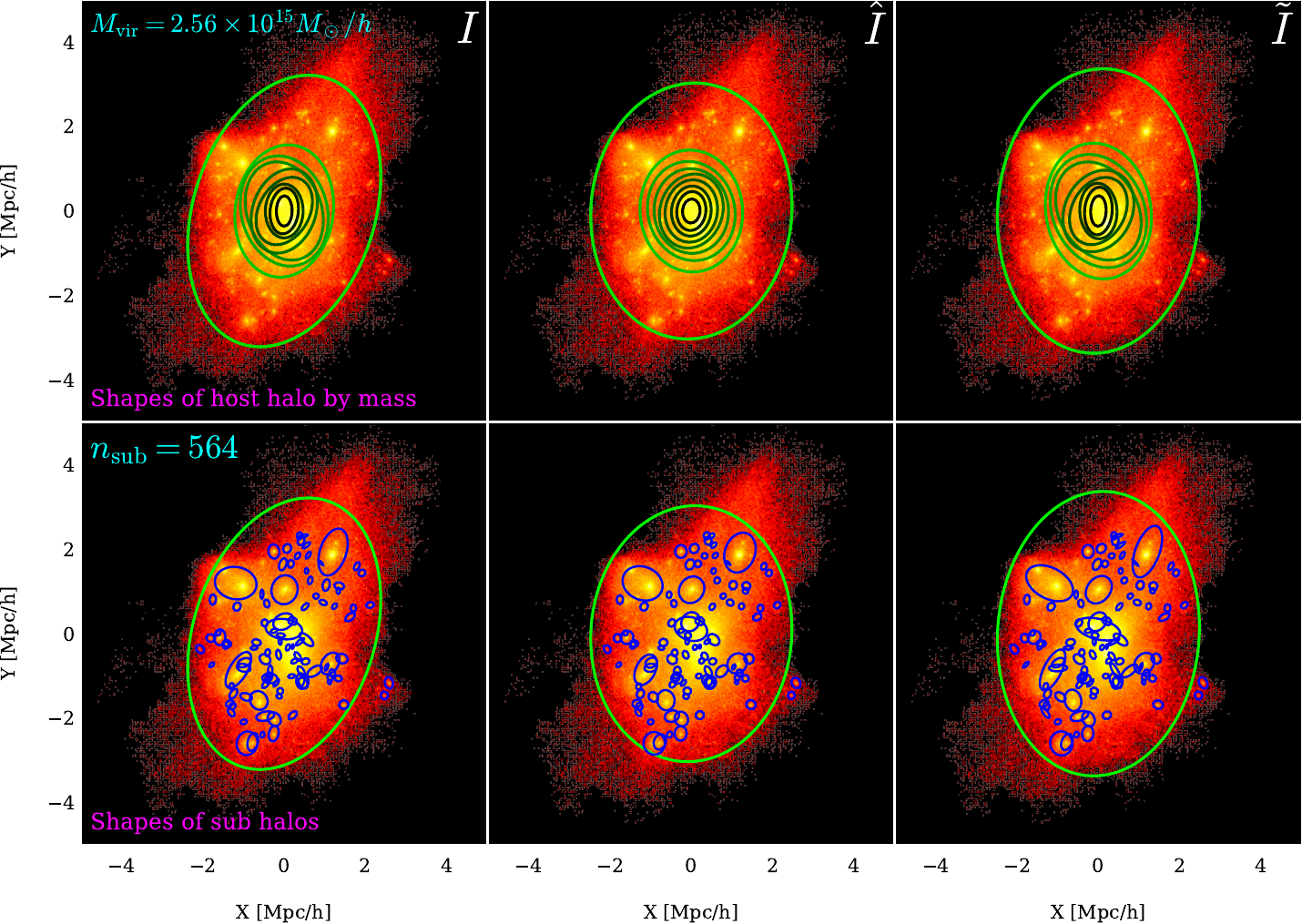}
  \caption{Projected image of a host halo and its subhalo shapes. The upper panels display the elliptical boundaries of the host halo shapes at various enclosed masses. The outermost boundary corresponds to $M_\mathrm{vir}$, with each subsequent inner boundary representing a 10\% decrement in mass, down to $0.3M_\mathrm{vir}$. The lower panels show the shapes of the top $100$ most massive subhalos within the host halo, with the boundary corresponding to $M_\mathrm{vir}$ for each subhalo. Each column represents a different weighting scheme used for the mass tensor calculations as indicated in the figure legend.
 {Alt text: Two-row by three-column gallery of projected halo-shape contours. Columns from left to right correspond to the unweighted, inverse-square-weighted, and ellipsoidal-distance-weighted schemes. The upper row shows host-halo isodensity ellipses at several enclosed masses, and the lower row shows the same schemes applied to the top one hundred subhalos within the host.}}
 \label{fig:halo_shape}
\end{figure}

Figure~\ref{fig:halo_shape} presents a two-dimensional projection of halo shapes as illustrated by elliptical boundaries, using the three different weighting schemes. The upper panels show ellipses at various radii for each scheme. The unweighted case, depicted in the left panel, appears the most stretched, particularly at the largest radius corresponding to $M_\mathrm{vir}$, as it is influenced significantly by contributions from the outer regions of the halo. In contrast, the other two estimators, which down-weight distant particles, produce shapes that appear more spherical overall. The two weighted schemes differ in that $\tilde{I}$ assigns equal weight to particles located on the same ellipsoidal contours. During the iterative estimation, $\tilde{I}$ is more sensitive to the iterative update than $\hat{I}$ because the weighting itself depends on the current axis estimate. This sensitivity is evident in the more intricate contour shapes observed at intermediate radii.

\begin{figure}[t]
 \centering
 \includegraphics[width=\linewidth]{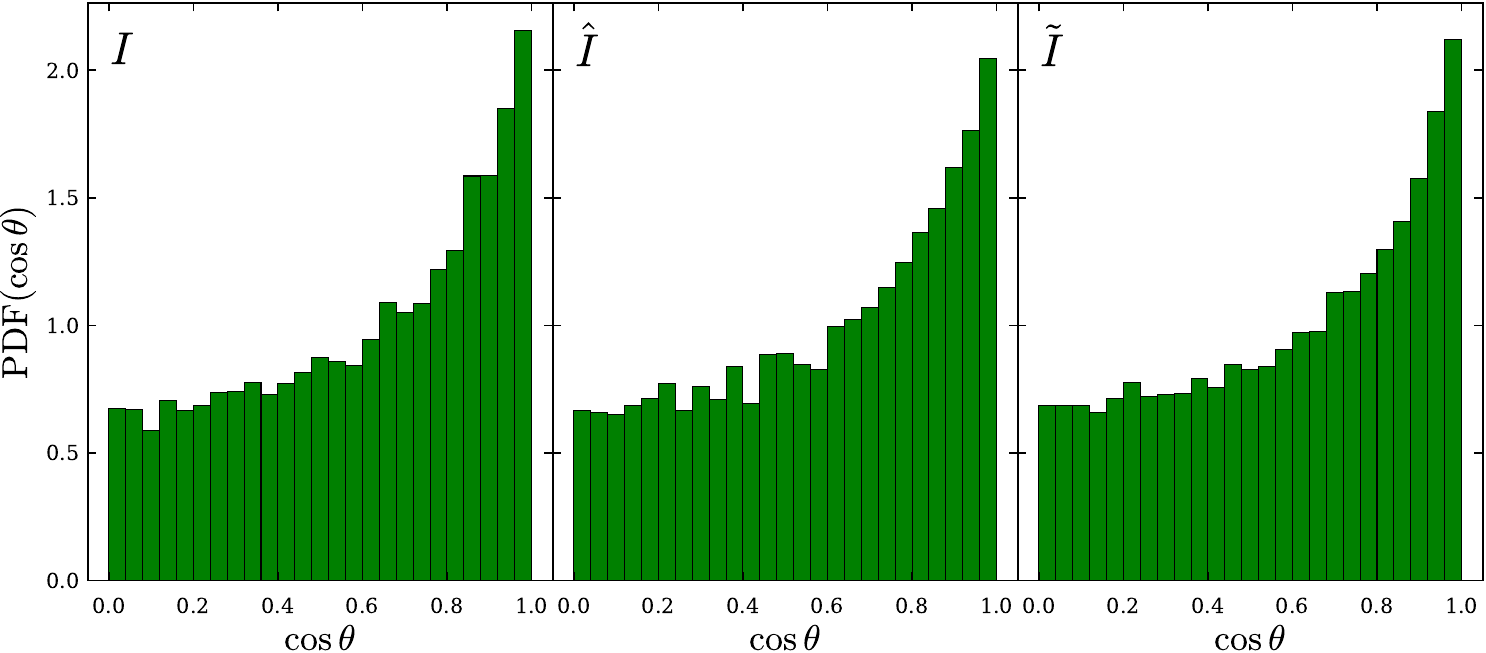}
  \caption{Distribution of the angle between the major axis of subhalo shapes and the direction toward the host halo center. We stack subhalos contained within the top $20$ most massive host halos in the simulation box ($M_\mathrm{vir} >2 \times 10^{15}\,h^{-1}\,M_\odot$). The increasing trend indicates that the major axes of subhalos are preferentially oriented toward the center of the host halo. Each column corresponds to a different weighting scheme as indicated in the figure legend.
 {Alt text: Three-panel plot of the PDF of the cosine of the angle between the subhalo major axis and the direction toward the host-halo center. Columns from left to right correspond to the unweighted, inverse-square-weighted, and ellipsoidal-distance-weighted schemes. All schemes show a distribution increasing with the cosine, indicating preferential alignment toward the host center.}}
 \label{fig:subhalo_angle}
\end{figure}

The lower panels depict the shapes of the same host halo as shown in the upper panels, along with its $100$ most massive subhalos at $M_\mathrm{vir}$. Notably, the major axes of the subhalos tend to align toward the center of the host halo, a trend consistent across all three estimators. This alignment is quantified in figure~\ref{fig:subhalo_angle}, where we stack subhalos within the $20$ most massive host halos in the simulation box. This figure shows the distribution of the cosine of the angle between the major axis of each subhalo and the direction toward the center of the host halo. For this analysis, we focus on massive halos with $M_\mathrm{vir}>2\times 10^{15}\, h^{-1} M_\odot$. A flat probability distribution function (PDF) would indicate random orientations, while the observed increase with $\cos\theta$ confirms that the major axes of the subhalos are preferentially oriented toward the center of the hosts. Although the axis ratios of the subhalos differ significantly depending on the weighting schemes, as we will discuss shortly, their orientations are notably robust to this choice.

\begin{figure}[t]
 \centering
 \includegraphics[width=\linewidth]{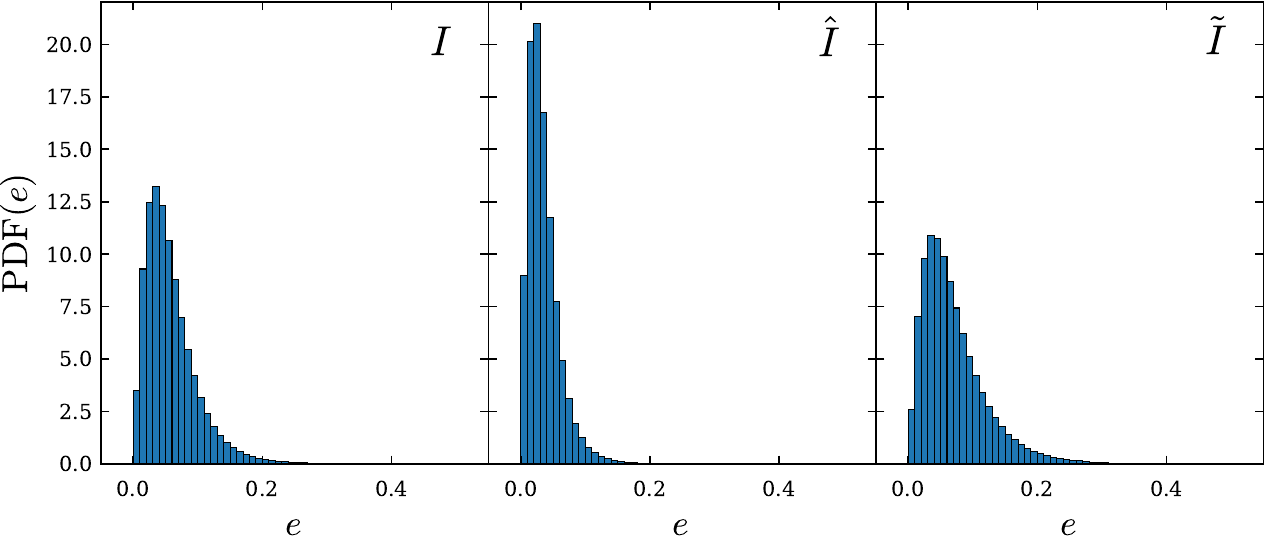}
  \caption{Distribution of ellipticities at $M_\mathrm{vir}$ for all halos more massive than $10^{11}\,h^{-1}\,M_\odot$ found by ROCKSTAR in a simulation box (without removing subhalos). Columns correspond to the unweighted $I$ (equation~\ref{eq:I_uw}), the inverse-square weighted $\hat{I}$ (equation~\ref{eq:I_sw}), and the ellipsoidal-distance weighted $\tilde{I}$ (equation~\ref{eq:I_ew}), as indicated in the figure legend.
 {Alt text: Three-panel histogram of halo ellipticity at the virial mass. Columns from left to right correspond to the unweighted, inverse-square-weighted, and ellipsoidal-distance-weighted schemes. The inverse-square-weighted (middle) panel shows the narrowest distribution centered on the smallest ellipticities, while the unweighted (leftmost) panel shows the broadest with the largest ellipticities.}}
 \label{fig:shape_ellipticity}
\end{figure}

Figure~\ref{fig:shape_ellipticity} shows the distribution of the ellipticity, as defined in equation~(\ref{eq:ellipticity}), for the three weighting schemes. We can see that the distributions clearly differ across the three weighting schemes. Among them, the shapes estimated using $\hat{I}$, which employs the inverse square of the distance from the halo center as the weight, exhibit the smallest ellipticities. On the other hand, while $\tilde{I}$ also downweights contributions from distant particles, its distribution is similar to the unweighted case, as it effectively gives larger relative weight to particles along elongated directions.

\begin{figure}[t]
 \centering
 \includegraphics[width=\linewidth]{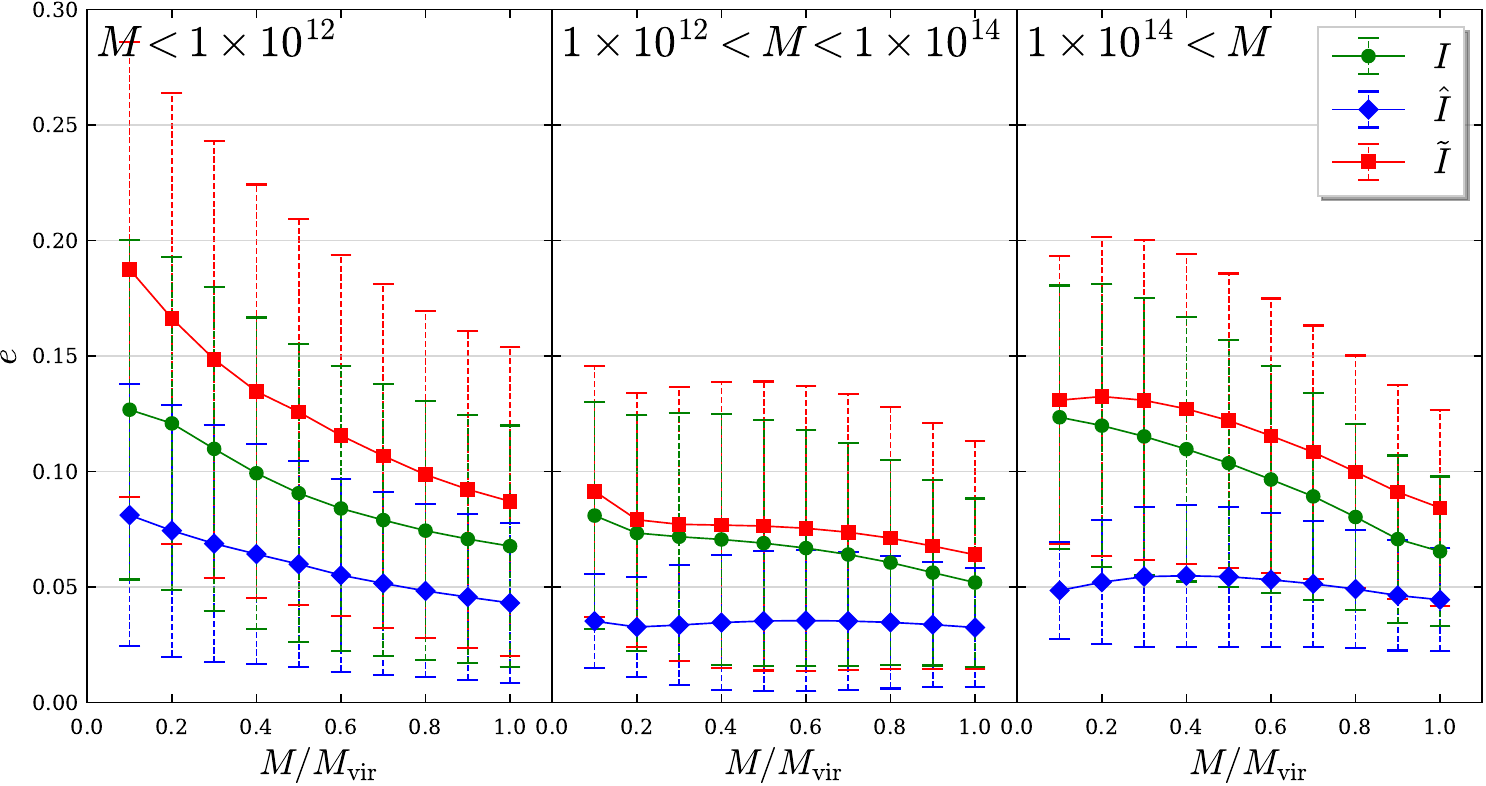}
  \caption{Dependence of the ellipticity on the distance from the center. We plot the mean and the standard deviation of the ellipticity as a function of the enclosed mass normalized by the virial mass. We consider the three weighting schemes for three different mass ranges as indicated in the figure legend.
 {Alt text: Three-panel plot of mean ellipticity with standard-deviation error bars against enclosed mass normalized by the virial mass. Columns from left to right correspond to the unweighted, inverse-square-weighted, and ellipsoidal-distance-weighted schemes, each showing three host-mass ranges. Ellipticity increases toward the halo center and varies notably between schemes.}}
 \label{fig:mass_ellipticity}
\end{figure}

We then show in figure~\ref{fig:mass_ellipticity} the dependence of ellipticity on the distance from the center. In the plot, we use the enclosing mass relative to the virial mass as a proxy for distance. Each data point corresponds to one of the enclosed-mass fractions used for the contours in the top panel of figure~\ref{fig:halo_shape}. The ellipticity is observed to be higher near the halo center, with this dependence being more pronounced for less massive halos. Additionally, we note that the choice of weighting scheme can significantly impact the estimated ellipticity, which is consistent with the visual impression from figure~\ref{fig:halo_shape}. The error bars represent the standard deviation across different halos, and we observe that the halo-to-halo scatter is quite large~\citep{Suto_2016}.

\begin{figure}[t]
 \centering
 \includegraphics[width=\linewidth]{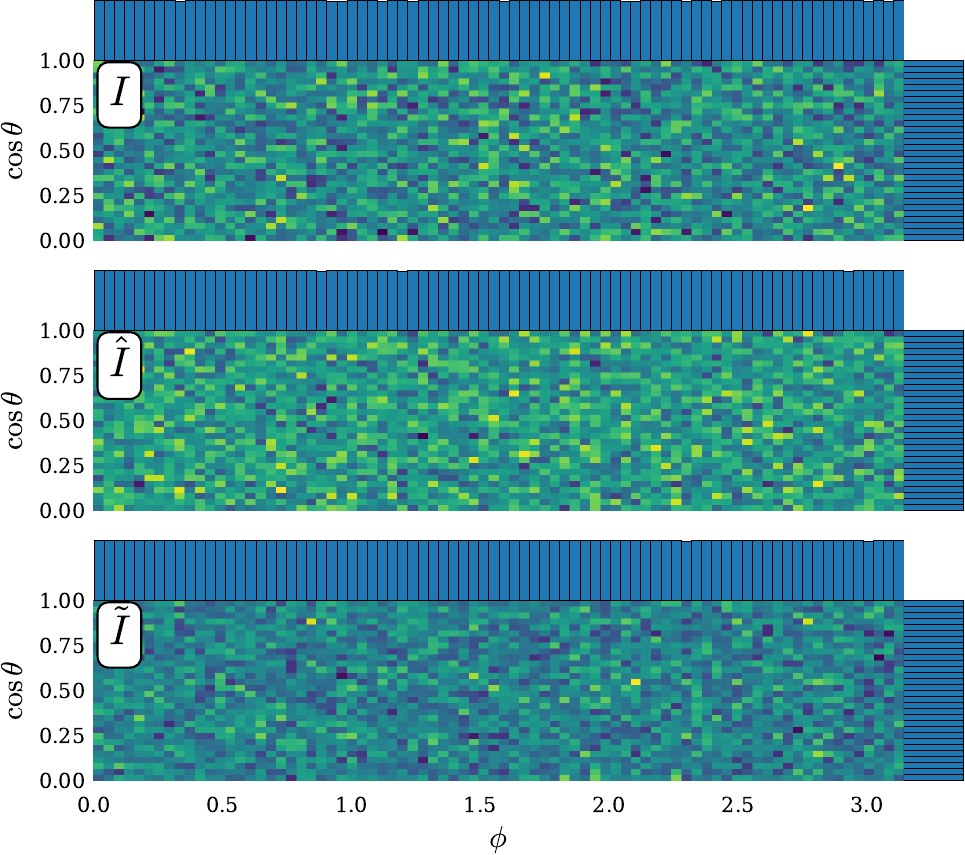}
  \caption{Joint distribution of the polar angle $\theta$ and the azimuthal angle $\phi$ of the eigenvector associated with the major axis, relative to the global Cartesian coordinates employed in the simulation. We use all halos more massive than $10^{11}\,h^{-1}\,M_\odot$ found by ROCKSTAR (without removing subhalos). The three panels show the results for the unweighted $I$, the inverse-square weighted $\hat{I}$, and the ellipsoidal-distance weighted $\tilde{I}$ schemes from top to bottom. The 1D projected distributions for $\phi$ and $\cos\theta$ are also shown by histograms on the top and right edges of each heatmap, respectively.
 {Alt text: Three two-dimensional heatmaps of the polar and azimuthal angles of the halo major-axis eigenvector, stacked vertically for the unweighted, inverse-square-weighted, and ellipsoidal-distance-weighted schemes. Each heatmap has one-dimensional marginal histograms attached to its upper and right edges. The distributions are consistent with isotropy across the simulation volume.}}
 \label{fig:shape_angle}
\end{figure}

As a final test of the validity of our implementation, we show the distribution of the direction of the major axis in the simulation box. In figure~\ref{fig:shape_angle}, we present a heatmap of the polar angle $\theta$ and the azimuthal angle $\phi$ of the eigenvectors associated with the largest eigenvalues. Along the upper and right boundary of the heatmap, we also show as histograms the 1D distributions of $\phi$ and $\cos\theta$, respectively. We do not observe any preferential direction in the figure, which confirms that the measurement is functioning correctly without being biased, for example, by the initial guess for the directions of the three axes in the iteration process.

\subsection{Spatial correlation analyses}
\label{subsec:spectra}

\begin{figure}[t]
 \centering
 \includegraphics[width=\linewidth]{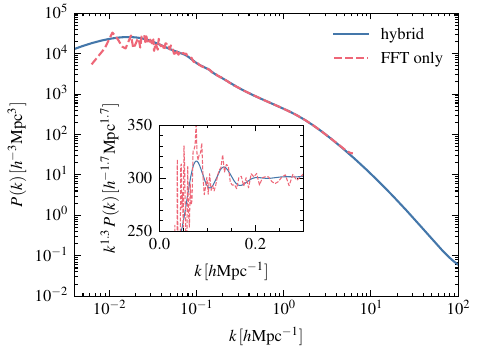}
 \caption{Matter power spectrum measured from a single simulation box with a comoving volume $(1\,\hiGpc)^3$ traced by $N_\mathrm{p}=3{,}000^3$ particles. The solid line represents the final estimate obtained using our method, while the dashed line corresponds to the standard FFT-based approach. The inset highlights a zoomed-in view of the BAO wiggles, illustrating the improved smoothness and precision of our method. {Alt text: Plot of the matter power spectrum measured from one simulation box, comparing the new hybrid estimator with the standard fast-Fourier-transform estimator. An inset zooms in on the baryon acoustic oscillation wiggles, showing improved smoothness from the new method.}\label{fig:power_measurement}}
\end{figure}

In this subsection we describe a measurement procedure designed to reduce these finite-volume and aliasing-related artifacts. As illustrated in figure~\ref{fig:power_measurement}, the standard FFT-based approach to measure the power spectrum exhibits noisy fluctuations, particularly on large scales, due to the finite number of Fourier modes available in the simulation box. Furthermore, the wavenumber range accessible by the standard method is constrained by the Nyquist wavenumber, beyond which the results are no longer reliable (indicated by the dashed line). Our new approach addresses these limitations, providing a smoother and more robust estimate of the power spectrum across an extended wavenumber range.

\subsubsection{Hybrid estimation of correlation functions}
\label{subsubsec:hybrid}
We measure the correlation function across the entire range of separations accessible from the simulation data.

To do this, we develop an efficient hybrid scheme to measure the correlation function, designed to exploit the full dynamical range of our simulations. Following the approach initially introduced in \citet{2019ApJ...884...29N}, we adopt a strategy analogous to the division of force into long- and short-range components in the simulation code. On large separations, the correlation function is computed using a grid-based method, leveraging FFT to achieve high computational efficiency. This approach is accurate as long as the pair separation significantly exceeds the grid spacing. For small separations, however, we switch to direct pair counting. Although a naive implementation of pair counting has a computational complexity of $\mathcal{O}(N^2)$, where $N$ is the number of particles or tracers being counted, setting a finite search radius makes the calculation feasible within a reasonable time frame. This hybrid method is employed to measure the correlation functions of halos, matter and their cross correlations.

We use $2,000^3$ FFT grids as the default setup for simulations with $N_\mathrm{p}=3,000^3$ particles. Note that this FFT grid is defined independently of the PM grid used inside the simulation code for force computation; we adopt a different mesh size here for post-processing efficiency. Particle masses are assigned to the FFT grid points using the Cloud-in-Cell (CIC) interpolation kernel \citep{hockney81}. The correlation function is measured by calculating the square norm, $\left|\delta_{\boldsymbol{k}}\right|^2$, on each grid point in Fourier space, followed by an inverse FFT to obtain the correlation function estimator $\xi(\boldsymbol{x})$. Denoting the pair separation by $x \equiv |\boldsymbol{x}|$, the angle-averaged estimate $\xi(x)$ is then obtained by binning radially in $x$, and the average within each bin provides the estimate of the correlation function on large scales.

The FFT-based method loses accuracy on scales approaching the grid spacing, because the CIC mass assignment kernel smooths the density field on scales comparable to $d_\mathrm{g} \equiv L_\mathrm{box}/N_\mathrm{FFT}^{1/3}$, suppressing the inferred correlation function there. We therefore switch to direct pair counting for pair separations smaller than $x_\mathrm{max} = 10\,d_\mathrm{g}$, which serves as the stitching scale between the two methods.

A naive direct pair counting scales as $\mathcal{O}(N^2)$ and is computationally prohibitive for $N_\mathrm{p} = 3{,}000^3$ matter particles. To reduce this cost, we first partition the simulation volume into a coarse grid with cell size $\sim x_\mathrm{max}$. For any given particle, only particles in the same cell or its immediate neighbors can lie within the search radius; all others are discarded without explicit distance evaluation. This neighbor-list approach reduces the effective cost to $\mathcal{O}(N \bar{n} x_\mathrm{max}^3)$, where $\bar{n}$ is the mean number density, making the computation tractable in practice. For each accepted pair at separation $x = |\boldsymbol{x}_i - \boldsymbol{x}_j| < x_\mathrm{max}$, the pair contribution is accumulated in radial bins. The angle-averaged count in each bin then yields the binned correlation function estimate on small scales, in exactly the same way as for the FFT-based estimate on large scales. The two estimates are stitched at $x_\mathrm{max} = 10\,d_\mathrm{g}$, where they are found to be consistent (see figure~\ref{fig:xicb_hybrid_scales}).

\subsubsection{Cleaning of the estimates}
\label{subsubsec:cleaning}

The raw estimate of the correlation function obtained from the procedure explained above still suffers from statistical error, often referred to as sample variance. This variance can be reduced only by averaging over many independent
realizations or by increasing the simulation volume. Here, to maximize the prediction accuracy from a small number of simulation realizations, we introduce a new cleaning scheme that works at the level of statistical quantities instead of random fields. To validate the accuracy of the procedure, we generate $100$ random realizations of smaller simulations with $N_\mathrm{p}=1024^3$ particles and the box size of $1024\,\hiMpc$ for the fiducial cosmology. We assess both the statistical and systematic errors by comparing the measurements with and without the procedure explained below.

\begin{figure}[t]
 \centering
 \includegraphics[width=\linewidth]{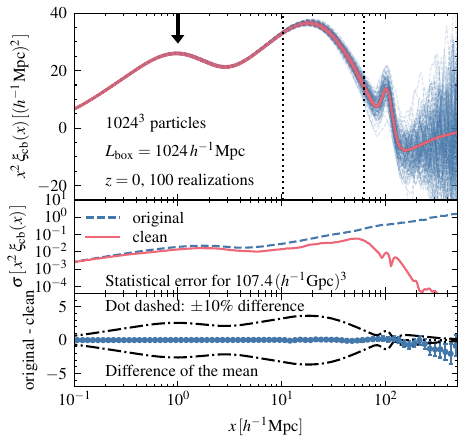}
 \caption{Comparison of the matter (cb) correlation function (scaled by separation squared) with and without the perturbation-theory based corrections. On the top panel, we show the measurements for the 100 realizations with (without) the corrections by the solid (dashed) lines. While $100$ solid lines are actually plotted, they look like a single bold line owing to their small scatter. The vertical arrow marks the FFT grid spacing. We adopt the FFT-based measurements above $\sim 10$ times the grid spacing, below which discrete sum of grid nodes as well as the mass assignment kernel can bias the result, and thus the direct pair counting method is adopted instead. The two vertical dotted lines show the two stitching scales, $x_1$ and $x_2$, explained in the text. The middle panel depicts the standard error of the two sets of measurements. These correspond to the statistical fluctuations on the mean of the $100$ realizations. The bottom panel shows the difference between the two results to assess the possible bias due to the corrections. We take the difference after averaging over the realizations. The difference is scaled by $x^2$ consistently with the other panels. We also show $\pm 10\%$ deviation by the dot-dashed lines. {Alt text: Three stacked panels of the matter correlation function multiplied by separation squared. The upper panel shows one hundred realizations with and without the cleaning procedure, the middle the standard error of the mean, and the lower the difference in the ensemble mean.} \label{fig:xicb_hybrid}}
\end{figure}

The basic idea behind this is that we know precisely the characteristics of the particular random realization under consideration in terms of its initial random phase and amplitude. Moreover, the nonlinear behavior of the two-point function around the baryon acoustic scale has been well studied in the literature (e.g., \cite{Meiksin99,eisenstein05,crocce08,matsubara08a}). Since the number of available modes around this scale is limited even for a $\sim\mathrm{Gpc}^3$ volume, as illustrated by the large scatter around the BAO peak in the top panel of figure~\ref{fig:xicb_hybrid}, we use the known perturbative description of BAO damping on these scales.

To do this, in addition to the measurement of the correlation function from a simulation by the hybrid pair counting-FFT based method, which we denote as $\hat{\xi}^\mathrm{sim}_\mathrm{cb}$, we compute three other estimates of the correlation function. First, we compute the linear correlation function, $\xi^\mathrm{L}_\mathrm{cb}$, based on the Fourier transform of the linear power spectrum:
\be
\xi_\mathrm{cb}^\mathrm{L}(x) = \frac{1}{2\pi^2} \int k^2 j_0(kx)P_\mathrm{cb}^\mathrm{L}(k)\mathrm{d}k,
\ee
where $j_0$ is the spherical Bessel function of order zero.
In addition, we measure the same quantity, but now with the random scatter corresponding to the particular random realization. To obtain this, denoted by $\hat{\xi}^{\mathrm{L}}_\mathrm{cb}$, we compute $|\delta_{\mathrm{cb},\bk}^{\mathrm{L}}|^2$ at three dimensional grid points, perform a three-dimensional inverse FFT, and finally take the average over the separation vectors with different directions in bins of $|\bx|$. This quantity shares a similar random noise pattern as $\hat{\xi}_\mathrm{cb}^\mathrm{sim}$ over the scales where nonlinearity corrections are small. Note that the quantities with a hat imply that they are a noisy estimate based on one random realization, while those without a hat are their expectation value, i.e., $\langle\hat{X} \rangle = X$. It is then easy to see that adding a term $\hat{\xi}^{\mathrm{L}}_\mathrm{cb}-\xi^{\mathrm{L}}_\mathrm{cb}$ to $\hat{\xi}_\mathrm{cb}^\mathrm{sim}$ partly cancels the sample variance while keeping the estimation unbiased, since $\langle\hat{\xi}^{\mathrm{L}}_\mathrm{cb}-\xi^{\mathrm{L}}_\mathrm{cb}\rangle = 0$. Multiplying instead by the factor $\xi^{\mathrm{L}}_\mathrm{cb}/\hat{\xi}^{\mathrm{L}}_\mathrm{cb}$ achieves a similar cancellation, but is unbiased only approximately---on scales where $\hat{\xi}^{\mathrm{L}}_\mathrm{cb}\simeq\xi^{\mathrm{L}}_\mathrm{cb}$---since $\langle\xi^{\mathrm{L}}_\mathrm{cb}/\hat{\xi}^{\mathrm{L}}_\mathrm{cb}\rangle \neq 1$ in general.

\begin{figure}[t]
 \centering
 \includegraphics[width=\linewidth]{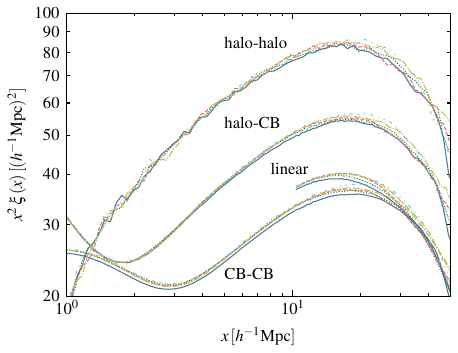}
 \caption{Correlation functions measured from five realizations randomly chosen from the 100 at $z=0$. We plot the halo auto, halo-cb cross and cb auto correlation functions from top to bottom. Also shown by the lines starting at around $10\,\hiMpc$ are the linear theory estimates with random noise corresponding to the same random realizations. We consider a halo sample with the number density $3.16\times 10^{-4}\,(\hiMpc)^{-3}$. The measurements from the same random realization are shown by the same line types. We can see how different correlation functions at different separations share the random noise. {Alt text: Single-panel plot of correlation functions measured from five random realizations, showing halo auto, halo-matter cross, matter auto, and the corresponding linear-theory estimates all overlaid. The realization-to-realization scatter shares the same ordering across different correlation functions on weakly nonlinear scales.} \label{fig:xi_realization}}
\end{figure}

We show in figure~\ref{fig:xi_realization} the measurements from five random realizations of comoving boxes with the side length $L_\mathrm{box} = 1024\,\hiMpc$. We show the auto and the cross correlation functions between the cb and halo density fields. Also shown by the lines starting at $\sim 10\,\hiMpc$ are the linear results with random scatter. This scale corresponds to $x_\mathrm{max} = 10\,d_\mathrm{g}$, the stitching scale introduced above, below which the direct pair counting method is used. We do not have a direct-pair-counting counterpart for $\hat{\xi}_\mathrm{cb}^\mathrm{L}$, since this linear field is constructed only on a grid and involves no particles. Different correlation functions from the same random realization are depicted by the same line style. We can see that the realization-to-realization scatter looks similar among different correlation functions. We also note that the ordering of different curves stays almost unchanged for the nonlinear ``cb-cb'' autocorrelation function. This suggests that the off-diagonal elements of the covariance matrix play a significant role. While they are noisier, the other correlation functions involving the halo density field show similar trends. As explicitly shown in what follows, we make use of this and estimate how the correlation function from a particular random realization deviates from its expectation value based on the comparison between $\xi_\mathrm{cb}^\mathrm{L}$ and $\hat{\xi}_\mathrm{cb}^\mathrm{L}$ even when we do not have multiple random realizations.

Finally, we compute another correlation function that accounts for the damping of the BAO feature. For each field $i$, this damping is characterized by the propagator, defined by the functional derivative of the nonlinear density contrast with respect to the initial linear cb density field \citep{crocce06b,crocce:2006uq}:
\be
\left\langle\frac{\delta \delta_{i,\bk}}{\delta \delta_{\mathrm{cb},\bk'}^{\mathrm{L}}}\right\rangle = (2\pi)^3\delta_\mathrm{D}^3(\bk-\bk')\,G_i(k).
\ee
Since we restrict ourselves to Gaussian random initial conditions, this can be estimated by the ratio of the auto and the cross spectra:
\be
G_i(k) = \frac{\left\langle\delta_{i,\bk}\,\delta_{\mathrm{cb},\bk}^{\mathrm{L}*}\right\rangle}{\left\langle|\delta_{\mathrm{cb},\bk}^{\mathrm{L}}|^2\right\rangle}.
\ee
Most of the sample variance is cancelled by taking the ratio of the two spectra, and the quantity $G_\mathrm{cb}$ is much less noisy compared to the power spectrum itself.
This function describes the ``loss of memory'' due to the large-scale random motion and exhibits a simple Gaussian-like damping behavior as a function of the wavenumber $k$, and approaches unity at the large scale limit~\citep{crocce:2006uq}. We can generalize this function to biased tracers by replacing $\delta_{\mathrm{cb},\bk}$ with the overdensity of any random field of interest, while keeping $\delta_{\mathrm{cb},\bk}^\mathrm{L}$ unchanged in the equations above. In that case, the large scale limit precisely corresponds to the linear bias factor.

\begin{figure}[t]
 \centering
 \includegraphics[width=\linewidth]{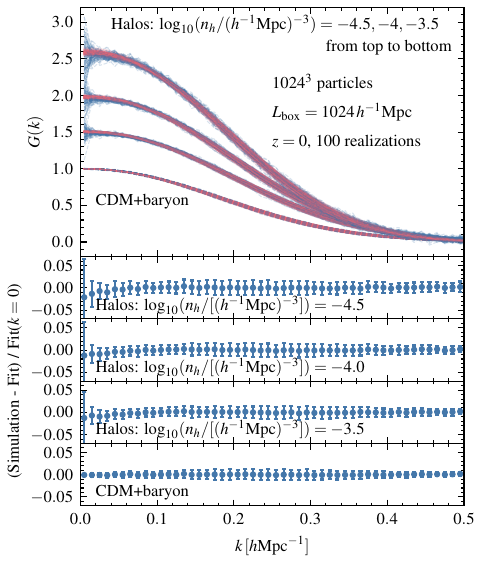}
 \caption{Propagators for the matter (cb) and halo samples with different number densities as indicated in the figure legend. The top panel compares the measurements from the 100 realizations (dashed) and fits to each of them (solid). The lower 3 panels show the residual of the fit normalized by the low-$k$ limit of the fit (i.e., the linear bias parameter). {Alt text: Four-panel plot. The top panel shows the propagator against wavenumber for matter and three halo samples, comparing measurements from one hundred realizations with their fits. The lower three panels show fit residuals normalized to the large-scale limit, scattering around zero.} \label{fig:prop}}
\end{figure}

We find that the propagator can be modeled by a simple four-parameter form, with the large-scale amplitude fixed to unity for the cb field:
\be
(a+bk^4)\exp\left\{-(ck)^d\right\}
\ee
where $a=1$ for the cb field, while for biased tracers $a$ corresponds to the large-scale linear bias. In the above, $c$ is roughly the rms displacement, and $d\simeq2$. The deviation from the simplest Gaussian form, which is exact for Zel'dovich approximation, can be accounted by introducing the kurtosis ($k^4$) term as well as allowing the index $d$ to depart from two. Note that we adopted a slightly different parameterization for the same function in DQ1, with a typical accuracy of the fit being $\sim 2\%$. This can be compared with the new fitting function demonstrated in figure~\ref{fig:prop} against $100$ random realizations in the fiducial cosmology. The figure shows that the fitting function typically achieves better accuracy than the DQ1 parameterization, both for matter (the cb fluid, more precisely) and halo fields, with no clear systematic trends as compared with the statistical error shown by the error bars corresponding to the $1024\,\hiMpc$ box.

Once the propagator is measured we can compute the BAO-damped correlation function between tracer $i$ and $j$ by
\be
\xi_{ij}^\mathrm{tree} = \mathcal{F}^{-1}\left[G_i(k)G_j(k)P_\mathrm{cb}^\mathrm{L}(k)\right],
\ee
where $\mathcal{F}^{-1}$ denotes the inverse Fourier transform. Note here that we use the linear power spectrum without sample variance, and a smooth curve for $G_i(k)$ obtained by the fit. Therefore, the quantity $\xi_\mathrm{cb}^\mathrm{tree}$ is little affected by the sample variance, and we omit the hat for the notation of this correlation function.

\begin{figure}[t]
 \centering
 \includegraphics[width=\linewidth]{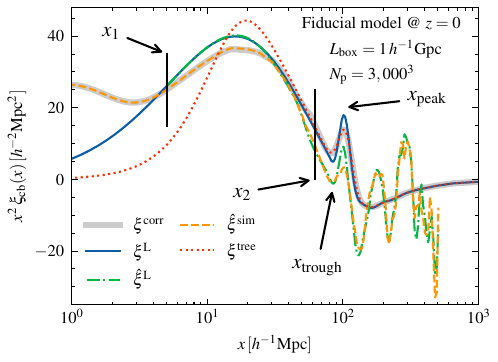}
 \caption{Our cleaning treatment on the cb correlation function. Various different estimates of the correlation function, multiplied by the separation squared, are plotted (see the legend for details) with the switching scales, $x_1$ and $x_2$, as well as $x_\mathrm{peak}$ and $x_\mathrm{trough}$ characterizing the BAO bump. The raw measurement from the simulation $\hat{\xi}_\mathrm{sim}$ is performed with the direct pair counting method below the scale $x_1$, and with the FFT method above it. The final estimate, $\xi^\mathrm{corr}$, closely follows one of the component estimates by construction. {Alt text: Plot of several correlation-function estimates multiplied by separation squared, against separation. Vertical lines mark the stitching scale $x_1$ between pair-counting and fast-Fourier-transform-based estimates and the BAO-region transition scale $x_2$; arrows with annotations identify the baryon-acoustic-oscillation peak and trough locations. The final estimate overlays the appropriate component in each regime.} \label{fig:xicb_hybrid_scales}}
\end{figure}

We compare the different correlation functions in figure~\ref{fig:xicb_hybrid_scales} measured from one random realization of our production run. On scales smaller than $x \sim 5\,\hiMpc$, nonlinear corrections are significant, and neither the linear nor the PT-based correlation function explains the increasing trend seen in $\hat{\xi}^\mathrm{sim}$ toward small separations. On such scales, we can rely only on the raw simulation result, $\hat{\xi}^\mathrm{sim}$. Even in such cases, as we saw in figure~\ref{fig:xi_realization}, the linear-theory ratio,  $\xi^\mathrm{L}_\mathrm{cb}/\hat{\xi}^\mathrm{L}_\mathrm{cb}$, can help to reduce the sample noise in the nonlinear correlation functions. We evaluate this ratio at the smallest scale at which the grid-based linear estimate is available, $x_1$, and multiply this to the nonlinear correlation function. We introduce a function $T_1(x; x_1,\Delta_1)$ to make a smooth transition from $\hat{\xi}^\mathrm{sim}$ to $(\xi^\mathrm{L}_\mathrm{cb}/\hat{\xi}^\mathrm{L}_\mathrm{cb})\hat{\xi}^\mathrm{sim}$ below $x_1$:
\be
\xi^\mathrm{corr}_{ij}(x) &=&
\left[1+\left(\frac{\xi^\mathrm{L}_\mathrm{cb}(x_1)}{\hat{\xi}^{\mathrm{L}}_\mathrm{cb}(x_1)}-1\right)T_1(x;x_1,\Delta_1)\right]\hat\xi_{ij}^\mathrm{sim}(x),\nonumber\\
&&\hspace{4cm}(x<x_1),\\
T_1(x;x_1,\Delta_1) &=& \exp\left[-\frac{(\log_{10}x-\log_{10}x_1)^2}{2\Delta_1^2}\right].
\ee
The function $T_1$ has a free parameter $\Delta_1$. We find that a value around unity, i.e., a slow transition over one order of magnitude in separation $x$, works best to suppress the sample noise. In practice, we adopt $\Delta_1 = 1$ for the auto correlation function of the cb density field. We instead adopt $\Delta_1 = 0.5$ for the cross correlation function of the halo and cb densities as well as the halo auto correlation function, whose noise is less strongly correlated with that of the linear cb correlation function (see figure~\ref{fig:xi_realization}).

On large scales, the two correlation functions, $\hat{\xi}^{\mathrm{L}}$ and  $\hat{\xi}^{\mathrm{sim}}$, exhibit similar random noise pattern. The departure of $\hat{\xi}^{\mathrm{L}}$ from its expectation value, $\xi^{\mathrm{L}}$, grows rapidly toward larger separation, indicating that $\hat{\xi}^{\mathrm{sim}}$ is also affected by a similar level of random scatter. On the other hand, the BAO-damped correlation function, $\xi^{\mathrm{tree}}$, is close to $\xi^{\mathrm{L}}$ on $x\gtrsim 40\hiMpc$ except that their difference looks prominent around the BAO bump at $\sim 100\,\hiMpc$. We consider a model to combine the simulation measurement, $\hat{\xi}^\mathrm{sim}$, with the linear-level noise correction, and the BAO-damped model, $\xi^\mathrm{tree}$. For biased tracers, we construct the corresponding linear estimates as $\hat{\xi}_{ij}^{\mathrm{L}} = b_i b_j \hat{\xi}_{\mathrm{cb}}^{\mathrm{L}}$ and $\xi_{ij}^{\mathrm{L}} = b_i b_j \xi_{\mathrm{cb}}^{\mathrm{L}}$, where $b_i$ denotes the large-scale bias inferred from the propagator fit. Then, our combined model is given by
\be
\xi^\mathrm{corr}_{ij}(x) &=& 
\bigl[1-T_2(x;x_2,\Delta_2)\bigr]\, \Bigl(\hat{\xi}_{ij}^\mathrm{sim}(x) -\hat{\xi}^\mathrm{L}_{ij}(x)+\xi_{ij}^\mathrm{L}(x)\Bigr)
\nonumber\\
&&\quad
+ T_2(x;x_2,\Delta_2)\,\xi^\mathrm{tree}_{ij}(x),\qquad(x>x_1),\\
T_2(x;x_2,\Delta_2) &=& \frac{1}{2}\left(1+\tanh\frac{\log_{10}x-\log_{10}x_2}{\Delta_2}\right).
\ee
Similarly to $T_1$, the function $T_2$ has parameters to specify the scale and width that determine how the transition occurs between the two estimates. We set the transition scale, $x_2$, at a scale somewhat smaller than the BAO bump:
\be
x_2 = x_\mathrm{trough} - 1.5 (x_\mathrm{peak}-x_\mathrm{trough}),
\ee
where $x_\mathrm{trough}$ and $x_\mathrm{peak}$ are the locations of the local minimum and maximum of the linear correlation function, $\xi_\mathrm{cb}^\mathrm{L}$. The transition width, $\Delta_2$, is set to $0.1$ for all the correlation functions.

\begin{figure}[t]
 \centering
 \includegraphics[width=\linewidth]{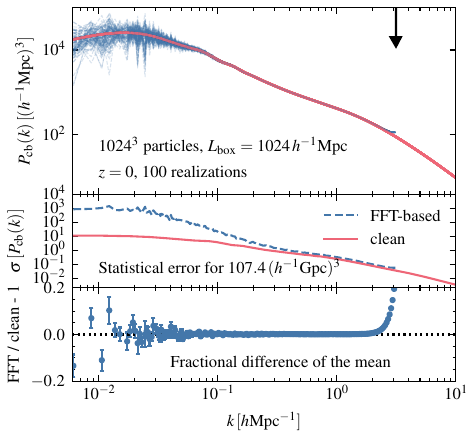}
 \caption{Matter (cb) power spectra measured with our new method (solid), as compared to the standard FFT-based method (dashed). We show the Nyquist wavenumber of the latter by the vertical arrow in the top panel. The middle panel compares the statistical error of the two methods. The bottom panel shows the fractional difference between the two sets of measurements. The FFT-based measurement starts to blow up near the Nyquist wavenumber, while the new method is not affected by this particular FFT-aliasing artifact. {Alt text: Three stacked panels of the matter power spectrum against wavenumber, comparing the new method with the standard fast-Fourier-transform-based method. The upper panel shows raw measurements, the middle the standard error, and the lower the fractional difference in the mean.} \label{fig:pcb_hybrid}}
\end{figure}

\begin{figure*}[t]
 \centering
 \includegraphics[width=0.4\linewidth]{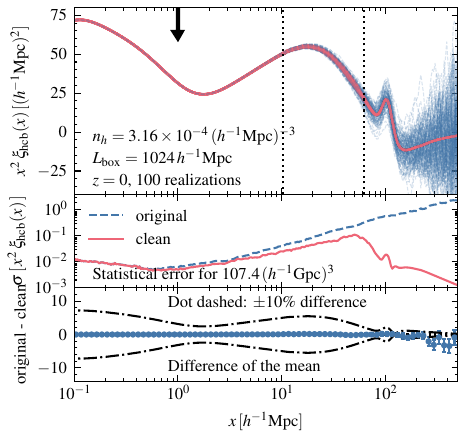}
 \qquad
 \includegraphics[width=0.4\linewidth]{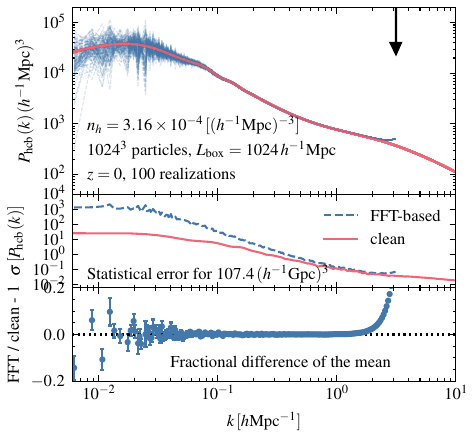}
 \caption{Same as figures~\ref{fig:xicb_hybrid} and \ref{fig:pcb_hybrid}, but for the halo-cb cross correlation and power spectrum. {Alt text: Two side-by-side three-panel groups: the left for the halo-matter cross correlation function and the right for the corresponding power spectrum. In both, the cleaning procedure reduces the scatter among realizations and the statistical error without introducing systematic bias.} \label{fig:hcb_hybrid}}
\end{figure*}

\begin{figure*}[t]
 \centering
 \includegraphics[width=0.4\linewidth]{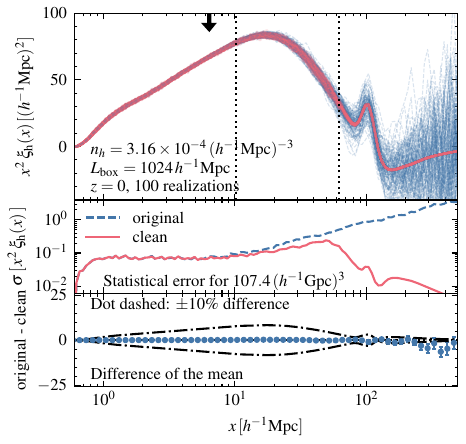}
 \qquad
 \includegraphics[width=0.4\linewidth]{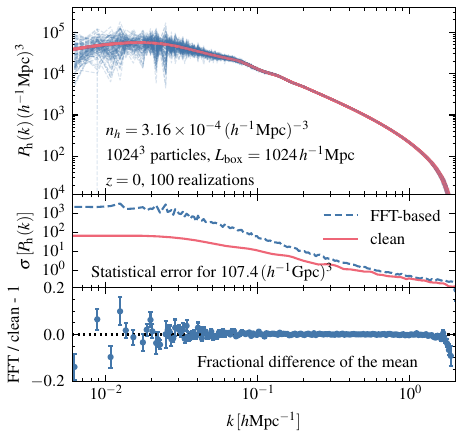}
 \caption{Same as figures~\ref{fig:xicb_hybrid} and \ref{fig:pcb_hybrid}, but for the halo auto correlation and power spectrum. We adopt a smaller FFT grid ($160^3$) for our new method here (vertical arrow in the left panel); the ``FFT-based'' reference curve in the right panel keeps the original grid. Direct pair counting is quicker for halos, whose number is much smaller than matter particles, so we can extend it up to $10\,d_\mathrm{g}$ in reasonable time. {Alt text: Two side-by-side three-panel groups for the halo auto correlation function on the left and power spectrum on the right. The cleaning procedure suppresses realization-to-realization scatter and statistical error, while the new fast-Fourier-transform grid spacing for halos is shown by a vertical marker in the left correlation-function panel.}\label{fig:hh_hybrid}}
\end{figure*}

Our final results after the cleaning procedure are shown for the cb auto correlation function by the thick solid line in figure~\ref{fig:xicb_hybrid_scales}. It follows different correlation-function estimates depending on scale and connects them smoothly. The results are also shown for the 100 random realizations in figure~\ref{fig:xicb_hybrid} for cb, as well as the halo--cb cross-correlation and halo auto-correlation functions in figures~\ref{fig:hcb_hybrid} and \ref{fig:hh_hybrid} (left panels). We plot on the top panels 100 dashed lines and 100 solid lines corresponding to the random realizations before and after the cleaning procedure, respectively, although the latter 100 are almost indistinguishable due to their small scatter. This clearly shows that the procedure helps to reduce the sample variance. A more quantitative assessment of the procedure can be done in the middle panels, where we compare the standard error of the mean for the two cases. The solid line for the cleaned correlation function shows a lower level of statistical error. It is interesting to note that the procedure is effective even on scales smaller than $x_1 = 10\,\hiMpc$ especially for the cb correlation function, although the procedure is based on linear theory. This improvement arises from the correlation of the noise at different scales and for different fields.

Finally, we plot the difference between the ensemble averaged correlation functions with and without the cleaning procedure to assess the systematic error in the bottom panels. We can see that the difference is negligibly small, typically much less than $10 \%$, except for the largest scales ($\gtrsim 200\,\hiMpc$), where the ensemble-averaged linear correlation function is expected to be more reliable than the noisy estimate from a finite number of realizations. Note that the size of the error bars corresponds to the standard error in the middle panel, which is the cosmic variance expected from a survey volume of $\sim 100\,(\hiGpc)^3$, comparable to upcoming Stage-IV galaxy surveys. Therefore, the possible systematic error originating from our procedure is negligible in practice.

\subsubsection{Predictions in Fourier space}
\label{subsubsec:fourier}
The procedure described so far is developed in terms of the correlation function instead of the power spectrum. We do this because the split of the short and long distance pairs as well as the modeling of the BAO damping in terms of the propagator-based method are more sharply localized in separation $x$ than in wavenumber $k$. We employ the FFTLog algorithm \citep{Talman78,Hamilton00} implemented in the \texttt{pyfftlog} Python package for efficient conversion.

We compare the power spectra obtained with the cleaning procedure followed by FFTLog with the standard FFT-based method in figure~\ref{fig:pcb_hybrid} for the cb fluctuations, and in figures~\ref{fig:hcb_hybrid} and \ref{fig:hh_hybrid} for halos (right panels). Similarly to figure~\ref{fig:xicb_hybrid} for the correlation function, we show the measurements from 100 random realizations and plot the measurements (upper), the standard error (middle) and the difference in the mean (lower). First, the upper panels suggest that our procedure helps to reduce the scatter among realizations. We again plot 100 solid lines for the measurements after the full procedure in these panels, which are almost indistinguishable. We can see this more quantitatively in the middle panel; the cleaning procedure reduces the noise by about two orders of magnitude on large scales. Secondly, the spectra obtained by our procedure show no statistically significant bias relative to the standard method over the range where the latter is reliable; the deviation is consistent with zero in the lower panels. One might notice that the fractional difference starts to blow up at $k\sim3\hMpci$. This is due to the aliasing effect in the FFT-based method. While several methods have been proposed to suppress this effect in FFT-based power-spectrum measurements, our method avoids it by reconstructing the power spectrum from a correlation function that uses direct pair counting on small
scales rather than relying on the FFT-based density field.

\section{Production runs}
\label{sec:productruns}

We now discuss our predictions based on the simulations with $N_\mathrm{p}=3,000^3$ particles listed in table~\ref{tab:product_runs}. These simulations are distinct from the LR/MR/HR validation runs used in section~\ref{sec:accuracy}; here we use the large-scale production suite that forms the basis of our emulator training. The simulation outputs are processed by the pipeline described in the previous section.

\begin{table*}
\tbl{Production runs presented in this paper. The physical density parameter of CDM $\omega_\mathrm{cdm}=\Omega_\mathrm{cdm}h^2$, the dark energy density parameter $\Omega_\mathrm{de}$, the sum of neutrino masses $M_\nu$, Hubble parameter $h$ in units of $100\,\mathrm{km/s/Mpc}$, the dark energy equation-of-state parameter $w_\mathrm{de}$, and the square of the effective sound speed of dark energy $c_\mathrm{s}^2$ are summarized. The first four models (Nu series) form a neutrino-mass sequence with $w_\mathrm{de}=-1$, and each is run in three comoving box sizes; Nu1 is our fiducial model, while Nonu ($M_\nu=0$) serves as the massless reference for the neutrino-induced suppression of the matter power spectrum. The last four models (DE series) form a clustering-dark-energy sequence with $w_\mathrm{de}=-0.6$, each run in a single $4\,h^{-1}\,\mathrm{Gpc}$ box. The corresponding numbers of random realizations are shown in the final column ($N_\mathrm{rand}$).}{
\begin{tabular}{lcccccccc}
\hline\noalign{\vskip3pt} 
\bfseries{Model} & 
$\omega_\mathrm{cdm}$ &
$\Omega_{\mathrm{de}}$ &
$M_\nu\,[\mathrm{eV}]$ &
$h$ &
$w_{\mathrm{de}}$ &
$c_\mathrm{s}^2$ &
$L_{\mathrm{box}} \, [h^{-1} \mathrm{Gpc}]$ &
$N_\mathrm{rand}$ \\ [2pt]
\hline\noalign{\vskip3pt} 
Nonu & $0.1198$ & $0.6843$ & $0$ & $0.6709$ & $-1$ & -- & $1, 2, 4$ & $3, 3, 4$ \\
Nu1 (fiducial) & $0.1198$ & $0.6843$ & $0.06$ & $0.6724$ & $-1$ & -- & $1, 2, 4$ & $3, 3, 4$ \\
Nu2 & $0.1198$ & $0.6843$ & $0.12$ & $0.6739$ & $-1$ & -- & $1, 2, 4$ & $3, 3, 4$ \\
Nu3 & $0.1198$ & $0.6843$ & $0.24$ & $0.6769$ & $-1$ & -- & $1, 2, 4$ & $3, 3, 4$ \\
\hline
DE0 & $0.06753$ & $0.8$ & $0$ & $0.6701$ & $-0.6$ & $1$ & $4$ & $1$ \\
DE1 & $0.06753$ & $0.8$ & $0$ & $0.6701$ & $-0.6$ & $0.01$ & $4$ & $1$ \\
DE2 & $0.06753$ & $0.8$ & $0$ & $0.6701$ & $-0.6$ & $0.001$ & $4$ & $1$ \\
DE3 & $0.06753$ & $0.8$ & $0$ & $0.6701$ & $-0.6$ & $0.0001$ & $4$ & $1$ \\ [2pt]
\hline\noalign{\vskip3pt} 
\end{tabular}}
\label{tab:product_runs}
\end{table*}

\begin{table}
\tbl{Parameters shared among the production runs: the physical baryon density parameter $\omega_\mathrm{b} = \Omega_\mathrm{b}h^2$, the primordial scalar amplitude parameter $\ln(10^{10}A_\mathrm{s})$, the spectral index $n_\mathrm{s}$, the number of particles $N_\mathrm{p}$.}{
\begin{tabular}{cccc}
\hline\noalign{\vskip3pt} 
$\omega_\mathrm{b}$ &
$\ln(10^{10}A_\mathrm{s})$ &
$n_\mathrm{s}$ &
$N_\mathrm{p}$ \\ [2pt]
\hline\noalign{\vskip3pt} 
$0.02225$ & $3.094$ & $0.9645$ & $3,000^3$\\ [2pt]
\hline\noalign{\vskip3pt} 
\end{tabular}}
\label{tab:product_runs2}
\end{table}

\subsection{Simulation suites}
\label{subsec:DQ2_suite}

Our primary production runs employ $N_\mathrm{p} = 3,000^3$ particles to model the cb density. Eight distinct cosmological models are considered, as detailed in tables~\ref{tab:product_runs} and \ref{tab:product_runs2}. In the first four models listed in table~\ref{tab:product_runs}, we vary the neutrino mass while keeping the total matter density parameter, $\Omega_\mathrm{m}$, fixed. The Hubble parameter, $h$, is adjusted slightly to maintain spatial flatness and constant physical densities for CDM and baryon. The other four models explore dark energy scenarios with an equation of state parameter $w_\mathrm{de}=-0.6$, while varying the effective sound speed parameter, $c_\mathrm{s}$. Although the value of $w_\mathrm{de}=-0.6$ is well beyond current observational bounds, it is used here to investigate extreme cases with substantial dark energy clustering. While we plan to extend this campaign by varying many other parameters, we begin with these two cases -- varying neutrino masses and dark-energy sound speed -- to validate our linear response approach.

For all simulations, we store particle snapshots at $11$ redshifts, evenly spaced in $\log(1+z)$ from $z=3$ to $z=0$: $z=3$, $2.4822$, $2.031$, $1.640$, $1.300$, $1$, $0.741$, $0.516$, $0.320$, $0.149$ and $0$. Although these outputs are somewhat coarser than those in DQ1 ($21$ snapshots between $z=0$ and $1.5$), we expect that emulation of summary statistics can still be performed accurately, provided that the quantities of interest evolve slowly and smoothly over time. We have extended the coverage up to $z=3$ to ensure usefulness for future surveys probing the higher-redshift Universe.

\subsection{Fiducial cosmology}
\label{subsec:fiducial}

We first consider the fiducial model, Nu1 in table~\ref{tab:product_runs}, whose parameters are consistent with the \textit{Planck} 2015 cosmology and include a minimal neutrino mass, $M_\nu=0.06\,\mathrm{eV}$. Together with the shared parameters in table~\ref{tab:product_runs2}, this model serves as the baseline for testing the residual box-size, resolution, and post-processing dependence of the DQ2 production configuration.

\subsubsection{Total matter power spectrum}

\begin{figure}[htbp]
 \centering
 \includegraphics[width=\linewidth]{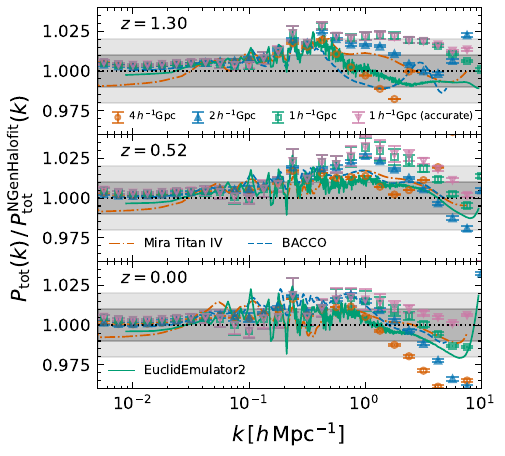}
 \caption{Comparison of the total matter power spectrum from the DQ2 production runs (symbols with error bars) against four published predictions, with the NgenHalofit prediction~\citep{Smith19} adopted as the reference (zero ratio). The other three models are shown by the dot-dashed (Mira-Titan Universe), solid (EuclidEmulator2), and dashed (BACCO) lines. The light (dark) shaded region marks the $2\%$ ($1\%$) deviation from the reference. The three panels correspond to $z = 1.3$, $0.52$, and $0$ from top to bottom. Our results are shown for three box sizes (circles, upward triangles, and squares for $L_\mathrm{box} = 1$, $2$, and $4\,\hiGpc$, respectively, all with the fiducial accuracy setting); downward triangles show $L_\mathrm{box} = 1\,\hiGpc$ with the ``accurate'' setting. {Alt text: Three stacked panels of the fractional difference of the total matter power spectrum against wavenumber, with rows from top to bottom corresponding to three different redshifts. Each panel compares four published predictions and our simulations in three box sizes; most differences sit within a two-percent band, with mild scale-dependent deviations attributed to differing neutrino treatments and mass resolutions.} \label{fig:ptot_emucomp}}
\end{figure}

The total matter power spectra from the DQ2 production runs are compared with published emulators and fitting formulas in figure~\ref{fig:ptot_emucomp}. We use the prediction by NgenHalofit~\citep{Smith19} as a reference and show our results with different box sizes using the three symbols with error bars. The other models are depicted by the dot-dashed line (Mira-Titan Universe: \cite{MiraTitan1,MiraTitan2,MiraTitan4}), the solid line (EuclidEmulator2: \cite{EuclidEmu1,EuclidEmu2}), and the dashed line (BACCO: \cite{BACCO}).

Overall, we can see that the agreement is typically within about $\pm 2\%$ (the light gray shade), with a tendency to better match toward larger scales. Our simulations with different box sizes agree within the error bars up to $k\sim 0.4\,\hMpci$ at all three redshifts, and a clear increasing trend with resolution can be seen on larger wavenumbers.
The mismatch on small scales originates from two different sources. First, as is clear from the difference among different symbols, the mass and spatial resolution of simulations can alter the power spectrum. The other arises from differences in the treatment of massive neutrinos. Notably, even with the smallest allowed value of the sum of three neutrino masses ($0.06\,\mathrm{eV}$), the difference in neutrino treatment matters for achieving percent-level accuracy.

NgenHalofit does not explicitly treat the effect of massive neutrinos. Nevertheless, we can feed the linear matter power spectrum computed for massive-neutrino cosmologies to this code to compute the matter power spectrum. This procedure effectively treats the total linear matter field as the input field to a CDM-like nonlinear mapping, without separately accounting for the distinct clustering behavior of the neutrino component. In fact, we adopted this procedure in DQ1. For a further test, we present a direct comparison between this and our new procedure in Appendix~\ref{sec:nu_approx}, where we can see that a similar increase in the ratio at $k\sim1\,\hMpci$ can be observed. Therefore, the discrepancy between NgenHalofit and our simulations is mainly due to this difference in neutrino treatment.

Our neutrino treatment is the most similar to that in EuclidEmulator2, which considers the gravitational force from linear neutrino perturbations in the course of $N$-body simulations. Indeed, their results are somewhere in between our middle- and highest-resolution simulations. This is reasonable because their simulation volume and number of particles are exactly the same as our highest-resolution simulations ($N_\mathrm{p} = 3,000^3$ and $L_\mathrm{box} = 1,000\,\hiMpc$).

The Mira-Titan Universe simulations employ a different procedure to treat neutrinos, as explained and tested by \citet{Upadhye14}. In that treatment, neutrinos contribute to the background expansion but not to the growth of cb perturbations during the simulation. The initial cb density
field is generated from the redshift-zero cb transfer function, rescaled back to the initial redshift using a scale-independent growth factor, and the linear neutrino contribution is added a posteriori when constructing the total matter power spectrum. 
They further combine the simulation results with the perturbative model (TimeRG: \cite{Pietroni:2008qy}) on large scales to suppress the sample variance in the simulation measurements. Apart from the sharp rise in the ratio at $z=1.3$, which is probably an artifact of the connection between the perturbative model and the simulation results, we see good agreement with our simulations. This prediction agrees best with our highest-resolution simulations, as fully expected from their simulation setups ($N_\mathrm{p} = 3,200^3$ and $L_\mathrm{box} = 2,100\,\mathrm{Mpc}$).

Finally, the BACCO emulator~\citep{BACCO}, shown by the dashed line, follows yet another strategy: its predictions are obtained by applying a cosmology-rescaling algorithm to a small number of high-resolution base simulations ($N_\mathrm{p} = 4,320^3$ in a box of $1,440\,\hiMpc$), with the contribution of massive neutrinos incorporated at the linear level. Its predictions are broadly consistent with our higher-resolution runs across the plotted range.

A more exhaustive comparison against these and other emulators and fitting formulas in the literature, together with the detailed characterization of the total matter power spectrum emulator we are constructing from the DQ2 production runs, is presented in the accompanying paper and is not repeated here.

\subsubsection{Halo mass function}

We compare the halo mass functions measured from the DQ2 production runs with the predictions of DarkEmulator~\citep{2019ApJ...884...29N}, which was trained on the DQ1 simulations. This comparison serves a dual purpose: validating the consistency of the DQ2 post-processing pipeline against the DQ1 baseline and quantifying the improvement brought by the updated mass correction formula and central-satellite separation criterion.

\begin{figure}[htbp]
 \centering
 \includegraphics[width=\linewidth]{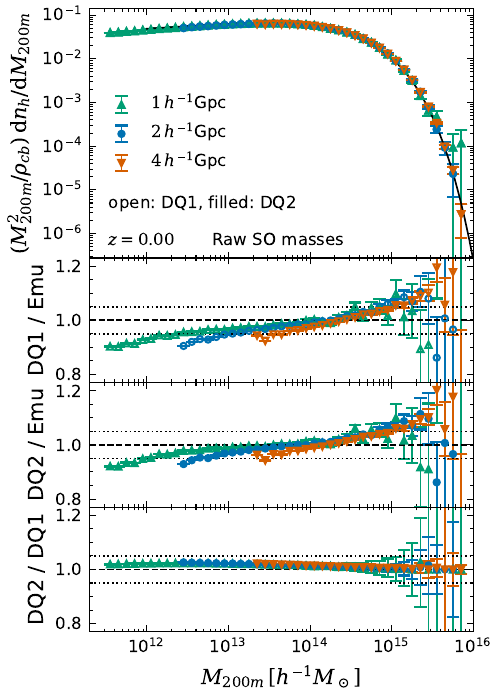}
 \caption{Comparison of the halo mass function at $z=0$ from simulations with three different box sizes (upward triangles, circles, and downward triangles for $L_\mathrm{box} = 1$, $2$, and $4\,\hiGpc$, respectively) and with two different criteria for the central/satellite split (DQ1 pipeline: open symbols; DQ2 pipeline: filled symbols). The upper panel shows the absolute mass function; the solid line is the DarkEmulator~\citep{2019ApJ...884...29N} prediction. The two middle panels show the ratio to DarkEmulator for each pipeline separately, and the bottom panel shows the DQ2-to-DQ1 ratio. No mass correction is applied. {Alt text: Four-panel halo mass function comparison at redshift zero for three box sizes. The top panel shows the absolute mass function alongside the DarkEmulator prediction, the middle two panels show the ratio to DarkEmulator for the two pipelines separately, and the bottom panel shows the ratio between pipelines.} \label{fig:HMF_S010_wocorr}}
\end{figure}

Figure~\ref{fig:HMF_S010_wocorr} shows the halo mass function at $z = 0$ before applying any mass correction, for simulations with the three box sizes used in the production runs ($L_\mathrm{box} = 1$, $2$, and $4\,\hiGpc$). The upper panel shows the absolute mass function alongside the DarkEmulator prediction (solid line). The two middle panels compare each pipeline separately against DarkEmulator: the upper middle panel shows the DQ1 pipeline (open symbols) and the lower middle panel the new DQ2 pipeline (filled symbols), while the bottom panel shows the DQ2-to-DQ1 ratio directly. In each panel, the lowest-mass bin for each box size corresponds to halos resolved by approximately $100$ simulation particles.

Both pipelines exhibit a similar trend relative to DarkEmulator: the ratio rises above unity at the high-mass end and falls below it at the low-mass end, remaining within $\pm 10\%$ over the plotted range. Near the low-mass cutoff the ratio drops to slightly above $0.9$, outside the $\pm 5\%$ band, as expected for halos resolved by the minimum particle count. The bottom panel shows that the DQ2-to-DQ1 ratio is highly consistent across box sizes: all three fall on a single curve with deviations of at most $\sim 2\%$ and sub-percent agreement at the high-mass end. This confirms that replacing the $M_{200}$-based central-satellite split of DQ1 with the $V_\mathrm{max}$-based criterion of DQ2 introduces no resolution-dependent bias in the halo abundance.

\begin{figure}[htbp]
 \centering
 \includegraphics[width=\linewidth]{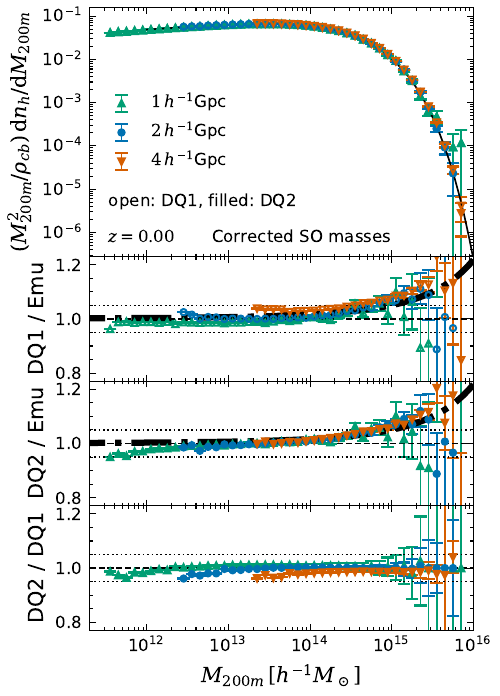}
 \caption{Same as figure~\ref{fig:HMF_S010_wocorr}, but with the respective mass correction factors applied (equation~\ref{eq:mass_correction1} for DQ1, equation~\ref{eq:mass_correction2} for DQ2). The bold dash-dotted lines in the middle panels show the ratio of the \citet{Tinker08} fitting formula with the exponential-damping parameter adjusted from $c=1.19$ to $c=1.175$, which captures the 
 high-mass upturn relative to DarkEmulator seen in the larger simulation boxes (see the main text for details). {Alt text: Same four-panel format as the previous figure, but after the empirical mass corrections are applied. Both pipelines agree with DarkEmulator within five percent for most of the mass range, with a high-mass upturn that grows with box size and is matched by a slightly renormalized Tinker formula.} \label{fig:HMF_S010_wcorr}}
\end{figure}

Figure~\ref{fig:HMF_S010_wcorr} repeats the comparison after applying the respective empirical mass corrections (equation~\ref{eq:mass_correction1} for DQ1 and equation~\ref{eq:mass_correction2} for DQ2). After correction, both pipelines agree with DarkEmulator to within $\pm 5\%$ up to several $\times 10^{14}\,h^{-1}M_\odot$. At higher masses the ratio rises systematically above unity, and the excess increases, rather than decreases, with simulation volume. We interpret this behavior primarily as a finite-volume effect in the halo-mass-function calibration of DarkEmulator, rather than as a deficiency of the DQ2 production runs. Although the original DQ1 simulation suite included both $1\,\hiGpc$ and $2\,\hiGpc$ boxes, the halo-mass-function module of DarkEmulator was calibrated using the $1\,\hiGpc$ boxes only. Long-wavelength modes beyond this scale are therefore absent from the calibration data, which can suppress the abundance of the rarest, most massive halos represented by the emulator. A hint of this effect was already visible in the DQ1 paper itself, where a cross-check against a small set of $2\,\hiGpc$ simulations showed a mild systematic offset at the high-mass end, though not clearly resolved within the statistical error bars available at that time. Extending the volume to $L_\mathrm{box} = 4\,\hiGpc$ in the DQ2 production runs makes the effect unambiguous: the additional large-scale power captured in the bigger boxes enhances the formation of the most massive halos relative to what the emulator---trained exclusively on $1\,\hiGpc$ realizations, and therefore missing these modes---predicts. As a phenomenological illustration, this effect can be mimicked by a mild renormalization of the \citet{Tinker08} fitting formula. The Tinker08 multiplicity function takes the form
\be
f(\sigma) = A\left[\left(\frac{\sigma}{b}\right)^{-a} + 1\right]\exp\left(-\frac{c}{\sigma^2}\right),
\label{eq:tinker08}
\ee
in which the parameter $c$ controls the strength of the exponential damping at the high-mass end (small $\sigma$); shifting $c$ is therefore approximately equivalent to rescaling $\sigma$ in that regime. We find that adjusting $c$ from its default value $c = 1.19$ to $c = 1.175$ (bold dash-dotted lines in the middle panels of figure~\ref{fig:HMF_S010_wcorr}) reproduces the observed box-size-dependent excess, consistent with the known sensitivity of the high-mass tail to large-scale power.

The primary improvement of the DQ2 pipeline over DQ1 is inter-resolution consistency of the corrected mass function. With the DQ1 correction (equation~\ref{eq:mass_correction1}), the three box sizes exhibit a residual spread of up to $\sim 3\%$ at the low-mass end after correction. The updated formula (equation~\ref{eq:mass_correction2}) reduces this to $\lesssim 1\%$, demonstrating its superior calibration across mass resolutions. Although the DQ2-to-DQ1 ratio acquires a mild mass dependence after correction---reflecting the different functional forms of the two correction formulas---it remains within $5\%$ throughout, consistent with the $\sim 2\%$ difference already present before correction. These results are shown at $z = 0$ as a representative case; the same conclusions hold at higher redshifts, consistent with the convergence tests presented in section~\ref{subsec:hmf}.

\subsection{Cosmologies with scale-dependent growth}
\label{subsec:sdg}

Having validated the code and post-processing pipeline at the fiducial cosmology, we now turn to the cosmologies in table~\ref{tab:product_runs} that feature scale-dependent growth. We demonstrate that GINKAKU, combined with the linear-response treatment of the external source term (in section~\ref{subsec:linear}), reproduces the expected nonlinear behavior in two qualitatively different regimes: small-scale suppression driven by the free streaming of massive neutrinos, and large-scale enhancement driven by the clustering of dark energy with non-unit sound speed. A quantitative, emulator-level characterization of these responses across the full DQ2 parameter space is the subject of the accompanying paper; here we restrict ourselves to a capability demonstration for the four neutrino and four dark-energy production-run models in table~\ref{tab:product_runs}.

\subsubsection{Massive neutrinos}

\begin{figure*}[t]
 \centering
 \includegraphics[width=8cm]{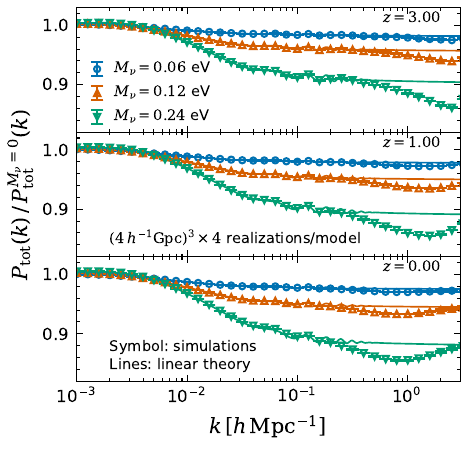}
 \includegraphics[width=8cm]{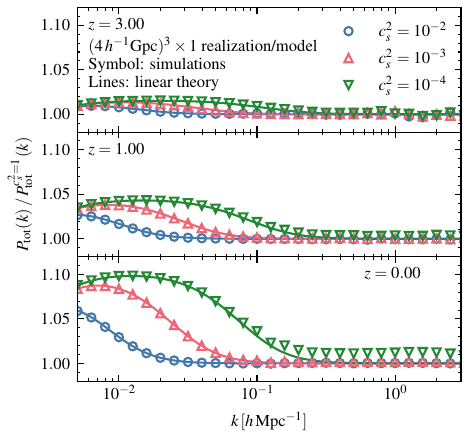}
 \caption{Ratio of the total matter power spectra for different cosmological models, measured at three representative redshifts ($z = 3$, $1$, and $0$ from top to bottom; see the figure legend). Left: models with different neutrino masses, $M_\nu$, normalized to the massless case. Right: models with different dark energy sound speeds, $c_\mathrm{s}$, normalized to the model with $c_\mathrm{s}=1$. In both panels, symbols denote the simulation measurements and solid lines the linear-theory predictions. The symbol assignment differs between the two panels: circles, upward triangles, and downward triangles correspond to $M_\nu = 0.06$, $0.12$, $0.24\,\mathrm{eV}$ in the left panel and to $c_\mathrm{s}^2 = 10^{-2}$, $10^{-3}$, $10^{-4}$ in the right panel (see the figure legend). {Alt text: Two side-by-side three-panel groups of the matter power spectrum ratio against wavenumber, with rows from top to bottom corresponding to $z = 3$, $1$, and $0$. The left group compares three massive-neutrino models to the massless reference, showing a spoon-shaped suppression. The right group compares three lower-sound-speed dark-energy models to the smooth-dark-energy reference, showing a growing enhancement at smaller wavenumber.} \label{fig:pk_ratio}}
\end{figure*}

The left panel of figure~\ref{fig:pk_ratio} shows the ratio of the total matter power spectrum measured from our $M_\nu = 0.06$, $0.12$, and $0.24\,\mathrm{eV}$ runs to that of the massless reference (model Nonu), at three representative redshifts. Symbols denote the simulation measurements, while solid lines show the corresponding linear-theory predictions.

In the linear regime, the ratios approach unity as $k\to0$, where neutrino free streaming is irrelevant, and begin to fall around $k\sim10^{-2}\,\hMpci$ toward the familiar free-streaming plateau set mainly by $f_\nu \equiv \Omega_\nu/\Omega_\mathrm{m}$.  This behavior agrees with linear theory. Moving to intermediate wavenumbers, the suppression deepens beyond the linear-theory prediction and develops the characteristic ``spoon'' shape that has been studied extensively in $N$-body simulations with explicit neutrino treatments~\citep{Brandbyge09,Bird12,Villaescusa13,2013MNRAS.428.3375A,2018MNRAS.481.1486B,2018JCAP...09..028B,EuclidEmu2,Elbers22}. This extra suppression arises because neutrino perturbations remain smooth on these scales, while the cb component continues to cluster nonlinearly. The neutrino contribution to the total matter perturbation therefore becomes progressively diluted relative to the growing cb component, enhancing the suppression already present in linear theory.

At still higher $k$, where the one-halo contribution dominates the total matter power spectrum, the ratio partially recovers, producing the well-known U-shaped feature discussed in the halo-model analyses of \citet{Saito08} and \citet{Massara14}. The turnaround is visible in our measurements at $k \gtrsim 1\,\hMpci$, and its amplitude and redshift dependence are consistent with those studies: the minimum of the ratio deepens and shifts to higher $k$ with decreasing redshift, reflecting the progression of nonlinear structure formation.

We do not separately present the halo mass function ratio here. The underlying physics is qualitatively the same amplification of the linear $\sigma^2(M)$ suppression, and nonlinear calibrations against neutrino simulations are already available in the literature~(e.g., \cite{Brandbyge10,Costanzi13,Biswas19}). A quantitative emulator-level study of the halo mass function response to massive neutrinos lies outside the scope of this code paper.

\subsubsection{Clustering dark energy} \label{subsubsec:clustering_de}

The right panel of figure~\ref{fig:pk_ratio} shows the analogous ratios for the four dark-energy production-run models. All four share $w_\mathrm{de} = -0.6$ and identical background expansion and matter content, differing only in the effective dark-energy sound speed, $c_\mathrm{s}^2 \in \{1, 10^{-2}, 10^{-3}, 10^{-4}\}$. We adopt the $c_\mathrm{s}^2 = 1$ case, in which dark energy is smooth on all sub-horizon scales, as the reference and plot each lower-$c_\mathrm{s}^2$ model relative to it.

The physical scale controlling the response is the dark-energy Jeans wavenumber, $k_\mathrm{J} \sim a H / c_\mathrm{s}$: on $k \gg k_\mathrm{J}$ pressure support prevents the dark-energy fluid from clustering, while on $k \ll k_\mathrm{J}$ it follows the gravitational potential of the cb fluid and provides an additional source for structure growth. Lowering $c_\mathrm{s}^2$ increases $k_\mathrm{J}$, allowing dark-energy perturbations to cluster down to smaller physical scales. The enhancement therefore grows with decreasing redshift and with decreasing $c_\mathrm{s}^2$, and its scale dependence is well described by linear theory, in agreement with the $N$-body studies of \citet{2019JCAP...08..013D}, \citet{Hassani19} and \citet{Hassani20}.

The only regime in which linear theory is noticeably insufficient is the $c_\mathrm{s}^2 = 10^{-4}$ model at low redshift. There the Jeans scale lies well within the mildly nonlinear regime, and the additional clustering propagates through mode coupling as an approximately scale-independent correction, boosting the matter power spectrum by about $1\%$ at $z=0$ across the measured range of $k$. The near-scale-independence of this residual is consistent with the enhancement entering chiefly through a rescaling of the late-time $\sigma_8$ of already-clustered matter rather than through a scale-dependent transfer-function response.

Taken together, the two panels of figure~\ref{fig:pk_ratio} show that GINKAKU with the linear-response source term successfully captures both mass-induced suppression (massive neutrinos) and sound-speed-dependent enhancement (clustering dark energy) through the fully nonlinear evolution. A detailed quantitative calibration of these responses over the DQ2 parameter space is the subject of the accompanying paper.

\section{Summary}
\label{sec:summary}

We have introduced GINKAKU, a new cosmological $N$-body code developed for the Dark Quest II (DQ2) simulation campaign, aimed at controlled ensemble production across the broader cosmological model space demanded by next-generation galaxy surveys. The code is built on the FDPS framework and couples a TreePM gravity solver with a linear-response treatment of an external source term in the Poisson equation. This combination preserves Newtonian particle dynamics while incorporating, at the linear level, physical effects that are difficult to represent with simulation particles --- most importantly massive-neutrino perturbations, general-relativistic corrections in the $N$-body gauge, radiation perturbations at early times, and clustering dark energy with non-unit effective sound speed.

Beyond the $N$-body solver itself, we have described an initial-condition generator that extends 2LPT to the scale-dependent linear growth required by the $N$-body gauge and massive neutrinos, and a renewed post-processing pipeline. The latter replaces the $M_{200}$-based central/satellite split of DQ1 with a $V_\mathrm{max}$-based criterion and introduces an updated empirical mass-correction formula (equation~\ref{eq:mass_correction2}) that substantially improves the inter-resolution consistency of the measured halo mass function. The pipeline also includes a halo-shape measurement extension supporting intrinsic-alignment statistics and a noise-reduced spatial-correlation estimator that stitches a hybrid pair-counting/FFT simulation estimate with linear theory and a propagator-damped linear template across the BAO region in configuration space, with Fourier-space predictions obtained via FFTLog.

The internal accuracy of the code has been validated through convergence studies with respect to the softening length, tree opening angle, PM grid, and time-step control parameters, as well as through direct cross-comparisons with L-GADGET2, GADGET-3, PKDGRAV3, and RAMSES on shared initial conditions. Across these tests GINKAKU in its fiducial configuration agrees with other TreePM and FMM codes at the $\lesssim 1\%$ level on the matter power spectrum up to the particle Nyquist wavenumber, with a mild excess only at $z=3$, while the ``accurate'' configuration tightens the agreement further at a larger wall-clock cost. Parallel scalability has been benchmarked on the Cray~XC50 and Fugaku platforms (Appendix~\ref{sec:scaling}): GINKAKU preserves close-to-ideal strong scaling up to $N_\mathrm{p}=7{,}500^3$ particles on Fugaku, indicating that production-scale runs well beyond the DQ2 baseline are feasible using only a limited fraction of the full machine.

Using this code we have carried out an initial set of DQ2 production runs: eight cosmological models with $N_\mathrm{p} = 3{,}000^3$ particles, spanning box sizes of $1$, $2$, and $4\,\hiGpc$ for the neutrino-mass sequence and $4\,\hiGpc$ for the clustering-dark-energy sequence, with snapshots stored at eleven redshifts between $z = 0$ and $z = 3$. For the fiducial cosmology, the total matter power spectrum from these runs is broadly consistent with the NgenHalofit, EuclidEmulator2, Mira-Titan Universe and BACCO predictions at the $\pm 2\%$ level across the plotted range of scales and redshifts, with the residual scale and neutrino-treatment dependence understood qualitatively. For the halo mass function, the measurements from both the DQ1 and DQ2 post-processing pipelines agree with DarkEmulator to within $\pm 5\%$ up to several $\times 10^{14}\,h^{-1}M_\odot$ after the respective empirical corrections are applied, and the updated DQ2 correction reduces the inter-resolution spread at the low-mass end to $\lesssim 1\%$. A mild high-mass excess over DarkEmulator, which grows with simulation box size, is traced to the limited volume of the DQ1 training set ($L_\mathrm{box} = 1\,\hiGpc$) and is reproduced by a small adjustment of the exponential-damping parameter in the \citet{Tinker08} fitting formula.

The scale-dependent-growth cosmologies further illustrate the capability of the linear-response method. For the massive-neutrino sequence, the total matter power spectrum ratios reproduce the characteristic nonlinear ``spoon'' suppression at intermediate wavenumbers and its partial recovery into a U-shape at the one-halo-dominated regime, consistent with previous $N$-body and halo-model studies. For the clustering-dark-energy sequence, the enhancement relative to the $c_\mathrm{s}^2 = 1$ reference is well described by linear theory, except for the $c_\mathrm{s}^2 = 10^{-4}$ model at low redshift, where an approximately scale-independent nonlinear correction of about $1\%$ at $z = 0$ propagates through mode coupling --- consistent with an effective late-time $\sigma_8$ rescaling.

A detailed quantitative characterization of the DQ2 suite, including the construction and validation of the total matter power spectrum emulator, is presented in the accompanying paper~\citep{Tanaka2026-emu}. Beyond this, simulation products from GINKAKU have already supported a range of published or submitted cosmological analyses, e.g., the Beyond-2pt mock data challenge~\citep{Beyond2pt24} and the small-scale HSC-Y3 cosmic-shear analysis based on DarkEmulator2~\citep{Terasawa24}. Therefore, the validation presented here documents the numerical reference for these applications and for the forthcoming emulator-scale campaigns. Future work will broaden the cosmological parameter space and extend the emulator effort to halo statistics, building on the code and pipeline infrastructure presented here.

\section*{Funding}
This work was supported in part by MEXT/JSPS KAKENHI Grant Numbers JP19H00677, JP20H05861, JP21H01081, JP21H01079, JP22K03634, JP24H00215, JP24H00221, JP25H00625, and JP26H00402. We also acknowledge financial support from the Japan Science and Technology Agency (JST) AIP Acceleration Research Grant Number JP20317829.

\begin{ack}
We thank Yosuke Kobayashi, Ken Osato, Masahiro Takada, Metin Ata, Tomoaki Ishiyama, St\'ephane Colombi and Joshua Barnes for useful discussions. Numerical simulations were carried out on the Cray XC50 at the Center for Computational Astrophysics, National Astronomical Observatory of Japan. Code development and tests were partly carried out at the Yukawa Institute Computer Facility, Cray XC40 and Yukawa-21. 
This research was supported by MEXT as “Program for Promoting Researches on the Supercomputer Fugaku” (Toward a unified view of the universe: from large scale structures to planets, JPMXP1020200109, project ID hp220173; Structure and Evolution of the Universe Unraveled by Fusion of Simulation and AI, JPMXP1020230406, project ID hp230204; Multi-wavelength Cosmological Simulations for Next-generation Surveys, JPMXP1020230407, project ID hp230202, hp240200 and hp250219) and used computational resources of supercomputer Fugaku provided by the RIKEN Center for Computational Science.
\end{ack}

\section*{Code and data availability}
GINKAKU is currently maintained as a production code for the Dark Quest II project and is not publicly released at this stage. The numerical methods, accuracy parameters, and post-processing prescriptions described in this paper are intended to be sufficient for reproducing the simulations presented here. The source code, parameter files, and derived simulation products are available from the corresponding author upon reasonable request.

\appendix

\section{Scaling}
\label{sec:scaling}

In this Appendix we examine the parallel scalability of GINKAKU on two different supercomputing systems. Appendix~\ref{subsec:scaling_xc50} compares its scaling against publicly available codes on the Cray~XC50 platform, while Appendix~\ref{subsec:scaling_fugaku} probes the large-$N$ regime on the Fugaku supercomputer.

\subsection{Cross-code scaling on Cray~XC50}
\label{subsec:scaling_xc50}

The tests in this subsection are run on the Cray~XC50 supercomputer at the Center for Computational Astrophysics, National Astronomical Observatory of Japan. Because PKDGRAV3 can additionally leverage GPUs, its absolute wall-clock time on a GPU-equipped system would differ substantially from what we measure here; we therefore restrict the comparison to CPU-only scaling rather than to absolute timing.

For all three codes we adopt the default accuracy parameters summarized in Table~\ref{tab:convergence_runs}, with the following PKDGRAV3-specific exceptions: a multipole expansion up to fourth order in the Fast Multipole Method and a time-stepping parameter $\eta=0.2$. Each code is run with $N_\mathrm{p} = 256^3$, $512^3$, $1{,}024^3$, $2{,}048^3$, and $3{,}072^3$ particles\footnote{We only consider GINKAKU and L-GADGET2 for $N_\mathrm{p}=3{,}072^3$.} in a $(500\,\hiMpc)^3$ comoving box, evolved from $z_\mathrm{ini}=49$ to $z=0$ in a flat $\Lambda$CDM background with $\Omega_\mathrm{m,0}=0.3$ and $\Omega_{\Lambda,0}=0.7$. The number of PM mesh points for both GINKAKU and L-GADGET2 is set to twice the number of particles per dimension.

\begin{figure*}[htbp]
 \centering
 \includegraphics[width=0.8\linewidth]{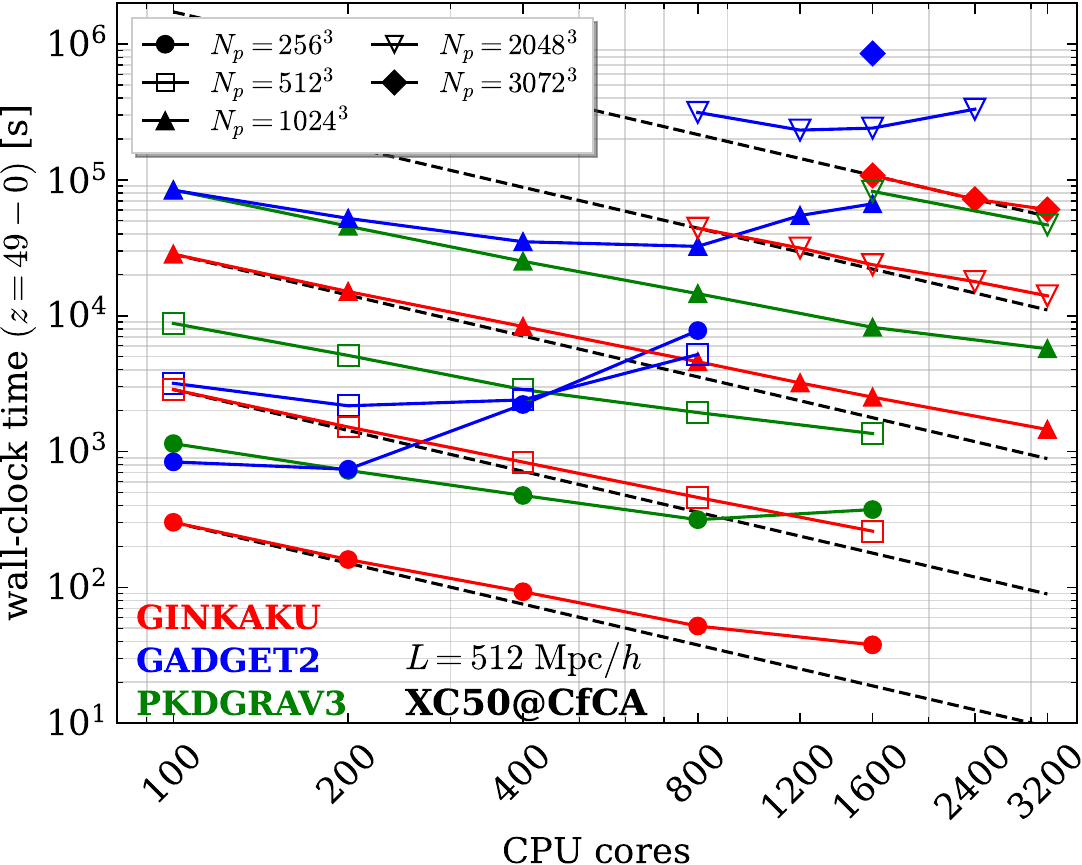}
 \caption{The wall-clock time of simulations from $z=49$ to $0$ as a function of number of CPU cores. Dashed lines show the ideal scaling for each size of the problem. Note that we show three, one, and zero measurements for the largest problem size of $3{,}072^3$ particles (diamonds) for GINKAKU, L-GADGET2 and PKDGRAV3, respectively.
 {Alt text: Plot of total wall-clock time against the number of central-processing-unit cores for three simulation codes at five problem sizes. Each code follows curves that roughly track the dashed ideal-scaling lines, with Ginkaku closest to the ideal across all sizes.}}
 \label{fig:scaling}
\end{figure*}

Figure~\ref{fig:scaling} reports the total wall-clock time as a function of the number of CPU cores. The three codes scale noticeably differently. GINKAKU performs the best across the resolutions tested: at fixed $N_\mathrm{p}$, its wall-clock time decreases with increasing core count along curves that closely track the dashed ideal-scaling lines. Quantitatively, GINKAKU achieves a strong-scaling efficiency of $89.2\%$ when going from $1{,}600$ to $3{,}200$ cores at $N_\mathrm{p}=3{,}072^3$, and a weak-scaling efficiency of $60.0\%$ when going from $400$ cores at $N_\mathrm{p}=1{,}024^3$ to $3{,}200$ cores at $N_\mathrm{p}=2{,}048^3$.

\subsection{Large-$N$ scaling on Fugaku}
\label{subsec:scaling_fugaku}

\begin{figure}[ht]
 \centering
 \includegraphics[width=\linewidth]{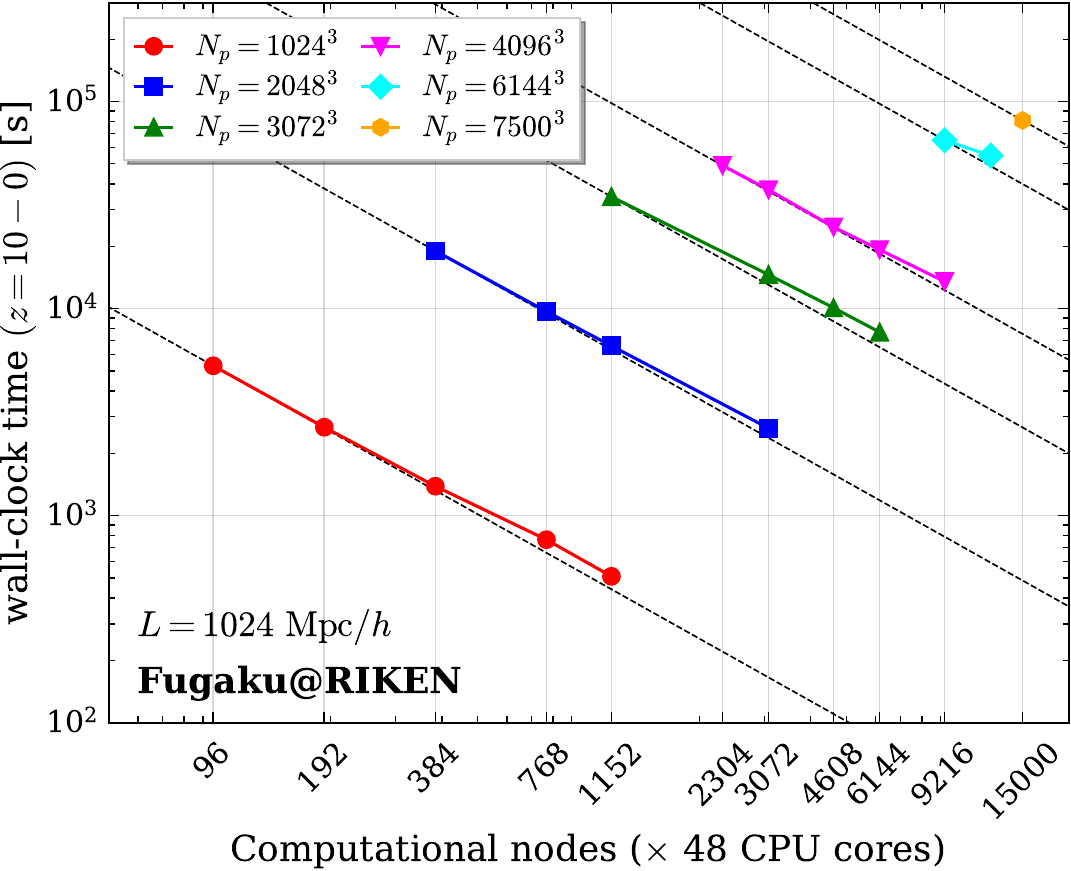}
 \caption{The wall-clock time of simulations from $z=10$ to $0$ as a function of the computational nodes of Fugaku. Dashed lines show the ideal scaling for each problem size. Only the $N_\mathrm{p}=7{,}500^3$ point is predicted from the computation time up to $z=0.5$.
 {Alt text: Plot of wall-clock time against the number of Fugaku compute nodes for several problem sizes. The measured timings closely track the dashed ideal-scaling lines, demonstrating that Ginkaku preserves its good scaling at much larger machine sizes.}}
\label{fig:large_num_scale}
\end{figure}

We next probe the large-$N$ regime on the Fugaku supercomputer of the RIKEN Center for Computational Science, whose nodes are based on the ARM~AArch64 architecture. The setup is otherwise identical to that of Appendix~\ref{subsec:scaling_xc50}, except that the simulations start at $z_\mathrm{ini}=10$ instead of $49$ and the PM mesh has the same number of points per dimension as the particles rather than twice as many. Figure~\ref{fig:large_num_scale} reports the wall-clock time as a function of the number of Fugaku nodes used, for problem sizes up to $N_\mathrm{p}=7{,}500^3$. Across the full range of node counts and problem sizes shown, the measured timings closely track the dashed ideal-scaling lines, indicating that the good scaling demonstrated on the Cray~XC50 in Appendix~\ref{subsec:scaling_xc50} is preserved when moving to the much larger Fugaku machine. As a concrete benchmark of the absolute performance, the $N_\mathrm{p}=6{,}144^3$ run in a $(500\,\hiMpc)^3$ box completed in less than a day on $6\%$ of the $158{,}976$ Fugaku nodes, demonstrating that GINKAKU can reach particle counts well in excess of the DQ2 production runs ($3{,}000^3$) on a small fraction of the machine.

\section{On the approximate neutrino treatment}
\label{sec:nu_approx}
In this paper, we employ the linear response method to include the effect of massive neutrinos. In our previous work (DQ1: \cite{2019ApJ...884...29N}), we instead treated this only at the level of the linear density transfer function of total matter perturbations. That is, we compute the transfer function at $z=0$ that includes the effect of massive neutrinos and then scale it to the starting redshift of the simulations by multiplying the scale-independent linear growth factor computed without taking into account massive neutrinos (i.e., the total matter backscaling method). $N$-body simulations are then performed without any distinction between CDM, baryon, and massive neutrinos: The simulation particles are considered to represent the density of \textit{total matter}. Although this method is guaranteed to give the correct linear power spectrum at $z=0$, we expect that the results at higher redshifts could be biased because of neglected scale-dependent growth. In DQ1, we fixed the sum of the neutrino masses to be $0.06\,\mathrm{eV}$, so its impact would be marginal. This was studied in Appendix~D of the DQ1 paper, but only at the linear level. Here, we test the potential impact of this effect more quantitatively at the nonlinear level.

\begin{figure}[t]
 \centering
 \includegraphics[width=\linewidth]{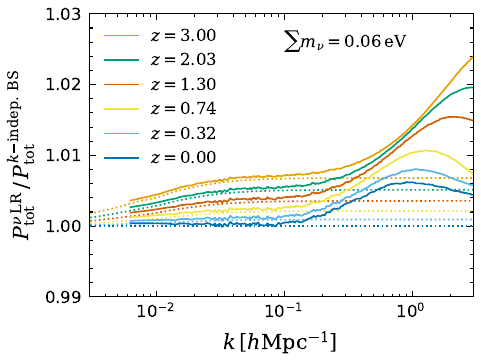}
 \caption{Impact of neglecting the scale-dependent growth due to massive neutrinos on the total matter power spectrum. We show the ratio of the spectrum with our linear response method to that with the backscaling method adopted in our previous study. The linear theory predictions and the simulation results are shown by the dotted and solid lines, respectively, at six redshifts ($z = 3$, $2.03$, $1.30$, $0.74$, $0.32$, and $0$) as indicated by the figure legend. {Alt text: Plot of the matter power spectrum ratio of the linear-response method to the backscaling method, against wavenumber. Solid lines show simulations and dotted lines linear theory at several redshifts. The simulation departure grows on small scales and peaks at mildly nonlinear wavenumbers.} \label{fig:nu_approx}}
\end{figure}

We perform a simulation following the backscaling method employed in DQ1 for the fiducial cosmological model presented in the main text with $3,000^3$ particles in a $1\,\hiGpc$ box. We match the random phases to one of the realizations to suppress the sample variance when the ratio of the power spectra of the matter density fields from the two methods is taken. Figure~\ref{fig:nu_approx} shows the results of the simulations (solid lines), which can be compared with the predictions of linear theory (dotted lines). As explained, the linear theory prediction is identical to unity at $z=0$ and gradually departs from unity as the redshift increases. The departure evaluated by linear theory is always below $1\%$, up to $z=3$ for the range of scales shown here. Although the simulation results basically follow the linear theory curves on large scales ($k\lesssim 0.1\,\hMpci$), the deviation from unity is enhanced on smaller scales due to nonlinear corrections. Quantitatively, the mismatch is peaked in $k$ around $k \sim 1\,\hMpci$ and decays monotonically with cosmic time: it is largest at the highest redshift shown ($z=3$) and remains at the $\sim 1\%$ level at $z \simeq 0.74$, a redshift representative of where the lensing kernel of typical Stage~IV weak-lensing surveys peaks. At $z=0$, where the backscaling method matches the proper linear-response scheme exactly at linear order by construction, a residual $\sim 0.6\%$ mismatch persists from nonlinear mode coupling. Even for the smallest sum of neutrino masses currently allowed by ground experiments, $\sum m_\nu = 0.06\,\mathrm{eV}$, these $\sim 1\%$-level biases at weak-lensing scales could be a non-negligible systematic at the precision targets of ongoing and upcoming Stage~IV surveys.

\section{Ultimate convergence among the codes}
\label{sec:ultimate}
We have made a comparison between different simulation codes in section~\ref{subsec:codes}, where we observe small but non-negligible discrepancies in terms of the matter power spectrum. We further investigate the reason for this in this Appendix.

We first investigate the difference between L-GADGET2 and GINKAKU. Both codes are based on the TreePM method and employ a similar parameterization for the accuracy setting. The fiducial setting for GINKAKU is indeed determined to roughly reproduce the L-GADGET2 runs with the parameters employed in DQ1.

\begin{figure}[htbp]
 \centering
 \includegraphics[width=\linewidth]{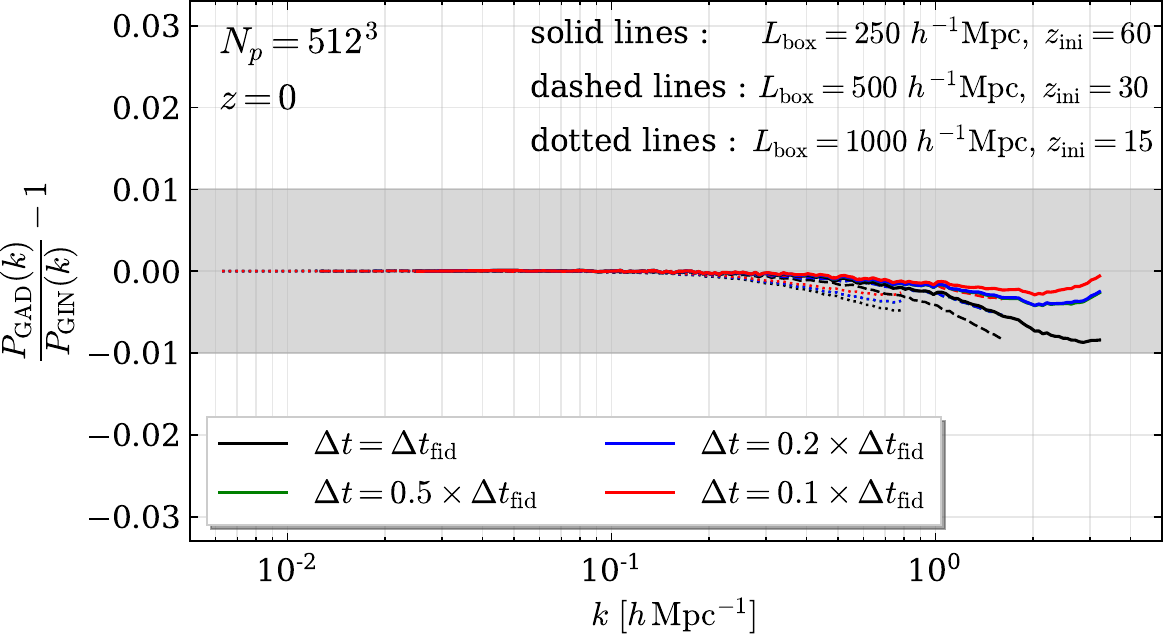}
 \caption{Convergence of the cb power spectrum from L-GADGET2 (with the DQ1 default setting) toward GINKAKU as the time-step parameter is tightened, shown for three resolutions and several time-step settings. {Alt text: Plot of the fractional power-spectrum difference between L-Gadget2 and Ginkaku against wavenumber, for three resolutions and several time-step settings. The differences shrink as the time-step parameter is tightened, indicating that the time-step scheme drives the residual code-to-code mismatch.} \label{fig:comp_fid_ref}}
\end{figure}

One notable difference between the two is in the timestepping: While L-GADGET2 uses individual timesteps for the simulation particles, GINKAKU employs a global timestep. We expect that this difference becomes subdominant when we employ more stringent timestep criteria. We test this by modifying L-GADGET2 to introduce a parameter corresponding to $\eta_\mathrm{global}$ in GINKAKU, which scales the Tree and PM timesteps. We show the fractional difference in power spectra between the two codes in figure~\ref{fig:comp_fid_ref}, where we adopt the same value of $\eta_\mathrm{global}$ for both codes and gradually decrease it. We see here that the fractional difference tends to decay with decreasing $\eta_\mathrm{global}$ for the three resolutions considered here. Therefore, the difference between the two codes originates from the timestep schemes.

\begin{figure*}[htbp]
 \centering
 \includegraphics[width=\linewidth]{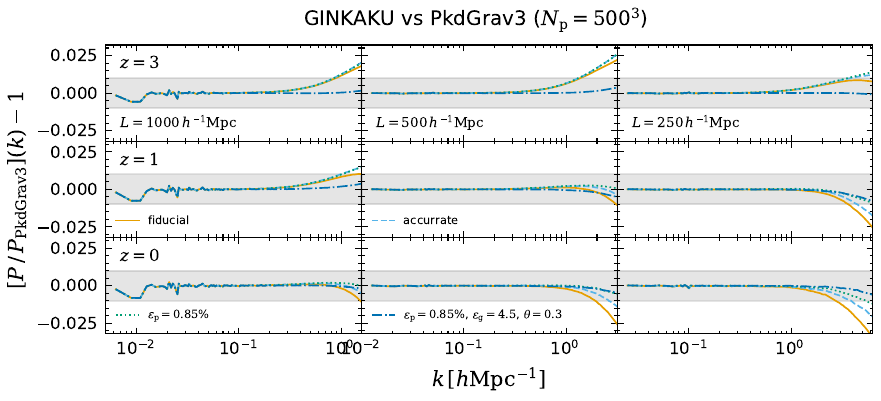}
 \caption{Code-to-code comparison between GINKAKU and PKDGRAV3. Each panel plots the fractional difference $P_GINKAKU/P_PKDGRAV3 - 1$ as a function of wavenumber. Rows correspond to $z=3$, $1$, and $0$ from top to bottom, and columns to $L_\mathrm{box}=1{,}000$, $500$, and $250\,\hiMpc$ from left to right. The four line styles in each panel show GINKAKU runs at progressively tighter accuracy settings, while the PKDGRAV3 reference always adopts its recommended accuracy parameters: the fiducial DQ2 setting (solid; Table~\ref{tab:ginkaku_models}), the accurate setting (dashed), a variant with smaller softening and reduced tree time step but otherwise fiducial parameters (dotted), and the most stringent variant which additionally tightens the tree opening angle to $\theta=0.3$ and shifts the force splitting to $\epsilon_\mathrm{g}=4.5$ (dot-dashed). {Alt text: Grid of fractional-difference panels comparing Ginkaku to Pkdgrav3 against wavenumber, with rows for three redshifts and columns for three box sizes. Four line styles show progressively tighter Ginkaku accuracy settings, with the most stringent variant showing a clear convergence trend toward Pkdgrav3 across the plotted range.} \label{fig:gin_vs_pkdgrav3}}
\end{figure*}

Figure~\ref{fig:gin_vs_pkdgrav3} compares the matter power spectrum from GINKAKU simulations against that from PKDGRAV3 run with its recommended accuracy parameters, using shared initial conditions and the three box sizes considered in the main text. We progressively tighten GINKAKU's accuracy parameters along the four configurations described in the figure caption, while keeping PKDGRAV3 fixed. The trend across the four configurations is the same in every panel: as the GINKAKU setting is tightened, the residual fractional difference with PKDGRAV3 shrinks toward zero. The most demanding variant (dot-dashed) --- in which the softening is reduced, the tree time step is shortened, the tree opening angle is tightened to $\theta=0.3$, and the force splitting is shifted to give the tree a wider dynamic range relative to the PM ($\epsilon_\mathrm{g}=4.5$) --- converges toward PKDGRAV3 at the sub-percent level across the bulk of the plotted range in all three redshifts and three box sizes.

We interpret this as evidence that the residual code-to-code differences identified in section~\ref{subsec:codes} between GINKAKU at its production-grade fiducial setting and PKDGRAV3 are not symptomatic of an underlying methodological disagreement between the TreePM and FMM approaches, but rather reflect the fact that the fiducial GINKAKU setting trades a small amount of accuracy at the deepest nonlinear scales for substantially lower wall-clock cost. With sufficiently strict accuracy parameters --- in particular a tighter tree opening angle and a force-splitting that places more of the dynamic range in the tree, mirroring what PKDGRAV3 effectively does as a predominantly tree-driven FMM code --- GINKAKU converges onto the PKDGRAV3 answer to well within $1\%$ across the scales and redshifts considered here.

\section{Mass definition}
\label{sec:mass_def}

\begin{figure*}[htbp]
 \centering
 \includegraphics[width=\linewidth]{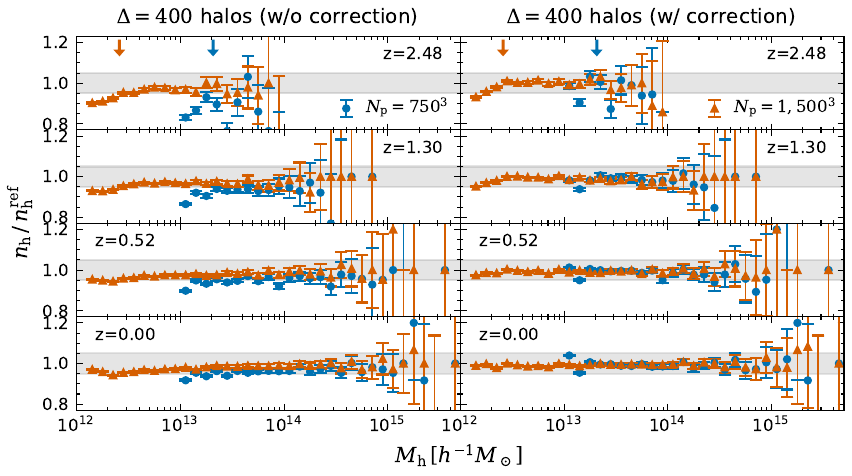}
 \caption{Same as figure~\ref{fig:HMF_th200}, but with $\Delta = 400$. {Alt text: Two-panel halo mass function ratio comparison for three resolutions, in the same format as the main-text figure for an overdensity threshold of two hundred but here for a threshold of four hundred. The empirical correction in the right panel brings all resolutions close to the five-percent band.} \label{fig:HMF_th400}}
\end{figure*}

\begin{figure*}[t]
 \centering
 \includegraphics[width=\linewidth]{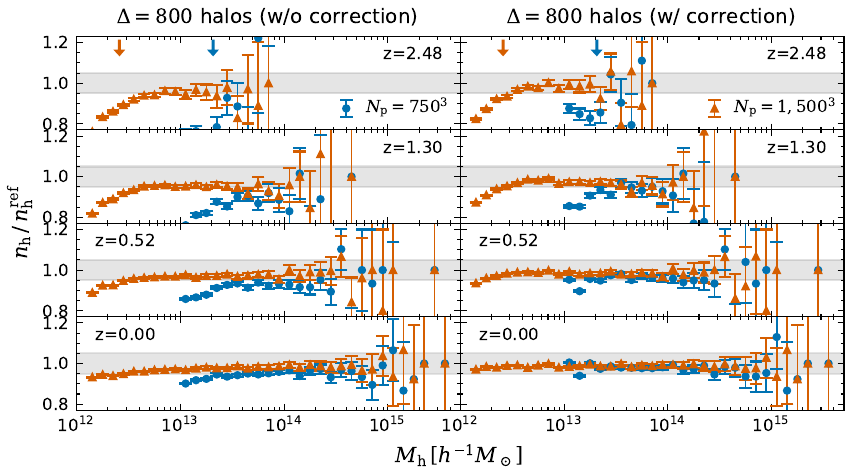}
 \caption{Same as figure~\ref{fig:HMF_th200}, but with $\Delta = 800$. {Alt text: Two-panel halo mass function ratio comparison for three resolutions, in the same format as for a threshold of two hundred but here for a threshold of eight hundred. Residuals exceed the five-percent band only at high redshift and the lowest mass bins.} \label{fig:HMF_th800}}
\end{figure*}

We have focused on halos with mass defined by the enclosed mass within the sphere of interior overdensity $\Delta=200$ with respect to the cosmic mean density. We show the halo mass function for different mass definitions in this Appendix using our fiducial cosmology. Figures~\ref{fig:HMF_th400} and \ref{fig:HMF_th800} show the convergence of the mass function against the mass resolution of the simulations for $\Delta=400$ and $\Delta=800$, respectively, in the same format as figure~\ref{fig:HMF_th200} for $\Delta=200$ in the main text: the left panels use the raw spherical-overdensity masses while the right panels apply the empirical mass correction in equation~(\ref{eq:mass_correction2}), and the shaded band in each subplot marks the target $\pm 5\%$ accuracy after correction.

For $\Delta=400$ (see figure~\ref{fig:HMF_th400}), the corrected mass function from the $1{,}500^3$ run shows a mild downward trend with decreasing halo mass below the $\sim 100$-particle threshold (vertical arrow at $M_\mathrm{h}\sim 3\times 10^{12}\,h^{-1}M_\odot$), which crosses the $\pm 5\%$ boundary in the lowest mass bin only at $z=2.48$. At $z=1.30$ the lowest mass bin sits just inside the band, while at $z=0.52$ and $z=0$ the residual is at most $\sim 2\%$ across the resolved mass range and shows no systematic trend with halo mass. The $750^3$ measurements develop a much larger scatter at the lowest masses and are inconclusive on this trend.

For $\Delta=800$ (see figure~\ref{fig:HMF_th800}) the redshift dependence of the residual is clearer: at $z=0$ the corrected $1{,}500^3$ measurements are essentially indistinguishable from the highest-resolution reference; at $z=0.52$ a weak downward trend appears toward the lowest bin, which falls marginally outside the band; at $z=1.30$ the three lowest bins are outside the band; and at $z=2.48$ the four lowest bins are outside, with the fourth bin corresponding roughly to the $100$-particle threshold. The empirical mass correction therefore preserves $\pm 5\%$ convergence well across the bulk of the resolved mass range at $\Delta=400$, and degrades only modestly at $\Delta=800$ when one combines a high overdensity definition, a marginal particle count, and a high redshift.

\begin{figure}[t]
 \centering
 \includegraphics[width=\linewidth]{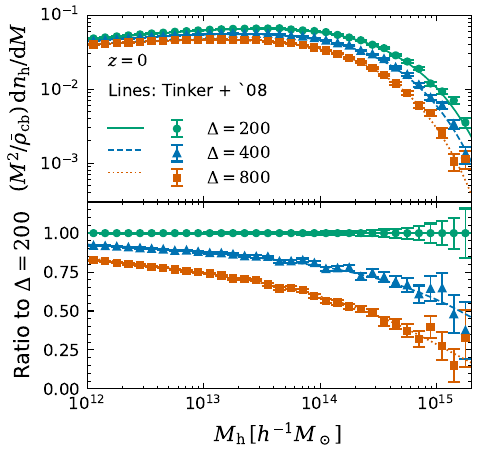}
 \caption{Halo mass function for different mass definitions, measured from the highest-resolution simulation after the mass correction in equation~(\ref{eq:mass_correction2}) is applied. Circles, triangles, and squares correspond to $\Delta = 200$, $400$, and $800$, respectively. The upper panel shows the mass function, and the lower panel shows the ratio to the $\Delta=200$ case. The Tinker08 fitting formula~\citep{Tinker08} is overplotted using the linear mass variance of the cb density. {Alt text: Two-panel halo mass function for several overdensity thresholds. The upper panel shows the mass function and the lower panel its ratio to the threshold of two hundred. Simulation measurements and the Tinker formula agree consistently across the resolved mass range.} \label{fig:HMF_massdef}}
\end{figure}

Next, we compare the measurements of the halo mass function from the highest-resolution simulation (after applying the mass correction) with the fitting formula by \citet{Tinker08}, which models the dependence of the halo mass function on the overdensity threshold. To compute the fitting formula we use the linear mass variance of the cb density rather than the total matter density, as suggested, e.g.\ by \citet{Ichiki12,Castorina14}: on cluster scales the massive-neutrino contribution to the variance is negligible compared with that of the cb fluid that drives halo formation, so it is the cb variance that controls halo abundance. Figure~\ref{fig:HMF_massdef} shows the mass function in the upper panel and its ratio to the $\Delta=200$ case in the lower panel. The simulation ratio is a monotonically decreasing function of halo mass: at $M_\mathrm{h}=10^{15}\,h^{-1}M_\odot$ the mass function for $\Delta=400$ ($800$) is suppressed by a factor of $\sim 2$ ($\sim 4$) relative to $\Delta=200$. The Tinker08 prediction reproduces this ratio consistently within the statistical error of our measurements across the entire mass range: the agreement is at the sub-percent level at the low-mass end ($M_\mathrm{h}\sim 10^{12}\,h^{-1}M_\odot$, where the error bars are themselves below $1\%$) and remains within the $\sim 15\%$ error bars at the high-mass end ($M_\mathrm{h}\sim 10^{15}\,h^{-1}M_\odot$). The standard Tinker08 prescription with cb linear variance is therefore consistent with our measurements across the overdensity-threshold range tested here.

\section{Implementation of Super Survey Modes}
\label{sec:SSM}
One can incorporate the effect of an over- or underdensity with a wavelength larger than the simulation box size --- a so-called super-survey mode (SSM) --- by employing the separate universe (SU) ansatz~(e.g., \cite{Sirko2005,Wagner15a,2016PhRvD..93f3507L}). Our implementation differs from the most common one in the literature in that we do not recast the SU dynamics as a rescaling of the cosmological parameters of the host model. Instead, we directly feed GINKAKU a tabulated expansion history that already encodes the SU evolution of the local patch, exploiting the fact that the kick and drift operators of the code (Eqs.~\ref{eq:kick} and \ref{eq:drift}) accept an arbitrary $\mathcal{H}(\tau)$ as input by construction.

The standard $\Lambda$CDM implementation exploits the fact that the dynamics of a local patch embedded in a uniform overdensity $\deltab$ are equivalent to those of a region of mean density in a slightly different $\Lambda$CDM cosmology. The most observationally relevant case starts from a flat geometry: a positive (negative) $\deltab$ is then reinterpreted as a closed (open) universe with appropriately rescaled $\Omega_\mathrm{m}$ and $H_0$. This duality allows one to use a standard simulation code without any modification, by simply feeding it the rescaled cosmological parameters together with an appropriately rescaled initial condition. The outputs --- snapshot scale factors and length scales --- must then be interpreted using the relation between the \textit{global} and the \textit{local} expansion histories.

The duality used in the standard implementation is special to flat $\Lambda$CDM, however, and does not in general extend to other dark-energy models. As shown by \citet{HuChiang16}, when the dark-energy fluid carries a finite Jeans scale --- as in $w$CDM with constant $w$ and effective sound speed $c_\mathrm{s}$ --- the local Friedmann equation in the SU patch picks up a non-adiabatic component that has no counterpart in the global cosmology, so that the local expansion history can no longer be reproduced by simply rescaling the global parameters within the same family. Our approach sidesteps this restriction: rather than mapping the SU patch onto a different model in the same family, we solve directly for the local expansion history $\mathcal{H}_W(\tau)$ dictated by the SU dynamics and pass it to the integrator as a numerical lookup table. The same strategy --- modifying the integrator's background expansion rather than rescaling cosmological parameters --- has been adopted by \citet{Akitsu_2021} in the closely related context of anisotropic SU simulations, where the modification is implemented inside GADGET; in our case the lookup-table interface to $\mathcal{H}(\tau)$ is built into GINKAKU from the outset (in section~\ref{subsec:nbody}), so no code modification is required to switch between the global and local expansion. We explain below our implementation of the SSM in more detail.

First, the key equation of the SSM valid for any cosmological model under the SU ansatz is that
\be
a_W = (1+\deltab)^{-1/3}a,\label{eq:aW}
\ee
where and hereafter, variables with a subscript $W$ indicate that the quantity is evaluated within the local (or, equivalently, the simulation) volume, which can be different from that for the global one. This equation simply states the mass conservation: a local volume embedded in an overdense region ($\deltab>0$) expands slower than the global average. The quantity $\deltab$, the background overdensity in terms of the cb fluid, is assumed to be in the growing mode, and its time evolution is governed by the spherical collapse model:
\be
\deltab'' +\left[1+\bigl(\ln\mathcal{H}(\tau)\bigr)'\right]\deltab'-\frac{4}{3}\frac{\deltab'^2}{1+\deltab} = \frac{3}{2}\Omega_\mathrm{cb}(\tau) (1+\deltab)\left[B\ast\deltab\right],\nonumber\\
\label{eq:deltab}
\ee
where the factor $B$ is the boost factor defined in equation~(\ref{eq:boost}), which takes account of the contributions from the perturbations of energy components other than the cb fluid, and the time derivative operator $'$ is with respect to $\tau = \log a$. We assume here that only the cb fluid is conserved in the simulated volume that follows the local background expansion $a_W$, the \textit{local} comoving frame, and the other contributions such as massive neutrinos are not. They simply follow the linear growth in the \textit{global}, instead of local, comoving frame. This is consistent with our implementation of the $N$-body code. Indeed, the above equation for $\deltab$ is identical to the evolution equation for the transfer function $\mathcal{M}_\mathrm{cb}$ in equation~(\ref{eq:Mcb_evo}), when the nonlinear terms are neglected and moving to the Fourier space to rewrite the convolution operator $\ast$ with a product.

We numerically solve equation~(\ref{eq:deltab}) to compute the nonlinear evolution of $\deltab$, and in turn obtain the scale factor, $a_W(t)$, via equation~(\ref{eq:aW}). We can then compute the local expansion rate $\mathcal{H}_W(t) = \dot{a}_W(t)$, which is passed to GINKAKU as a numerical table. This table is used to compute the integrals in the kick and drift operator (Equations~\ref{eq:kick} and \ref{eq:drift}). Physically, this is the friction due to cosmic expansion, which slows down the growth of structure.

To test our implementation, we measure the ``growth-only'' responses from a set of simulations with different values of $\deltabzero$, the value of $\deltab$ linearly extrapolated to the present day. We adopt $\deltabzero = 0, \pm 0.2, \pm 0.4, \pm 0.8$ and $\pm 1.2$, and generate one random realization per each (nine simulations in total). These simulations are initialized in the same Lagrangian volume of $2\,\hiGpc$ with $750^3$ particles with the same random phases. The responses are defined by
\be
G_n(k) = \left.\frac{1}{P_\mathrm{cb}(k)}\frac{\mathrm{d}^n \tilde{P}_\mathrm{cb}(\tilde{k},\deltabzero)}{\mathrm{d}\deltabzero^n}\right|_{\tilde{k}},
\ee
where $\tilde{P}_\mathrm{cb}(\tilde{k},\deltabzero)$ is the nonlinear power spectrum measured in the local comoving frame and evaluated at the local comoving wavenumber $\tilde{k}$. That is, although the physical box sizes of two simulations with different values of $\deltabzero$ differ at a fixed time, we always compare them at the same wavenumber as measured in their respective local comoving frames. With this convention the geometric dilation contribution to the response cancels by construction, leaving only the genuine growth response that we are interested in --- the so-called \textit{growth-only} response defined and measured in this way by \citet{Wagner15a,Wagner15b}. In practice, at each wavenumber $k$ we evaluate the derivatives in the definition above by fitting the dependence of $\tilde{P}_\mathrm{cb}(\tilde{k},\deltabzero)$ on $\deltabzero$ with an eighth-order polynomial across the nine simulations --- a fit that exactly interpolates the nine sampling points --- and reading off $G_n(k)$ as $n!$ times the coefficient of $\deltabzero^n$ in the resulting Taylor expansion about $\deltabzero=0$.

\begin{figure}[t]
 \centering
 \includegraphics[width=\linewidth]{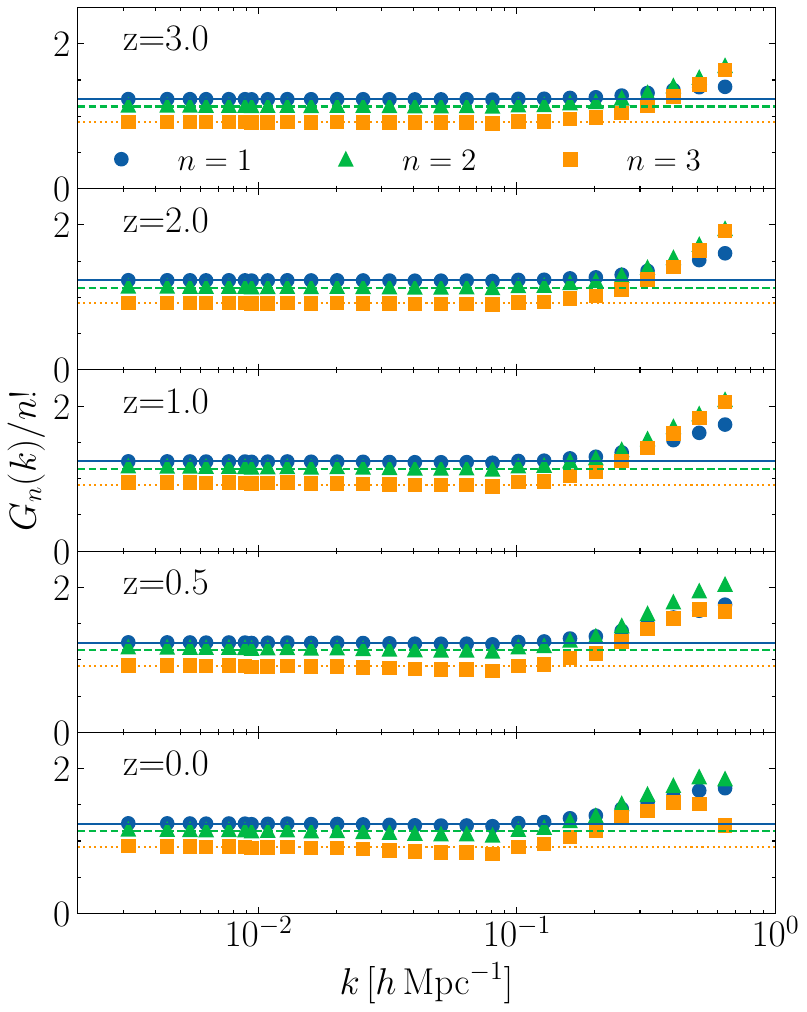}
 \caption{First three \textit{growth-only} response functions $G_n$, scaled by $1/n!$, defined as the $n$-th derivative of the matter power spectrum with respect to the background linear density contrast, measured from SU simulations performed with our code. The five panels correspond to $z = 3$, $2$, $1$, $0.5$, and $0$ from top to bottom; within each panel, $G_n/n!$ for $n = 1$, $2$, and $3$ are overplotted. The horizontal lines indicate the corresponding large-scale linear-theory limits. {Alt text: Five-panel plot of the first three growth-only response functions, scaled by one over n-factorial, against wavenumber, measured from separate-universe simulations. Panels correspond to five redshifts from z equals three at the top to z equals zero at the bottom; within each panel, curves for n equals 1, 2, and 3 approach their large-scale linear-theory limits, indicated by horizontal lines, at the smallest wavenumbers and depart from them at larger wavenumbers due to nonlinear mode coupling.} \label{fig:SU_response}}
\end{figure}

Figure~\ref{fig:SU_response} compares the first three growth-only responses measured from the simulations with their large-scale theoretical expectations, shown as horizontal lines. These limits are the linear-theory predictions for $G_n(k\to 0)$, obtained by perturbatively expanding the modified linear growth factor in powers of $\deltabzero$ at fixed local wavenumber, and are scale-independent in the linear regime. For an Einstein--de Sitter background, and to high accuracy in $\Lambda$CDM, \citet[their equation~(2.15)]{Wagner15b} give them in closed form as
\be
G_1^\mathrm{lin} = \frac{26}{21},\quad
G_2^\mathrm{lin} = \frac{3002}{1323},\quad
G_3^\mathrm{lin} = \frac{240272}{43659},
\ee
i.e., approximately $1.238$, $2.269$, and $5.504$ for $n=1$, $2$, and $3$, respectively. The simulation measurements approach these limits at the smallest wavenumbers probed and depart from them on smaller scales as nonlinear mode coupling injects additional $\deltab$ dependence into $\tilde{P}_\mathrm{cb}$. The good agreement at low $k$ for all three orders confirms that the SU evolution is being correctly propagated by our integrator when the local expansion history is supplied as a lookup table, and validates the tabulated-expansion implementation as an alternative to cosmology-rescaling SU schemes for cosmologies in which the latter is not directly available.

\bibliographystyle{apj}
\bibliography{lssref}

\end{document}